\documentclass[11pt]{article}
\usepackage{amsmath,amssymb}
\usepackage[usenames]{color}
\usepackage{pstricks}
\numberwithin{equation}{section}

\newcommand{\nc}{\newcommand}
\def\vvdots{\mathinner{\mkern1mu\raise1pt\vbox{\kern7pt\hbox{.}}\mkern2mu
  \raise4pt\hbox{.}\mkern2mu\raise7pt\hbox{.}\mkern1mu}}
\nc{\fh}{\hat{f}}
\nc{\muh}{\hat{\mu}}
\nc{\nuh}{\hat{\nu}}
\nc{\bib}{\bibitem}
\nc{\al}{\alpha}
\nc{\g}{\gamma}
\nc{\G}{\Gamma}
\nc{\D}{\Delta}
\nc{\eps}{\epsilon}
\nc{\la}{\lambda}
\nc{\La}{\Lambda}
\nc{\var}{\varphi}
\nc{\pa}{\partial}
\nc{\nn}{\nonumber \\ }
\nc{\hf}{\frac{1}{2}}
\nc{\dz}{\frac{dz}{2\pi i}}
\nc{\bin}[2]{\left(\!\!\!\begin{array}{c} {#1}\\ {#2} \end{array}\!\!\!\right)}
\nc{\be}{\begin{equation}}
\nc{\ee}{\end{equation}}
\nc{\bea}{\begin{eqnarray}}
\nc{\eea}{\end{eqnarray}}
\nc{\bra}[1]{\langle {#1}|}
\nc{\ket}[1]{|{#1}\rangle}
\nc{\ketw}[1]{({#1})^{\phantom{a}}_{{\cal W}}}
\nc{\chit}{\raisebox{0.25ex}{$\chi$}}
\nc{\chih}{\raisebox{0.25ex}{$\hat\chi$}}
\nc{\Db}{\mbox{\boldmath $D$}}
\nc{\Hb}{\mbox{\boldmath $H$}}
\nc{\Hc}{\mathcal{H}}
\nc{\Lc}{\mathcal{L}}
\nc{\Nc}{\mathcal{N}}
\nc{\Pc}{\mathcal{P}}
\nc{\Rc}{\mathcal{R}}
\nc{\Vc}{\mathcal{V}}
\nc{\Ib}{\mbox{\boldmath $I$}}
\nc{\qb}{\bar{q}}
\nc{\Bh}{\hat{B}}
\nc{\Gh}{\hat{G}}
\nc{\gh}{\hat{g}}
\nc{\hh}{\hat{h}}
\nc{\Jh}{\hat{J}}
\nc{\Nh}{\hat{N}}
\nc{\Nch}{\hat{\mathcal{N}}}
\nc{\Ph}{\hat{P}}
\nc{\Pch}{\hat{\mathcal{P}}}
\nc{\Qh}{\hat{Q}}
\nc{\Xh}{\hat{X}}
\nc{\Yh}{\hat{Y}}
\nc{\vh}{\hat{v}}
\nc{\Wh}{\hat{W}}
\nc{\It}{\tilde{I}}
\nc{\Jt}{\tilde{J}}
\nc{\Qt}{\tilde{Q}}
\nc{\vt}{\tilde{v}}
\nc{\wt}{\tilde{w}}
\nc{\vb}{\mathbf{v}}
\nc{\Ac}{\mathcal{A}}
\nc{\Bc}{\mathcal{B}}
\nc{\Cc}{\mathcal{C}}
\nc{\Dc}{\mathcal{D}}
\nc{\Ec}{\mathcal{E}}
\nc{\Fc}{\mathcal{F}}
\nc{\Ic}{\mathcal{I}}
\nc{\Jc}{\mathcal{J}}
\nc{\Oc}{\mathcal{O}}
\nc{\Qc}{\mathcal{Q}}
\nc{\Sc}{\mathcal{S}}
\nc{\Tc}{\mathcal{T}}
\nc{\Wc}{\mathcal{W}}
\nc{\Xc}{\mathcal{X}}
\nc{\Yc}{\mathcal{Y}}
\nc{\Zc}{\mathcal{Z}}
\nc{\fus}{\mbox{}\,\hat\otimes\,\mbox{}}
\nc{\Sct}{\tilde{\mathcal{S}}}
\nc{\Ih}{\hat{I}}
\nc{\ch}{{\rm ch}}
\nc{\R}{{\cal R}}
\nc{\dkk}{\delta_{j,\{k,k'\}}^{(2)}}
\nc{\ddkk}{\delta_{j,\{k,k'\}}^{(4)}}
\nc{\dddkk}{\delta_{j,\{k,k'\}}^{(8)}}
\nc{\dnn}{\delta_{j,\{n,n'\}}^{(2)}}
\nc{\ddnn}{\delta_{j,\{n,n'\}}^{(4)}}
\nc{\dddnn}{\delta_{j,\{n,n'\}}^{(8)}}
\def\vvdots{\mathinner{\mkern1mu\raise1pt\vbox{\kern7pt\hbox{.}}\mkern2mu
  \raise4pt\hbox{.}\mkern2mu\raise7pt\hbox{.}\mkern1mu}}
\nc{\gauss}[2]{\left[\!\!\begin{array}{c} {#1}\\ {#2} \end{array}\!\!\right]}
\nc{\sbin}[2]{\left\{\!\!\!\begin{array}{c} {#1}\\ {#2} 
\end{array}\!\!\!\right\}}
\nc{\sbinlr}[2]{\Big\langle\!\!\begin{array}{c} {#1}\\ {#2} 
\end{array}\!\!\Big\rangle}
\nc{\bino}[2]{\left(\!\!\begin{array}{c} {#1}\\ {#2} \end{array}\!\!\right)}

\definecolor{lightblue}{rgb}{.61,.61,1}
\definecolor{midblue}{rgb}{.7,.7,1}
\definecolor{lightlightblue}{rgb}{.85,.85,1}
\definecolor{lightestblue}{rgb}{.96,.96,1}

\begin{document}

\topmargin -5mm
\oddsidemargin 5mm

\setcounter{page}{1}

\mbox{}\vspace{-16mm}
\thispagestyle{empty}

\begin{center}
{\huge {\bf Graph fusion algebras of ${\cal WLM}(p,p')$}}

\vspace{7mm}
{\Large J{\o}rgen Rasmussen}
\\[.3cm]
{\em Department of Mathematics and Statistics, University of Melbourne}\\
{\em Parkville, Victoria 3010, Australia}\\[.4cm]
{\tt J.Rasmussen\!\!\ @\!\!\ ms.unimelb.edu.au}

\end{center}

\vspace{8mm}
\centerline{{\bf{Abstract}}}
\vskip.4cm
\noindent
We consider the ${\cal W}$-extended logarithmic minimal model ${\cal WLM}(p,p')$. 
As in the rational minimal models, the so-called fundamental fusion algebra of ${\cal WLM}(p,p')$
is described by a simple graph fusion algebra. 
The fusion matrices in the regular representation thereof are mutually commuting, but in
general not diagonalizable. Nevertheless, we show that they can be brought simultaneously to
block-diagonal forms whose blocks are upper-triangular matrices of dimension 1, 3, 5 or 9.
The directed graphs associated with the two fundamental modules are described in detail.
The corresponding adjacency matrices share a complete set of common generalized 
eigenvectors organized as a web constructed by interlacing the Jordan chains of the two matrices. 
This web is here called a Jordan web and it consists of connected 
subwebs with 1, 3, 5 or 9 generalized eigenvectors.
The similarity matrix, formed by concatenating these vectors, simultaneously brings
the two fundamental adjacency matrices to Jordan canonical form modulo permutation similarity. 
The ranks of the participating Jordan blocks are 1 or 3, and the corresponding eigenvalues 
are given by $2\cos \frac{j\pi}{\rho}$ where $j=0,\ldots,\rho$ and $\rho=p,p'$. 
For $p>1$, only some of the modules in the fundamental fusion algebra of ${\cal WLM}(p,p')$ 
are associated with boundary conditions within our lattice approach. 
The regular representation of the corresponding fusion subalgebra has features similar to the
ones in the regular representation of the fundamental fusion algebra, but with dimensions 
of the upper-triangular blocks and connected Jordan-web components given by 1, 2, 3 or 8. 
Some of the key results are illustrated for ${\cal W}$-extended critical percolation 
${\cal WLM}(2,3)$.
\newpage
\renewcommand{\thefootnote}{\arabic{footnote}}
\setcounter{footnote}{0}

\tableofcontents

\newpage

\section{Introduction}
\label{SecIntro}

A central question of much current interest is whether an extended symmetry algebra 
${\cal W}$~\cite{Zam85,BS9210} exists for logarithmic conformal 
field theories~\cite{Gur9303,Flo0111,Gab0111,Kaw0204} like the logarithmic minimal models 
${\cal LM}(p,p')$~\cite{PRZ0607,RP0706,RP0707}. These models contain a countably 
{\em infinite} number of inequivalent Virasoro modules which the extended symmetry should 
reorganize into a {\em finite} number of ${\cal W}$-extended modules closing under fusion. 
In the case of the logarithmic minimal models ${\cal LM}(1,p')$, the existence and properties of 
such an extended ${\cal W}$-symmetry, including the associated fusion rules, are by now largely 
understood~\cite{GK9606,FHST0306,FGST0504,CF0508,GR0707,AM0707,GT0711,PRR0803}. 
The works~\cite{FGST0606a,FGST0606b} strongly indicate the existence of a  
${\cal W}_{p,p'}$ symmetry algebra for general augmented minimal models, but offer only very 
limited insight into the associated fusion algebras. Recently, a detailed description of these fusion 
algebras has been provided in~\cite{RP0804,Ras0805,Ras0812} generalizing the approach
of~\cite{PRR0803}. Extending ideas originating with 
Cardy~\cite{Cardy86,Cardy89}, this approach uses a strip-lattice implementation of fusion to 
obtain the fusion rules of the entire series of logarithmic minimal models ${\cal LM}(p,p')$ in the 
${\cal W}$-extended picture where they are denoted by ${\cal WLM}(p,p')$. It is stressed,
that the extended picture is described by the {\em same} lattice model as the Virasoro picture. 

Contrary to the situation in the Virasoro picture, for $p>1$, there is no identity 
nor a pair of so-called fundamental modules in the lattice approach to ${\cal WLM}(p,p')$.
In~\cite{Ras0812}, we found that one can supplement the set of indecomposable modules 
associated with boundary conditions by a set of reducible yet indecomposable rank-1 modules.
This algebraically enlarged set was shown to yield a well-defined fusion algebra called the
{\em fundamental fusion algebra}. This algebra is so named since it is 
generated from repeated fusions of the two {\em fundamental modules} 
$\ketw{2,1}$ and $\ketw{1,2}$ in addition to the identity $\ketw{1,1}$ which is now
present for all $p$. It was also found
that the fusion algebra generated by modules associated with boundary conditions 
is an {\em ideal} of the fundamental fusion algebra. Further algebraic extensions exist. 
In particular, for $p>1$, there are additional {\em irreducible} modules not associated with 
boundary conditions. Their fusion properties have been systematically examined only very 
recently~\cite{GRW0905,Ras0906,Wood0907}. Here we restrict ourselves to the 
modules generating the fundamental fusion algebra. 

The fusion matrices of a standard rational conformal field theory are diagonalizable.
This is made manifest by the Verlinde formula~\cite{Ver88} where the diagonalizing
similarity matrix is the modular $S$-matrix of the characters in the theory.
In a logarithmic conformal field theory, on the other hand, there are typically 
more linearly independent representations
than linearly independent characters due to the presence of indecomposable modules
of rank greater than 1. Consequently, there is no Verlinde formula in the usual sense
and the fusion matrices may not all be diagonalizable. This is indeed the situation for the 
${\cal W}$-extended logarithmic minimal models ${\cal WLM}(p,p')$ analyzed in the present work.

In the regular representation of a fusion algebra, the fusion matrices are mutually commuting.
Viewing the fusion matrices as adjacency matrices of graphs, the fusion rules are succinctly 
encoded in these fusion graphs. In this context, the regular representation of a fusion algebra is 
referred to as the {\em graph fusion algebra}. Fusion graphs have been the key to the 
classification of rational conformal field theories on the cylinder~\cite{BPPZ9809,BPPZ9908} 
and on the torus~\cite{Ocn99,Ocn00,PZ0011,PZ0101}. 
In the rational $A$-type theories, the Verlinde algebra yields a diagonal modular
invariant, while $D$- and $E$-type theories are related to non-diagonal
modular invariants. The Ocneanu algebras arise when considering fusion on the
torus, with left and right chiral halves of the theory, and involve Ocneanu graphs.
We refer to~\cite{Kos88,DiFZ90,DiF92,PZ9510} for earlier results on the interrelation
between fusion algebras, graphs and modular invariants.
It is our hope that the present work will be a step in the direction of
extending these fundamental insights to the logarithmic conformal field theories.

As already indicated, the fusion matrices in the regular representation of the fundamental
fusion algebra are mutually commuting, but in general not diagonalizable. 
Nevertheless, we show that they can be brought simultaneously to
block-diagonal forms whose blocks are upper-triangular matrices of dimension 1, 3, 5 or 9.
The directed graphs associated with the two fundamental modules are described in detail.
They consist of a number of connected components of which there are two prototypes.
The adjacency matrices of these {\em tadpole} graphs and {\em eye-patch} graphs
are Jordan decomposed explicitly.
Combining them, the adjacency matrices $X$ and $Y$ of the two fundamental graphs
are found to share a complete set of common generalized 
eigenvectors organized as a web constructed by interlacing the Jordan chains of $X$ and $Y$. 
This web is here called a {\em Jordan web} and it consists of connected 
subwebs with 1, 3, 5 or 9 generalized eigenvectors.
The similarity matrix, formed by concatenating these vectors, {\em simultaneously} brings 
$X$ and $Y$ to Jordan canonical form {\em modulo} permutation similarity. 
For $p>1$, it is simply not possible to properly Jordan decompose them simultaneously.
The ranks of the participating Jordan blocks are 1 or 3, and the corresponding eigenvalues 
are given by $2\cos \frac{j\pi}{\rho}$ where $j=0,\ldots,\rho$ and $\rho=p,p'$. 

For $p=1$, the fundamental fusion graph with adjacency matrix $Y$ is given by a {\em single} 
eye-patch graph and is thus connected. The fundamental fusion matrix $X$ acts
as a conjugation on this eye-patch graph. In contrast to the situation for $p>1$, 
as demonstrated in~\cite{Ras0908}, these simple properties allow for the existence of
a similarity matrix which simultaneously brings all fusion matrices of the fundamental
fusion algebra of ${\cal WLM}(1,p')$ to Jordan form.
The two fundamental fusion matrices, in particular, are both brought to Jordan {\em canonical}
form by this similarity transformation. The present work is an extension of the
paper~\cite{Ras0908} on ${\cal WLM}(1,p')$ to the general series of ${\cal W}$-extended
logarithmic minimal models ${\cal WLM}(p,p')$.

For $p>1$, only some of the modules in the fundamental fusion algebra of ${\cal WLM}(p,p')$ 
are associated with boundary conditions within our lattice approach. The fusion matrices in 
the regular representation of the corresponding fusion subalgebra 
have features similar to the ones for the larger fundamental fusion algebra. From~\cite{Ras0812},
we know that the modules associated with boundary conditions form an ideal of the 
fundamental fusion algebra. Their matrix realizations $\Nh_\mu$ therefore follow from the 
realizations 
$N_\mu$ of the generators of the fundamental fusion algebra by elimination of the rows
and columns corresponding to the modules not associated with boundary conditions.
According to~\cite{Ras0812}, every fusion matrix $N_\mu$
can be written as a polynomial in the fundamental fusion matrices $X$ and $Y$.
Likewise, every fusion matrix $\Nh_\mu$
can be written as a polynomial in the {\em auxiliary fusion matrices} $\Xh$ and $\Yh$
obtained from $X$ and $Y$ by the aforementioned elimination procedure.
For $p>1$, the matrix $\Yh$ does {\em not} correspond to
a module associated with a boundary condition and is, in this sense, auxiliary. For $p>2$,
this applies to both $\Xh$ and $\Yh$.
Despite their auxiliary status, the matrices $\Xh$ and $\Yh$ are very useful
in the description of the spectral decomposition of the fusion matrices $\Nh_\mu$.
We refer to the corresponding directed graphs as {\em auxiliary fusion graphs}.
As in the case of the fundamental fusion graphs, the auxiliary fusion graphs consist of a 
certain number of connected components of which there are two prototypes:
{\em cycle} graphs and the eye-patch graphs above. 
We show that the auxiliary adjacency matrices $\Xh$ and $\Yh$ share a complete set of 
common generalized eigenvectors, and that the corresponding Jordan web consists of connected 
subwebs with 1, 2, 3 or 8 generalized eigenvectors.
We subsequently show that the fusion matrices $\Nh_\mu$ can be brought simultaneously to
block-diagonal forms whose blocks are upper-triangular matrices of dimension 1, 2, 3 or 8.

The remaining part of this paper is organized as follows.
Section~\ref{SecWLM} briefly reviews some basics of ${\cal WLM}(p,p')$ and its fundamental 
fusion algebra. The associated graph fusion algebras are formally introduced and the fusion rules
involving the fundamental modules are summarized. 
The cycle, tadpole and eye-patch graphs are defined in Section~\ref{SecAdjacency},
and the spectral decompositions of their adjacency matrices are worked out in detail.
These results are conveniently expressed in terms of Chebyshev polynomials.
Using the summarized fusion rules just mentioned, in Section~\ref{SecFundAux},
we determine the fundamental and auxiliary fusion graphs as well as their adjacency matrices.
We recall that the connected components of these graphs are of the form discussed 
in Section~\ref{SecAdjacency}.
In Section~\ref{SecSpectral}, we work out the spectral decompositions of the fundamental
and auxiliary fusion matrices. In both cases, we determine a complete set of common 
generalized eigenvectors and describe the corresponding Jordan web and its connected 
components. The Jordan canonical forms of the fundamental and auxiliary fusion matrices
follow readily. Arising as the result of the 
simultaneous similarity transformation of the general fusion matrices, 
we also present explicit expressions for the block-diagonal forms of these matrices.
Section~\ref{SecConclusion} contains some concluding remarks and indications of future
work, while Appendix~\ref{AppJordan} provides elementary examples demonstrating that two 
commuting matrices may not share a complete set of common generalized eigenvectors nor 
necessarily be brought simultaneously to Jordan form. In Appendix~\ref{AppJordanWeb},
the Jordan subwebs formed by the common 
generalized eigenvectors of $X$ and $Y$ are collected in table form with respect to
the corresponding eigenvalues. At various places in the paper, some of the key results 
are illustrated for ${\cal W}$-extended critical percolation ${\cal WLM}(2,3)$.

\section{${\cal W}$-extended logarithmic minimal models}
\label{SecWLM}

A logarithmic minimal model ${\cal LM}(p,p')$ is defined~\cite{PRZ0607,RP0707} 
for every coprime pair of positive integers $p<p'$. The model has central charge
\be
 c\;=\;1-6\frac{(p'-p)^2}{pp'}
\label{c}
\ee
and conformal weights
\be
 \D_{\rho,\sigma}\;=\;\frac{(\rho p'-\sigma p)^{2}-(p'-p)^2}{4pp'},\hspace{1.2cm} 
  \rho,\sigma\in\mathbb{N}
\label{D}
\ee
Its ${\cal W}$-extension ${\cal WLM}(p,p')$ is discussed 
in~\cite{PRR0803,RP0804,Ras0805,Ras0812} and briefly reviewed in the following.
Throughout, we are using the following notation and conventions
\be
 \mathbb{Z}_{n,m}\;=\;\mathbb{Z}\cap[n,m],\qquad
 \eps(n)\;=\;\frac{1-(-1)^n}{2},\qquad
 n\cdot m\;=\;1+\eps(n+m),\qquad n,m\in\mathbb{Z}
\label{eps}
\ee
and
\be
 \kappa,\kappa'\in\mathbb{Z}_{1,2},\qquad
 a\in\mathbb{Z}_{1,p-1},\qquad
 b\in\mathbb{Z}_{1,p'-1},\qquad
 r\in\mathbb{Z}_{1,p},\qquad
 s\in\mathbb{Z}_{1,p'}
\label{kkabrs}
\ee

\subsection{Modules associated with boundary conditions}

The indecomposable modules in ${\cal WLM}(p,p')$, which can be associated with 
Yang-Baxter integrable boundary conditions on the strip lattice and ${\cal W}$-invariant
boundary conditions in the continuum scaling limit,
were identified in~\cite{RP0804,Ras0805} by extending constructions in~\cite{PRR0803}
pertaining to the case $p=1$. The set of these modules is given by
\be
 \big\{ \Wc(\D_{\kappa p,b}),\Wc(\D_{a,\kappa p'}),\Wc(\D_{\kappa p,p'}),
   \ketw{\R_{\kappa p,s}^{a,0}},\ketw{\R_{r,\kappa p'}^{0,b}},\ketw{\R_{\kappa p,p'}^{a,b}}\big\}
\label{JWout}
\ee
and is of cardinality
\be
 6pp'-2p-2p'
\label{6pp}
\ee
Here we have adopted the notation of~\cite{GRW0905} denoting a
${\cal W}$-irreducible module of conformal weight $\D$ by $\Wc(\D)$.
Thus, there are $2p+2p'-2$ irreducible (hence indecomposable rank-1) modules
\be
 \big\{\Wc(\D_{\kappa p,s}),\Wc(\D_{r,\kappa p'})\big\}
\label{r1}
\ee
where the two modules $\Wc(\D_{\kappa p,p})=\Wc(\D_{p,\kappa p'})$ are listed twice,
in addition to $4pp'-2p-2p'$ indecomposable rank-2 modules
\be
 \big\{\ketw{\R_{\kappa p,s}^{a,0}}, \ketw{\R_{r,\kappa p'}^{0,b}}\big\}
\label{r2}
\ee
and $2(p-1)(p'-1)$ indecomposable rank-3 modules
\be
 \big\{\ketw{\R_{\kappa p,\kappa' p'}^{a,b}}\big\}
 \hspace{1.2cm}\mathrm{subject\ to}\ \ \ 
  \ketw{\R_{p,2p'}^{a,b}}\equiv\ketw{\R_{2p,p'}^{a,b}}\quad
   \mathrm{and}\quad \ketw{\R_{2p,2p'}^{a,b}}\equiv\ketw{\R_{p,p'}^{a,b}}
\label{r3}
\ee
The associative and commutative fusion algebra of the modules (\ref{JWout}) was determined
in~\cite{Ras0805,Ras0812}. There is no algebra unit or identity for $p>1$, while, for
$p=1$, the irreducible module $\Wc(\D_{1,1})$ is the identity.

\subsection{Fundamental fusion algebra}
\label{SecFundFus}

In~\cite{Ras0812}, we found that one can supplement the set of indecomposable
modules (\ref{JWout}) by the $(p-1)(p'-1)$ reducible yet indecomposable rank-1 modules 
\be
 \big\{\ketw{a,b}\big\},\qquad\quad \D(\ketw{a,b})\;=\;\D_{a,b}
\label{ab}
\ee
with conjectured embedding patterns given by
\be
 \mbox{
 \begin{picture}(100,60)(-30,0)
    \unitlength=0.8cm
  \thinlines
\put(-3.3,1){$\ketw{a,b}:$}
\put(-0.7,1.9){$\ketw{\D_{2p-a,b}}$}
\put(2.7,0){$\ketw{\D_{a,b}}$}
\put(2.6,0.6){\vector(-4,3){1.2}}
\put(5,1){$=$}
 \end{picture}
}
\hspace{2cm}
 \mbox{
 \begin{picture}(100,60)(-30,0)
    \unitlength=0.8cm
  \thinlines
\put(-0.7,1.9){$\ketw{\D_{a,2p'-b}}$}
\put(2.7,0){$\ketw{\D_{a,b}}$}
\put(2.6,0.6){\vector(-4,3){1.2}}
 \end{picture}
}
\label{abemb}
\ee
Their characters read
\be
 \chit[\ketw{a,b}](q)\;=\;\frac{1}{\eta(q)}\sum_{k\in\mathbb{Z}}(k^2-1)
    \Big(q^{(ap'+bp+2kpp')^2/4pp'}-q^{(ap'-bp+2kpp')^2/4pp'}\Big)
\label{abchar}
\ee
where $\eta(z)$ is the Dedekind eta function
\be
  \eta(q)\;=\;q^{\frac{1}{24}} \prod_{n=1}^\infty (1-q^n)
\label{eta}
\ee

The cardinality of the enlarged set of indecomposable modules is readily seen to be given by
\be
 7pp'-3p-3p'+1
\label{CardJFund}
\ee
and this set was shown in~\cite{Ras0812} to yield a well-defined fusion algebra called the
{\em fundamental fusion algebra}
\be
 \mathrm{Fund}[{\cal WLM}(p,p')]\;=\;\big\langle\ketw{1,1},\ketw{2,1},\ketw{1,2}\big\rangle
\ee 
This algebra is so named since it is 
generated from repeated fusions of the two {\em fundamental modules} 
$\ketw{2,1}$ and $\ketw{1,2}$ in addition to the identity $\ketw{1,1}$ which is now
present for all $p$. 
The module $\ketw{1,1}$ is irreducible for $p=1$ in which case $\ketw{1,1}=\Wc(\D_{1,1})$. 
The module $\ketw{2,1}$ is irreducible for $p=1,2$ in which case $\ketw{2,1}=\Wc(\D_{2,1})$. 
The module $\ketw{1,2}$ is irreducible for $p'=2$ in which case $\ketw{1,2}=\Wc(\D_{1,2})$.
{}From~\cite{Ras0812}, 
we know that the fusion algebra generated by the modules (\ref{JWout}) is an {\em ideal}
of the fundamental fusion algebra.
To simplify the notation, we sometimes write $\ketw{\R_{r,s}^{0,0}}=\ketw{r,s}$, or
$\ketw{r,s}=\Wc(\D_{r,s})$ if $\ketw{r,s}$ happens to be irreducible.

Further algebraic extensions exist. 
In particular, for $p>1$, there are irreducible modules not associated with boundary conditions
as the ones in (\ref{JWout}). Their
fusion properties have been systematically examined only very 
recently~\cite{GRW0905,Ras0906,Wood0907}. Here we restrict ourselves to the 
modules generating the fundamental fusion algebra.

\subsection{Fusion products of fundamental modules}
\label{SecFusProducts}

Since the associative and commutative
fundamental fusion algebra is generated from repeated fusions of the two
fundamental modules $\ketw{2,1}$ and $\ketw{1,2}$, the complete set of fusion rules can 
be reconstructed from knowledge of the basic fusion products involving these two modules.
Here we list all such fusion products. For $p=1$, we have
\bea
 \ketw{2,1}\otimes\ketw{\kappa,s}&=&\ketw{2\cdot\kappa,s}\nn
 \ketw{2,1}\otimes\ketw{\R_{1,\kappa p'}^{0,b}}&=&\ketw{\R_{1,(2\cdot\kappa)p'}^{0,b}}
\label{21a}
\eea
while for $p>1$, we have
\bea
 \ketw{2,1}\otimes\ketw{a,b}&=&\big(1-\delta_{a,1}\big)\ketw{a-1,b}\oplus\ketw{a+1,b}\nn
 \ketw{2,1}\otimes\ketw{\kappa p,s}&=&\ketw{\R_{\kappa p,s}^{1,0}}\nn
 \ketw{2,1}\otimes\ketw{a,\kappa p'}
   &=&\big(1-\delta_{a,1}\big)\ketw{a-1,\kappa p'}\oplus\ketw{a+1,\kappa p'}  \nn
 \ketw{2,1}\otimes\ketw{\R_{\kappa p,s}^{a,0}}&=&2\delta_{a,1}\ketw{\kappa p,s}
   \oplus2\delta_{a,p-1}\ketw{(2\cdot\kappa)p,s} \nn
  &&\oplus \big(1-\delta_{a,1}\big)\ketw{\R_{\kappa p,s}^{a-1,0}}    
    \oplus\big(1-\delta_{a,p-1}\big)\ketw{\R_{\kappa p,s}^{a+1,0}}
    \nn
 \ketw{2,1}\otimes\ketw{\R_{r,\kappa p'}^{0,b}}&=&\delta_{r,1}\ketw{\R_{2,\kappa p'}^{0,b}}
   \oplus\delta_{r,p}\ketw{\R_{\kappa p,p'}^{1,b}}\nn
 &&\oplus\big(1-\delta_{r,1}\big)\big(1-\delta_{r,p}\big)\big(\ketw{\R_{r-1,\kappa p'}^{0,b}}
    \oplus\ketw{\R_{r+1,\kappa p'}^{0,b}}\big)\nn
 \ketw{2,1}\otimes\ketw{\R_{\kappa p,p'}^{a,b}}&=&2\delta_{a,1}\ketw{\R_{p,\kappa p'}^{0,b}}
  \oplus2\delta_{a,p-1}\ketw{\R_{p,(2\cdot\kappa)p'}^{0,b}}\nn
 &&
  \oplus \big(1-\delta_{a,1}\big)\ketw{\R_{\kappa p,p'}^{a-1,b}}
   \oplus\big(1-\delta_{a,p-1}\big)\ketw{\R_{\kappa p,p'}^{a+1,b}}
\label{21}
\eea
Since $p'>p\geq1$, we simply have
\bea
 \ketw{1,2}\otimes\ketw{a,b}&=&\big(1-\delta_{b,1}\big)\ketw{a,b-1}\oplus\ketw{a,b+1}\nn
 \ketw{1,2}\otimes\ketw{\kappa p,b}
   &=&\big(1-\delta_{b,1}\big)\ketw{\kappa p,b-1}\oplus\ketw{\kappa p,b+1}\nn
 \ketw{1,2}\otimes\ketw{r,\kappa p'}&=&\ketw{\R_{r,\kappa p'}^{0,1}}\nn
 \ketw{1,2}\otimes\ketw{\R_{\kappa p,s}^{a,0}}&=&\delta_{s,1}\ketw{\R_{\kappa p,2}^{a,0}}
   \oplus\delta_{s,p'}\ketw{\R_{\kappa p,p'}^{a,1}}\nn
  &&\oplus\big(1-\delta_{s,1}\big)\big(1-\delta_{s,p'}\big)\big(\ketw{\R_{\kappa p,s-1}^{a,0}}
    \oplus\ketw{\R_{\kappa p,s+1}^{a,0}}\big)\nn
 \ketw{1,2}\otimes\ketw{\R_{r,\kappa p'}^{0,b}}&=&2\delta_{b,1}\ketw{r,\kappa p'}
   \oplus2\delta_{b,p'-1}\ketw{r,(2\cdot\kappa)p'}\nn
  &&\oplus\big(1-\delta_{b,1}\big)\ketw{\R_{r,\kappa p'}^{0,b-1}}\oplus
   \big(1-\delta_{b,p'-1}\big)\ketw{\R_{r,\kappa p'}^{0,b+1}}\nn
 \ketw{1,2}\otimes\ketw{\R_{\kappa p,p'}^{a,b}}&=&2\delta_{b,1}\ketw{\R_{\kappa p,p'}^{a,0}}
   \oplus2\delta_{b,p'-1}\ketw{\R_{(2\cdot\kappa)p,p'}^{a,0}}\nn
  &&\oplus\big(1-\delta_{b,1}\big)\ketw{\R_{\kappa p,p'}^{a,b-1}}\oplus\big(1-\delta_{b,p'-1}\big)
    \ketw{\R_{\kappa p,p'}^{a,b+1}}
\label{12}
\eea
for all $p\in\mathbb{N}$.

\subsection{Graph fusion algebras}

Let $\Ic_f$ denote the set of indecomposable modules appearing in the fundamental fusion
algebra of ${\cal WLM}(p,p')$. In the regular representation 
\be
 N_\mu N_\nu\;=\;\sum_{\la\in\Ic_f} N_{\mu,\nu}{}^\la N_\la,\qquad\quad
  \mu,\nu\in\Ic_f
\label{NNN}
\ee
of this fusion algebra, the fusion matrices $N_\mu$ are mutually commuting, but in general 
not diagonalizable. Viewing the fusion matrices as 
adjacency matrices of graphs, the fusion rules are neatly encoded in the graphs.
In this context, (\ref{NNN}) is referred to as the {\em graph fusion algebra} of ${\cal WLM}(p,p')$,
in this case corresponding to the fundamental fusion algebra.

As demonstrated in Section~\ref{SecFundFusGra}, the {\em fundamental fusion graphs}, 
the ones associated to the two fundamental modules, have two 
particular types of connected and directed components. 
In Section~\ref{SecAdjacency}, we discuss the 
spectral decomposition of the adjacency matrices of these subgraphs. 
The adjacency matrices of the fundamental fusion graphs themselves are given by the 
matrix realizations of the two fundamental modules. We use 
\be
 X\;=\;(1+\delta_{p,1})N_{\ketw{2,1}},\qquad\quad
 Y\;=\;N_{\ketw{1,2}}
\label{XNYN}
\ee
as a convenient shorthand for these matrices. The normalization of $X$ is 
chosen to ensure universality of notation in the following. In Section~\ref{SecSpectral}, 
we will demonstrate that $X$ and $Y$ can be {\em simultaneously} brought 
to Jordan form, {\em modulo} permutation similarity, by a common similarity transformation. 
It is recalled that two matrices $A$ and $B$ are {\em permutation similar} if for some
permutation matrix $P$,
\be
 A\;=\;P^{-1}BP
\ee

In~\cite{Ras0812}, we found that the fundamental fusion algebra is isomorphic to
the polynomial ring
\be
 \mathbb{C}[X,Y]\big/\big(P_{p}(X),P_{p'}(Y),P_{p,p'}(X,Y)\big)
\label{FundPol}
\ee
where
\be
 P_n(x)\;=\;2\big(T_{2n}(\tfrac{x}{2})-1\big)U_{n-1}(\tfrac{x}{2}),\qquad
 P_{n,n'}(x,y)\;=\;\big(T_n(\tfrac{x}{2})-T_{n'}(\tfrac{y}{2})\big)
   U_{n-1}(\tfrac{x}{2})U_{n'-1}(\tfrac{y}{2})
\label{Pn}
\ee
Here $T_n(z)$ and $U_n(z)$ denote the Chebyshev polynomials of the first and second kind, 
respectively. The isomorphism is given by
\bea
 \ketw{a,b}&\leftrightarrow&
      U_{a-1}\big(\tfrac{X}{2}\big)U_{b-1}\big(\tfrac{Y}{2}\big)\nn
 \Wc(\D_{\kappa p,s})&\leftrightarrow&
      \tfrac{1}{\kappa}U_{\kappa p-1}\big(\tfrac{X}{2}\big)U_{s-1}\big(\tfrac{Y}{2}\big)\nn
 \Wc(\D_{a,\kappa p'})&\leftrightarrow&
      \tfrac{1}{\kappa}U_{a-1}\big(\tfrac{X}{2}\big)U_{\kappa p'-1}\big(\tfrac{Y}{2}\big)\nn
 \ketw{\R_{\kappa p,s}^{a,0}}&\leftrightarrow&
     \tfrac{2}{\kappa}T_a\big(\tfrac{X}{2}\big)U_{\kappa p-1}\big(\tfrac{X}{2}\big)
        U_{s-1}\big(\tfrac{Y}{2}\big)\nn
 \ketw{\R_{r,\kappa p'}^{0,b}}&\leftrightarrow&
     \tfrac{2}{\kappa}U_{r-1}\big(\tfrac{X}{2}\big) 
       T_b\big(\tfrac{Y}{2}\big)U_{\kappa p'-1}\big(\tfrac{Y}{2}\big)\nn
 \ketw{\R_{\kappa p,p'}^{a,b}}&\leftrightarrow&
    \tfrac{4}{\kappa}T_a\big(\tfrac{X}{2}\big)
    U_{\kappa p-1}\big(\tfrac{X}{2}\big)
    T_b\big(\tfrac{Y}{2}\big)U_{p'-1}\big(\tfrac{Y}{2}\big)
\label{abR4}
\eea
where it is noted that
\be
 U_{\kappa p-1}\big(\tfrac{X}{2}\big)U_{p'-1}\big(\tfrac{Y}{2}\big)
  \;\equiv\; U_{p-1}\big(\tfrac{X}{2}\big)U_{\kappa p'-1}\big(\tfrac{Y}{2}\big)\qquad
    (\mathrm{mod}\ P_{p,p'}(X,Y))
\ee
Identifying the formal entities $X$ and $Y$ appearing in (\ref{FundPol}) with the
two fundamental matrices (\ref{XNYN}) of matching notation, we obtain the regular 
representation (\ref{NNN}) of the fundamental fusion algebra.

Letting $\Ic_b$ denote the set of indecomposable modules (\ref{JWout}) associated with 
boundary conditions, the regular representation of the corresponding fusion algebra is given by 
\be
 \Nh_\mu \Nh_\nu\;=\;\sum_{\la\in\Ic_b} \Nh_{\mu,\nu}{}^\la \Nh_\la,\qquad\quad
  \mu,\nu\in\Ic_b
\label{NNNb}
\ee
Since this fusion algebra is an ideal of the fundamental fusion algebra, $\Nh_\mu$, for every 
$\mu\in\Ic_b$, is obtained from $N_\mu$ by elimination of the rows and columns corresponding 
to the $(p-1)(p'-1)$ modules (\ref{ab}) {\em not} associated with boundary conditions.
Indeed, ordering the elements of $\Ic_f$ according to $\Ic_f=(\Ic_f\setminus\Ic_b)\cup\Ic_b$
yields
\be
 N_\mu\;=\;\left(\!\!
\begin{array}{c|c}
  \ast&\ast \\
\hline
  0&\ast 
\end{array}
\!\!\right),\quad \mu\in\Ic_f\setminus\Ic_b;\qquad\quad
 N_\mu\;=\;\left(\!\!
\begin{array}{c|c}
  0&\ast \\
\hline
\\[-.4cm]
  0&\Nh_\mu 
\end{array}
\!\!\right),\quad \mu\in\Ic_b
\ee
Utilizing this block-triangular structure for $X$ and $Y$
\be
 X\;=\;\left(\!\!
\begin{array}{c|c}
  \ast&\ast \\
\hline
\\[-.4cm]
  0&\Xh
\end{array}
\!\!\right),\qquad\quad
 Y\;=\;\left(\!\!
\begin{array}{c|c}
  \ast&\ast \\
\hline
\\[-.4cm]
  0&\Yh
\end{array}
\!\!\right) 
\ee
we have
\be
 N_\mu\;=\;\mathrm{pol}_\mu(X,Y)\;=\;\left(\!\!
\begin{array}{c|c}
  \ast&\ast \\
\hline
\\[-.4cm]
  0&\mathrm{pol}_\mu(\Xh,\Yh)
\end{array}
\!\!\right) 
\label{NpolXY}
\ee
where $\mathrm{pol}_\mu(X,Y)$ is the polynomial appearing in (\ref{abR4}) for $\mu\in\Ic_f$.
It follows that we can express the fusion matrices $\Nh_\mu$ in terms of the
matrices $\Xh$ and $\Yh$
\be
 \Nh_\mu\;=\;\mathrm{pol}_\mu(\Xh,\Yh),\qquad \mu\in\Ic_b
\ee
using the {\em same} polynomial as in the description of $N_\mu$ in terms of $X$ and $Y$
(\ref{NpolXY}).
For $p>2$, we have $\ketw{2,1},\ketw{1,2}\in\Ic_f\setminus\Ic_b$, in which case
the matrices $\Xh$ and $\Yh$ should be thought of as {\em auxiliary} matrices.
Similarly, for $p=2$, we have $\ketw{1,2}\in\Ic_f\setminus\Ic_b$.
Despite their auxiliary status, the matrices $\Xh$ and $\Yh$ are very useful
in the description of the spectral decomposition of the fusion matrices $\Nh_\mu$.
We refer to the corresponding directed graphs as {\em auxiliary fusion graphs}.
As in the case of the fundamental fusion graphs, the auxiliary ones consist of
two particular types of connected and directed components,
one of which also appears as subgraphs of the fundamental fusion graphs.

\section{Spectral decomposition of adjacency matrices}
\label{SecAdjacency}

\subsection{Cycle, tadpole and eye-patch graphs}

As already mentioned, a fundamental (or auxiliary) fusion graph consists of a number of 
connected components. 
There are three prototypes: {\em cycle} graphs, {\em tadpole} graphs and {\em eye-patch} graphs,
and they are the topic of the present section.
The connected subgraphs of the fundamental fusion graphs are all tadpole
or eye-patch graphs, while the connected subgraphs of the auxiliary fusion graphs are 
all cycle or eye-patch graphs.
All of these connected graphs depend on a single integer {\em order parameter} $\rho\geq2$.

We refer to a connected and directed graph of the type
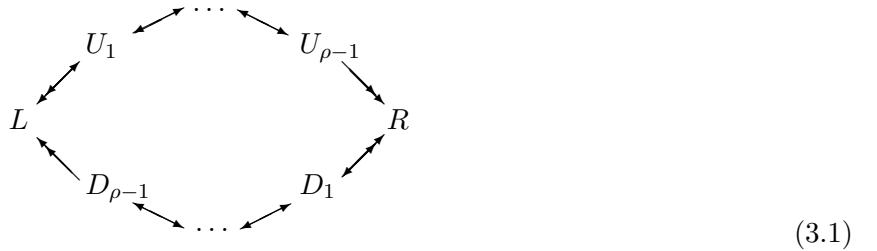
\be
 \mbox{
 \begin{picture}(100,100)(160,-45)
    \unitlength=0.75cm
  \thinlines
\put(6.3,-0.1){$L$}
\put(7.45,1){\vector(-1,-1){0.5}}
\put(7.3,0.85){\vector(-1,-1){0.5}}
\put(7.05,0.6){\vector(1,1){0.5}}
\put(7.65,1.25){$U_1$}
\put(8.8,1.75){\vector(2,1){0.6}}
\put(9,1.85){\vector(-2,-1){0.5}}
\put(9.55,2){$\ldots$}
\put(10.7,1.85){\vector(2,-1){0.55}}
\put(10.9,1.75){\vector(-2,1){0.55}}
\put(11.45,1.25){$U_{\rho-1}$}
\put(12.2,1.1){\vector(1,-1){0.6}}
\put(12.35,0.95){\vector(1,-1){0.6}}
\put(7.55,-1){\vector(-1,1){0.6}}
\put(7.4,-0.85){\vector(-1,1){0.6}}
\put(7.65,-1.25){$D_{\rho-1}$}
\put(8.8,-1.55){\vector(2,-1){0.6}}
\put(9,-1.65){\vector(-2,1){0.5}}
\put(9.6,-1.9){$\ldots$}
\put(10.95,-1.6){\vector(-2,-1){0.55}}
\put(10.75,-1.7){\vector(2,1){0.55}}
\put(11.45,-1.25){$D_1$}
\put(12.3,-0.85){\vector(1,1){0.5}}
\put(12.45,-0.7){\vector(1,1){0.5}}
\put(12.7,-0.45){\vector(-1,-1){0.5}}
\put(13,-0.1){$R$}
 \end{picture}
}
\label{GCp}
\ee
as a {\em cycle graph} with order parameter $\rho$. 
Its order is $2\rho$ and the labeling of the $2\rho$ vertices has been chosen to reflect their 
position in the graph. The cycle graph with order parameter $\rho=2$ is given by
\be
 \mbox{
 \begin{picture}(100,80)(120,-30)
    \unitlength=0.75cm
  \thinlines
\put(6.3,-0.1){$L$}
\put(7.45,1){\vector(-1,-1){0.5}}
\put(7.3,0.85){\vector(-1,-1){0.5}}
\put(7.05,0.6){\vector(1,1){0.5}}
\put(7.7,1.25){$U_1$}
\put(8.4,1.1){\vector(1,-1){0.6}}
\put(8.55,0.95){\vector(1,-1){0.6}}
\put(7.55,-1){\vector(-1,1){0.6}}
\put(7.4,-0.85){\vector(-1,1){0.6}}
\put(7.7,-1.25){$D_1$}
\put(8.5,-0.85){\vector(1,1){0.5}}
\put(8.65,-0.7){\vector(1,1){0.5}}
\put(8.9,-0.45){\vector(-1,-1){0.5}}
\put(9.2,-0.1){$R$}
 \end{picture}
}
\label{GC2}
\ee
In the ordered basis
\be
 \big\{ L,U_1,\ldots,U_{\rho-1},R,D_1,\ldots ,D_{\rho-1}\big\}
\ee
the adjacency matrix associated to the cycle graph (\ref{GCp}) is given by
\be
 \Cc_\rho\;=\; 
\left(\!\!
\begin{array}{c|ccccc|c|ccccc}
  0&1&&&&&&&&&&\\
\hline
  2&0&1&&&&&&&&&\\
  &1&0&&&&&&&&&\\
  &&&\ddots&&&&&&&&\\
  &&&&0&1&&&&&&\\
  &&&&1&0&2&&&&&\\
\hline
  &&&&&0&0&1&&&&\\
\hline
  &&&&&&2&0&1&&&\\
  &&&&&&&1&0&&&\\
  &&&&&&&&&\ddots&&\\
  &&&&&&&&&&0&1\\
  2&&&&&&&&&&1&0  
\end{array}
\!\!\right)
\label{MCp}
\ee
The first and $(\rho+1)$'th rows and columns (corresponding to $L$ and $R$) 
are emphasized to signal their special status. For $\rho=2$, the adjacency matrix is
\be
 \Cc_2\;=\;
\left(\!\!
\begin{array}{cccc}
 0&1&0&0\\
 2&0&2&0\\
 0&0&0&1\\
 2&0&2&0
\end{array}
\!\!\right)
\label{MC2}
\ee

We refer to a connected and directed graph of the type
\be
 \mbox{
 \begin{picture}(100,100)(100,-45)
    \unitlength=0.75cm
  \thinlines
\put(-0.5,0){$L_{1}$}
\put(0.8,0.15){\vector(-1,0){0.4}}
\put(0.8,0.15){\vector(1,0){0.35}}
\put(1.6,0){$\ldots$}
\put(3,0.15){\vector(-1,0){0.4}}
\put(3,0.15){\vector(1,0){0.35}}
\put(3.5,0){$L_{\rho-1}$}
\put(4.85,0.15){\vector(1,0){0.85}}
\put(6,0){$L_{\rho}$}
\put(7.45,1){\vector(-1,-1){0.5}}
\put(7.3,0.85){\vector(-1,-1){0.5}}
\put(7.05,0.6){\vector(1,1){0.5}}
\put(7.65,1.25){$U_1$}
\put(8.8,1.75){\vector(2,1){0.6}}
\put(9,1.85){\vector(-2,-1){0.5}}
\put(9.55,2){$\ldots$}
\put(10.7,1.85){\vector(2,-1){0.55}}
\put(10.9,1.75){\vector(-2,1){0.55}}
\put(11.45,1.25){$U_{\rho-1}$}
\put(12.2,1.1){\vector(1,-1){0.6}}
\put(12.35,0.95){\vector(1,-1){0.6}}
\put(7.55,-1){\vector(-1,1){0.6}}
\put(7.4,-0.85){\vector(-1,1){0.6}}
\put(7.65,-1.25){$D_{\rho-1}$}
\put(8.8,-1.55){\vector(2,-1){0.6}}
\put(9,-1.65){\vector(-2,1){0.5}}
\put(9.6,-1.9){$\ldots$}
\put(10.95,-1.6){\vector(-2,-1){0.55}}
\put(10.75,-1.7){\vector(2,1){0.55}}
\put(11.45,-1.25){$D_1$}
\put(12.3,-0.85){\vector(1,1){0.5}}
\put(12.45,-0.7){\vector(1,1){0.5}}
\put(12.7,-0.45){\vector(-1,-1){0.5}}
\put(13,-0.1){$R$}
 \end{picture}
}
\label{GTp}
\ee
as a {\em tadpole graph} with order parameter $\rho$. 
Its order is $3\rho-1$ and the labeling of the $3\rho-1$ vertices has been chosen to reflect their 
position in the graph. The tadpole graph with order parameter $\rho=2$ is given by
\be
 \mbox{
 \begin{picture}(100,80)(100,-30)
    \unitlength=0.75cm
  \thinlines
\put(4,0){$L_{1}$}
\put(4.95,0.15){\vector(1,0){0.75}}
\put(6,0){$L_{2}$}
\put(7.45,1){\vector(-1,-1){0.5}}
\put(7.3,0.85){\vector(-1,-1){0.5}}
\put(7.05,0.6){\vector(1,1){0.5}}
\put(7.7,1.25){$U_1$}
\put(8.4,1.1){\vector(1,-1){0.6}}
\put(8.55,0.95){\vector(1,-1){0.6}}
\put(7.55,-1){\vector(-1,1){0.6}}
\put(7.4,-0.85){\vector(-1,1){0.6}}
\put(7.7,-1.25){$D_1$}
\put(8.5,-0.85){\vector(1,1){0.5}}
\put(8.65,-0.7){\vector(1,1){0.5}}
\put(8.9,-0.45){\vector(-1,-1){0.5}}
\put(9.2,-0.1){$R$}
 \end{picture}
}
\label{GT2}
\ee
In the ordered basis
\be
 \big\{ L_1,\ldots, L_{\rho-1},L_\rho,U_1,\ldots,U_{\rho-1},R,D_1,\ldots ,D_{\rho-1}\big\}
\ee
the adjacency matrix associated to the tadpole graph (\ref{GTp}) is given by
\be
 \Tc_\rho\;=\; 
\left(\!\!
\begin{array}{ccccc|c|ccccc|c|ccccc}
  0&1&&&&&&&&&&&&&&&\\
  1&0&&&&&&&&&&&&&&&\\
  &&\ddots&&&&&&&&&&&&&&\\
  &&&0&1&&&&&&&&&&&&\\
  &&&1&0&1&&&&&&&&&&&\\
\hline
  &&&&0&0&1&&&&&&&&&&\\
\hline
  &&&&&2&0&1&&&&&&&&&\\
  &&&&&&1&0&&&&&&&&&\\
  &&&&&&&&\ddots&&&&&&&&\\
  &&&&&&&&&0&1&&&&&&\\
  &&&&&&&&&1&0&2&&&&&\\
\hline
  &&&&&&&&&&0&0&1&&&&\\
\hline
  &&&&&&&&&&&2&0&1&&&\\
  &&&&&&&&&&&&1&0&&&\\
  &&&&&&&&&&&&&&\ddots&&\\
  &&&&&&&&&&&&&&&0&1\\
  &&&&&2&&&&&&&&&&1&0  
\end{array}
\!\!\right)
\label{MTp}
\ee
The $\rho$'th and $(2\rho)$'th rows and columns (corresponding to $L_\rho$ and $R$) 
are emphasized to signal their special status. For $\rho=2$, the adjacency matrix is
\be
 \Tc_2\;=\;
\left(\!\!
\begin{array}{ccccc}
 0&1&0&0&0\\
 0&0&1&0&0\\
 0&2&0&2&0\\
 0&0&0&0&1\\
 0&2&0&2&0
\end{array}
\!\!\right)
\label{MT2}
\ee

We also introduce what we call an {\em eye-patch graph} with order parameter $\rho$
\be
 \mbox{
 \begin{picture}(100,100)(160,-45)
    \unitlength=0.75cm
  \thinlines
\put(-0.5,0){$L_{1}$}
\put(0.8,0.15){\vector(-1,0){0.4}}
\put(0.8,0.15){\vector(1,0){0.35}}
\put(1.6,0){$\ldots$}
\put(3,0.15){\vector(-1,0){0.4}}
\put(3,0.15){\vector(1,0){0.35}}
\put(3.5,0){$L_{\rho-1}$}
\put(4.85,0.15){\vector(1,0){0.85}}
\put(6,0){$L_{\rho}$}
\put(7.45,1){\vector(-1,-1){0.5}}
\put(7.3,0.85){\vector(-1,-1){0.5}}
\put(7.05,0.6){\vector(1,1){0.5}}
\put(7.65,1.25){$U_1$}
\put(8.8,1.75){\vector(2,1){0.6}}
\put(9,1.85){\vector(-2,-1){0.5}}
\put(9.55,2){$\ldots$}
\put(10.7,1.85){\vector(2,-1){0.55}}
\put(10.9,1.75){\vector(-2,1){0.55}}
\put(11.45,1.25){$U_{\rho-1}$}
\put(12.2,1.1){\vector(1,-1){0.6}}
\put(12.35,0.95){\vector(1,-1){0.6}}
\put(7.55,-1){\vector(-1,1){0.6}}
\put(7.4,-0.85){\vector(-1,1){0.6}}
\put(7.65,-1.25){$D_{\rho-1}$}
\put(8.8,-1.55){\vector(2,-1){0.6}}
\put(9,-1.65){\vector(-2,1){0.5}}
\put(9.6,-1.9){$\ldots$}
\put(10.95,-1.6){\vector(-2,-1){0.55}}
\put(10.75,-1.7){\vector(2,1){0.55}}
\put(11.45,-1.25){$D_1$}
\put(12.3,-0.85){\vector(1,1){0.5}}
\put(12.45,-0.7){\vector(1,1){0.5}}
\put(12.7,-0.45){\vector(-1,-1){0.5}}
\put(13,0){$R_{\rho}$}
\put(14.6,0.15){\vector(-1,0){0.75}}
\put(15,0){$R_{\rho-1}$}
\put(16.8,0.15){\vector(-1,0){0.4}}
\put(16.8,0.15){\vector(1,0){0.35}}
\put(17.55,0){$\ldots$}
\put(18.9,0.15){\vector(-1,0){0.4}}
\put(18.9,0.15){\vector(1,0){0.35}}
\put(19.5,0){$R_{1}$}
 \end{picture}
}
\label{GPp}
\ee
which, for $\rho=2$, reduces to
\be
 \mbox{
 \begin{picture}(100,80)(120,-30)
    \unitlength=0.75cm
  \thinlines
\put(4,0){$L_{1}$}
\put(4.95,0.15){\vector(1,0){0.75}}
\put(6,0){$L_{2}$}
\put(7.45,1){\vector(-1,-1){0.5}}
\put(7.3,0.85){\vector(-1,-1){0.5}}
\put(7.05,0.6){\vector(1,1){0.5}}
\put(7.7,1.25){$U_1$}
\put(8.4,1.1){\vector(1,-1){0.6}}
\put(8.55,0.95){\vector(1,-1){0.6}}
\put(7.55,-1){\vector(-1,1){0.6}}
\put(7.4,-0.85){\vector(-1,1){0.6}}
\put(7.7,-1.25){$D_1$}
\put(8.5,-0.85){\vector(1,1){0.5}}
\put(8.65,-0.7){\vector(1,1){0.5}}
\put(8.9,-0.45){\vector(-1,-1){0.5}}
\put(9.2,0){$R_{2}$}
\put(10.9,0.15){\vector(-1,0){0.75}}
\put(11.2,0){$R_{1}$}
 \end{picture}
}
\label{GP2}
\ee
The order of the graph (\ref{GPp}) is $4\rho-2$, and the labeling of the $4\rho-2$ vertices has 
been chosen to reflect their position in the graph.
In the ordered basis
\be
 \big\{ L_1,\ldots, L_{\rho-1},L_\rho,U_1,\ldots,U_{\rho-1},
     R_1,\ldots, R_{\rho-1},R_\rho,D_1,\ldots ,D_{\rho-1}\big\}
\ee
the adjacency matrix associated to the graph (\ref{GPp}) is given by
\be
 \Ec_{\rho}\;=\; 
\left(\!\!
\begin{array}{ccccc|c|ccccc|ccccc|c|ccccc}
  0&1&&&&&&&&&&&&&&&&&&&&\\
  1&0&&&&&&&&&&&&&&&&&&&&\\
  &&\ddots&&&&&&&&&&&&&&&&&&&\\
  &&&0&1&&&&&&&&&&&&&&&&&\\
  &&&1&0&1&&&&&&&&&&&&&&&&\\
\hline
  &&&&0&0&1&&&&&&&&&&&&&&&\\
\hline
  &&&&&2&0&1&&&&&&&&&&&&&&\\
  &&&&&&1&0&&&&&&&&&&&&&&\\
  &&&&&&&&\ddots&&&&&&&&&&&&&\\
  &&&&&&&&&0&1&&&&&&&&&&&\\
  &&&&&&&&&1&0&0&&&&&2&&&&&\\
\hline
  &&&&&&&&&&0&0&1&&&&&&&&&\\
  &&&&&&&&&&&1&0&&&&&&&&&\\
  &&&&&&&&&&&&&\ddots&&&&&&&&\\
  &&&&&&&&&&&&&&0&1&&&&&&\\
  &&&&&&&&&&&&&&1&0&1&&&&&\\
\hline
  &&&&&&&&&&&&&&&0&0&1&&&&\\
\hline
  &&&&&&&&&&&&&&&&2&0&1&&&\\
  &&&&&&&&&&&&&&&&&1&0&&&\\
  &&&&&&&&&&&&&&&&&&&\ddots&&\\
  &&&&&&&&&&&&&&&&&&&&0&1\\
  &&&&&2&&&&&&&&&&&&&&&1&0
\end{array}
\!\!\right)
\label{MPp}
\ee
The $\rho$'th and $(3\rho-1)$'th rows and columns (corresponding to $L_\rho$ and $R_\rho$) 
are emphasized to signal their special status.
For $\rho=2$, the adjacency matrix (\ref{MPp}) is meant to reduce to
\be
 \Ec_2\;=\;
\left(\!\!
\begin{array}{cccccc}
 0&1&0&0&0&0\\
 0&0&1&0&0&0\\
 0&2&0&0&2&0\\
 0&0&0&0&1&0\\
 0&0&0&0&0&1\\
 0&2&0&0&2&0
\end{array}
\!\!\right)
\label{MP2}
\ee

Extending the definition of the order parameter to $\rho=1$, we let a cycle,
tadpole or eye-patch graph collapse to the following directed order-2 graph
\be
  \mbox{
 \begin{picture}(60,20)(0,0)
    \unitlength=0.75cm
  \thinlines
\put(0,0){$L$}
\put(1.3,.15){\vector(1,0){0.35}}
\put(1.3,.15){\vector(1,0){0.6}}
\put(1.3,.15){\vector(-1,0){0.45}}
\put(1.3,.15){\vector(-1,0){0.7}}
\put(2.05,0){$R$}
 \end{picture}
}
\label{Gp1}
\ee
This type of graph is relevant only when considering the series ${\cal WLM}(1,p')$.
The corresponding adjacency matrix is
\be
 \Cc_1\;=\;\Tc_1\;=\;\Ec_1\;=\;
\left(\!\!
\begin{array}{cc}
 0&2\\
 2&0
\end{array}
\!\!\right)
\label{MT1P1}
\ee

\subsection{Spectral decompositions}

In preparation for the spectral decomposition of the adjacency matrices (\ref{MCp}), 
(\ref{MTp}) and (\ref{MPp}), we recall that canonical Jordan blocks of rank-2 or -3 
associated to the eigenvalue $\la$ of a matrix $A$ are given by
\be
   \Jc_{\la,2}\;=\;\left(\!\!\begin{array}{cc} \la&1 \\ 0&\la \end{array}\!\!\right),\qquad\quad
   \Jc_{\la,3}\;=\;\left(\!\!\begin{array}{ccc} \la&1&0 \\ 0&\la&1 \\ 0&0&\la \end{array}\!\!\right)
\label{Jb}
\ee 
They appear in the Jordan decomposition of $A$ if the eigenvalue $\la$ gives rise to a 
Jordan chain of length 2 or 3, where a Jordan chain of length 3, in particular, is given by
\be
 Av^{(2)}\;=\;\la v^{(2)}+v^{(1)},\qquad \quad
 Av^{(1)}\;=\;\la v^{(1)}+v^{(0)},\qquad \quad
 Av^{(0)}\;=\;\la v^{(0)}
\label{Av}
\ee
This chain of relations implies that
\be
 \big(A-\la I\big) v^{(2)}\;=\;v^{(1)},\qquad\quad
 \big(A-\la I\big) v^{(1)}\;=\;v^{(0)},\qquad\quad   
 \big(A-\la I\big)^{\ell+1} v^{(\ell)}\;=\;0,\quad \ell\in\mathbb{Z}_{0,2}
\label{Alv}
\ee
indicating that the vectors are {\em generalized eigenvectors}. 
A proper eigenvector is merely a special type of generalized eigenvector.
We say that an upper-triangular (square) matrix with identical entries 
$\la$ on the diagonal is a Jordan block if the geometric multiplicity of (the single eigenvalue)
$\la$ is 1. It is a Jordan {\em canonical} block (as in (\ref{Jb}))
if the entries on the super-diagonal are 1 while all entries above the super-diagonal are 0.
A block-diagonal matrix is of Jordan (canonical) form if every block is a Jordan
(canonical) block.

Following~\cite{Ras0908}, we introduce the $2\rho-1$ functions 
$f_h(x)$, $h\in\mathbb{Z}_{1,2\rho-1}$, defined by
\be
 f_k(x)\;=\;U_{k-1}(\tfrac{x}{2}),\qquad\quad
 f_\rho(x)\;=\;U_{\rho-1}(\tfrac{x}{2}),\qquad \quad
 f_{\rho+k}(x)\;=\;2T_k(\tfrac{x}{2})U_{\rho-1}(\tfrac{x}{2})
\label{f}
\ee
Here, and in the following, we are using the convention
\be
 k\in\mathbb{Z}_{1,\rho-1}
\ee
Certain useful properties of $f_h(x)$ are listed here, while further details can be
found in~\cite{Ras0908}.
For $\rho>2$, the functions satisfy recursive relations allowing us to express 
$x f_h(x)$, for $h\in\mathbb{Z}_{1,2\rho-2}$, as
\bea
 f_2(x)&=&x f_1(x)  \nn
 f_{h-1}(x)+f_{h+1}(x)&=&x f_h(x),\qquad\quad h\in\mathbb{Z}_{2,\rho-1}\nn
 f_{\rho+1}(x)&=&x f_\rho(x)\nn
 2f_\rho(x)+f_{\rho+2}(x)&=&x f_{\rho+1}(x)\nn
 f_{h-1}(x)+f_{h+1}(x)&=&x f_h(x),\qquad\quad h\in\mathbb{Z}_{\rho+2,2\rho-2}
\label{ff}
\eea
It follows that
\be
 \begin{array}{rll}
 f_2'(x)&=\quad\!\!x f_1'(x)+f_1(x),\qquad\qquad\qquad\qquad\quad\ \ 
      f_2''(x)&=\quad\!\! xf_1''(x)+2f_1'(x)  \\[3pt]
 f_{h-1}'(x)+f_{h+1}'(x)&=\quad\!\! x f_h'(x)+f_h(x),\qquad\qquad\ \!
     f_{h-1}''(x)+f_{h+1}''(x)&=\quad\!\!xf_h''(x)+2f_h'(x)\\[3pt]
 f_{\rho+1}'(x)&=\quad\!\!x f_\rho'(x)+f_\rho(x),\qquad\qquad\qquad\qquad\ \
      f_{\rho+1}''(x)&=\quad\!\!x f_\rho''(x)+2f_\rho'(x)\\[3pt]
 2f_\rho'(x)+f_{\rho+2}'(x)&=\quad\!\!x f_{\rho+1}'(x)+f_{\rho+1}(x),\qquad\ \ \!
    2f_\rho''(x)+f_{\rho+2}''(x)&=\quad\!\!x f_{\rho+1}''(x)+2f_{\rho+1}'(x)\\[3pt]
 f_{h-1}'(x)+f_{h+1}'(x)&=\quad\!\!x f_h'(x)+f_h(x),\qquad\qquad
     f_{h-1}''(x)+f_{h+1}''(x)&=\quad\!\!x f_h''(x)+2f_h'(x)
 \end{array}
\label{fdiff}
\ee
with the conditions on $h$ adopted from (\ref{ff}).
It is noted that we have not included any relations involving $xf_{2\rho-1}(x)$ for general $x$. 
Instead, we focus on evaluations at $x=\la_j$, for $j\in\mathbb{Z}_{0,\rho}$, where
\bea
 2(-1)^if_\rho(\la_j)+f_{2\rho-2}(\la_j)&=&\la_j f_{2\rho-1}(\la_j),
    \qquad\qquad\qquad  j\in\mathbb{Z}_{1,\rho-1}  \ \mathrm{or}\ i=j\in\{0,\rho\}\nn
 2(-1)^kf_\rho'(\la_k)+f_{2\rho-2}'(\la_k)&=&\la_k f_{2\rho-1}'(\la_k)+f_{2\rho-1}(\la_k)\nn
 2(-1)^kf_\rho''(\la_k)+f_{2\rho-2}''(\la_k)&=&\la_k f_{2\rho-1}''(\la_k)+2f_{2\rho-1}'(\la_k)
\label{f2p}
\eea
We also note that
\be
 f_h(\la_k)\;=\;0,\qquad h\in\mathbb{Z}_{\rho,2\rho-1}
\label{fhlk}
\ee
A convenient notation to be used below is
\be
 \tfrac{1}{\ell!}f_h^{(\ell)}(x)\;=\;\tfrac{1}{\ell!}\pa_x^\ell f_h(x),
   \qquad \ell\in\mathbb{Z}_{0,2},
   \quad h\in\mathbb{Z}_{1,2\rho-1}
\ee
where $0!=1$.

Since the spectral decompositions of $\Cc_\rho$, $\Tc_\rho$ and $\Ec_{\rho}$ are worked out 
in Section~\ref{SecCycle}, \ref{SecTad} and \ref{SecEye} for $\rho\geq2$, we here discuss 
the rather trivial spectral decomposition of $\Cc_1=\Tc_1=\Ec_1$ (\ref{MT1P1}) in the 
framework employed in those sections. 
With $\rho=1$, the eigenvalues are 
\be
 \la_j\;=\;2\cos\frac{j\pi}{\rho},\qquad j\in\mathbb{Z}_{0,\rho}
\label{laj}
\ee
with corresponding eigenvectors given by
\be
 V_0\;=\;\left(\!\!\!\begin{array}{c}  
  f_{\rho}(\la_0)
  \\[4pt]
  (-1)^0 f_{\rho}(\la_0)
  \end{array}\!\!\!\right)\;=\;
 \left(\!\!\begin{array}{c}  
 2 \\[4pt]  2
  \end{array}\!\!\right),\qquad
 V_\rho\;=\;\left(\!\!\!\begin{array}{c}  
  f_{\rho}(\la_\rho)
  \\[4pt]
  (-1)^\rho f_{\rho}(\la_\rho)
  \end{array}\!\!\!\right)\;=\;
 \left(\!\!\!\begin{array}{c}  
  -2 \\[4pt]   2
  \end{array}\!\!\!\right)
\label{V}
\ee
The minimal and characteristic polynomials of $\Cc_1$ are given by
\be
  m(\Cc_1)\;=\;(\Cc_1-\la_0I)(\Cc_1-\la_\rho I)\;=\;\Cc_1^2-4I,\qquad
 \det(\la I-\Cc_1)\;=\;(\la-\la_0)(\la-\la_\rho)\;=\;\la^2-4
\label{charC1}
\ee
while the similarity matrix constructed by concatenating the two eigenvectors
\be
 Q_{\Cc_1}\;=\;\big(V_0\ V_\rho\big)
\ee
diagonalizes $\Cc_1$
\be
 Q_{\Cc_1}^{-1}\Cc_1 Q_{\Cc_1}
   \;=\;\mathrm{diag}\big(\la_0,\la_\rho\big)\;=\;\mathrm{diag}\big(2,-2\big)
\label{JC1}
\ee

\subsubsection{Cycle graphs}
\label{SecCycle}

It follows from the explicit construction of generalized eigenvectors below that 
the eigenvalues of $\Cc=\Cc_\rho$, $\rho>1$, are given by (\ref{laj}),
while the minimal and characteristic polynomials of $\Cc$ are given by
\bea
  &m(\Cc)\;=\;(\Cc-\la_0I)(\Cc-\la_\rho I)\displaystyle{\prod_{k=1}^{\rho-1}(\Cc-\la_k I)^2}
   \;=\;(\Cc^2-4I)U_{\rho-1}^2(\tfrac{\Cc}{2})\nn
 &\det(\la I-\Cc)\;=\;(\la-\la_0)(\la-\la_\rho)\displaystyle{\prod_{k=1}^{\rho-1}(\la-\la_k)^2}
   \;=\; (\la^2-4)U_{\rho-1}^2(\tfrac{\la}{2})
\label{charC}
\eea
This implies that the Jordan canonical form of $\Cc$ consists of $\rho-1$ rank-2 blocks
associated to the eigenvalues $\la_k$,  
and two rank-1 blocks associated to the eigenvalues $\la_0=2$, $\la_\rho=-2$.
The number of linearly independent eigenvectors of $\Cc$ is thus $\rho+1$. Since 
the null-space of $\Cc$ is empty for $\rho$ odd but one-dimensional for $\rho$ even, 
the rank of $\Cc$ is
\be
 \mathrm{rank}(\Cc)\;=\;2\rho-\eps(\rho-1)\;=\;2\rho-1+\eps(\rho)
\label{rankC}
\ee

To establish these results on $\Cc$, we now discuss the two eigenvectors 
corresponding to $\la_0=2$, $\la_\rho=-2$, and the $\rho-1$ Jordan chains of length 2 
associated to $\la_k$. Using the various properties of the functions $f_h(x)$ discussed 
above, it is straightforward to verify that 
\be
 C_0\;=\;\left(\!\!\!\begin{array}{c}  
  f_{\rho}(\la_0)\\ \vdots\\ f_{2\rho-1}(\la_0)
  \\[.15cm] \hline\\[-.3cm]  
  (-1)^0 f_{\rho}(\la_0)\\ \vdots\\ (-1)^0 f_{2\rho-1}(\la_0)
  \end{array}\!\!\!\right),\qquad
 C_\rho\;=\;\left(\!\!\!\begin{array}{c}  
  f_{\rho}(\la_\rho)\\ \vdots\\ f_{2\rho-1}(\la_\rho)
  \\[.15cm] \hline\\[-.3cm]  
  (-1)^\rho f_{\rho}(\la_\rho)\\ \vdots\\ (-1)^\rho f_{2\rho-1}(\la_\rho)
  \end{array}\!\!\!\right)
\label{Cj}
\ee
are eigenvectors of $\Cc$
\be
 \Cc C_0\;=\;\la_0 C_0,\qquad 
 \Cc C_\rho\;=\;\la_\rho C_\rho 
\ee
For every $k\in\mathbb{Z}_{1,\rho-1}$, it is likewise verified that 
\be
 C_k^{(0)}\;=\;\left(\!\!\!\begin{array}{c}  
  f'_{\rho}(\la_k)\\ \vdots\\ f'_{2\rho-1}(\la_k)
  \\[.15cm] \hline\\[-.3cm]  
  (-1)^k f'_{\rho}(\la_k)\\ \vdots\\ (-1)^k f'_{2\rho-1}(\la_k)
  \end{array}\!\!\!\right),\qquad
 C_k^{(1)}\;=\;\left(\!\!\!\begin{array}{c}  
  \tfrac{1}{2}f''_{\rho}(\la_k)\\ \vdots\\ \tfrac{1}{2}f''_{2\rho-1}(\la_k)
  \\[.15cm] \hline\\[-.3cm]  
  \tfrac{1}{2}(-1)^k f''_{\rho}(\la_k)\\ \vdots\\ \tfrac{1}{2}(-1)^k f''_{2\rho-1}(\la_k)
  \end{array}\!\!\!\right)
\label{Ck12}
\ee
form the Jordan chain 
\be
 \Cc C_k^{(0)}\;=\;\la_k C_k^{(0)},\qquad
 \Cc C_k^{(1)}\;=\;\la_k C_k^{(1)}+C_k^{(0)}
\ee

Finally, the $2\rho$-dimensional matrix $Q_\Cc$ is constructed by concatenating 
the generalized eigenvectors (\ref{Cj}) and (\ref{Ck12})
\be
 Q_\Cc\;=\;\left(\!\!\begin{array}{c|cc|c|cc|c} C_0&C_1^{(0)}&C_1^{(1)}&\ldots&
    C_{\rho-1}^{(0)}&C_{\rho-1}^{(1)}&C_\rho  \end{array}\!\!\right)
\label{QC}
\ee
By a similarity transformation, this matrix converts $\Cc$ into its Jordan canonical form
\be
 J_\Cc\;=\;Q_\Cc^{-1}\Cc Q_\Cc
   \;=\;\mathrm{diag}\big(\la_0,\Jc_{\la_1,2},\ldots,\Jc_{\la_{\rho-1},2},\la_\rho\big)
\label{JC}
\ee

\subsubsection{Tadpole graphs}
\label{SecTad}

It follows from the explicit construction of generalized eigenvectors below that 
the eigenvalues of $\Tc=\Tc_\rho$, $\rho>1$, are given by (\ref{laj}), 
while the minimal and characteristic polynomials of $\Tc$ are given by
\bea
  &m(\Tc)\;=\;(\Tc-\la_0I)(\Tc-\la_\rho I)\displaystyle{\prod_{k=1}^{\rho-1}(\Tc-\la_k I)^3}
   \;=\;(\Tc^2-4I)U_{\rho-1}^3(\tfrac{\Tc}{2})\nn
 &\det(\la I-\Tc)\;=\;(\la-\la_0)(\la-\la_\rho)\displaystyle{\prod_{k=1}^{\rho-1}(\la-\la_k)^3}
   \;=\; (\la^2-4)U_{\rho-1}^3(\tfrac{\la}{2})
\label{charT}
\eea
This implies that the Jordan canonical form of $\Tc$ consists of $\rho-1$ rank-3 blocks
associated to the eigenvalues $\la_k$,  
and two rank-1 blocks associated to the eigenvalues $\la_0=2$, $\la_\rho=-2$.
The number of linearly independent eigenvectors of $\Tc$ is thus $\rho+1$. Since 
the null-space of $\Tc$ is empty for $\rho$ odd but one-dimensional for $\rho$ even, 
the rank of $\Tc$ is
\be
 \mathrm{rank}(\Tc)\;=\;3\rho-1-\eps(\rho-1)\;=\;3\rho-2+\eps(\rho)
\label{rankT}
\ee

To establish these results on $\Tc$, we now discuss the two eigenvectors 
corresponding to $\la_0=2$, $\la_\rho=-2$, and the $\rho-1$ Jordan chains of length
3 associated to $\la_k$. Using the various properties of the functions $f_h(x)$ discussed 
above, it is straightforward to verify that 
\be
 T_j\;=\;\left(\!\!\!\begin{array}{c}  
  f_1(\la_j)\\ \vdots\\ f_{\rho-1}(\la_j)
  \\[.15cm] \hline\\[-.3cm]  
  f_{\rho}(\la_j)\\ \vdots\\ f_{2\rho-1}(\la_j)
  \\[.15cm] \hline\\[-.3cm]  
  (-1)^j f_{\rho}(\la_j)\\ \vdots\\ (-1)^j f_{2\rho-1}(\la_j)
  \end{array}\!\!\!\right),\quad j=0,\rho;\qquad
 T_k^{(0)}\;=\;\left(\!\!\!\begin{array}{c}  
  f_1(\la_k)\\ \vdots\\ f_{\rho-1}(\la_k)
  \\[.15cm] \hline\\[-.3cm]  
  f_{\rho}(\la_k)\\ \vdots\\ f_{2\rho-1}(\la_k)
  \\[.15cm] \hline\\[-.3cm]  
  (-1)^k f_{\rho}(\la_k)\\ \vdots\\ (-1)^k f_{2\rho-1}(\la_k)
  \end{array}\!\!\!\right)
  \;=\;\left(\!\!\!\begin{array}{c}  
  f_1(\la_k)\\ \vdots\\ f_{\rho-1}(\la_k)
  \\[.15cm] \hline\\[-.3cm]  
  0\\ \vdots\\ 0
  \\[.15cm] \hline\\[-.3cm]  
  0\\ \vdots\\ 0
  \end{array}\!\!\!\right)
\label{Tj}
\ee
are eigenvectors of $\Tc$
\be
 \Tc T_0\;=\;\la_0 T_0,\qquad 
 \Tc T_k^{(0)}\;=\;\la_k T_k^{(0)},\qquad 
 \Tc T_\rho\;=\;\la_\rho T_\rho 
\ee
For every $k\in\mathbb{Z}_{1,\rho-1}$, it is likewise verified that $T_k^{(0)}$ together with 
\be
 T_k^{(1)}\;=\;\left(\!\!\!\begin{array}{c}  
  f'_1(\la_k)\\ \vdots\\ f'_{\rho-1}(\la_k)
  \\[.15cm] \hline\\[-.3cm]  
  f'_{\rho}(\la_k)\\ \vdots\\ f'_{2\rho-1}(\la_k)
  \\[.15cm] \hline\\[-.3cm]  
  (-1)^k f'_{\rho}(\la_k)\\ \vdots\\ (-1)^k f'_{2\rho-1}(\la_k)
  \end{array}\!\!\!\right),\qquad
 T_k^{(2)}\;=\;\left(\!\!\!\begin{array}{c}  
  \tfrac{1}{2}f''_1(\la_k)\\ \vdots\\ \tfrac{1}{2}f''_{\rho-1}(\la_k)
  \\[.15cm] \hline\\[-.3cm]  
  \tfrac{1}{2}f''_{\rho}(\la_k)\\ \vdots\\ \tfrac{1}{2}f''_{2\rho-1}(\la_k)
  \\[.15cm] \hline\\[-.3cm]  
  \tfrac{1}{2}(-1)^k f''_{\rho}(\la_k)\\ \vdots\\ \tfrac{1}{2}(-1)^k f''_{2\rho-1}(\la_k)
  \end{array}\!\!\!\right)
\label{Tk12}
\ee
form the Jordan chain 
\be
 \Tc T_k^{(0)}\;=\;\la_k T_k^{(0)},\qquad
 \Tc T_k^{(1)}\;=\;\la_k T_k^{(1)}+T_k^{(0)},\qquad
 \Tc T_k^{(2)}\;=\;\la_k T_k^{(2)}+T_k^{(1)}
\ee

Finally, the $(3\rho-1)$-dimensional matrix $Q_\Tc$ is constructed by concatenating 
the generalized eigenvectors (\ref{Tj}) and (\ref{Tk12})
\be
 Q_\Tc\;=\;\left(\!\!\begin{array}{c|ccc|c|ccc|c} T_0&T_1^{(0)}&T_1^{(1)}&T_1^{(2)}&\ldots&
    T_{\rho-1}^{(0)}&T_{\rho-1}^{(1)}&T_{\rho-1}^{(2)}&T_\rho  \end{array}\!\!\right)
\label{QT}
\ee
By a similarity transformation, this matrix converts $\Tc$ into its Jordan canonical form
\be
 J_\Tc\;=\;Q_\Tc^{-1}\Tc Q_\Tc
   \;=\;\mathrm{diag}\big(\la_0,\Jc_{\la_1,3},\ldots,\Jc_{\la_{\rho-1},3},\la_\rho\big)
\label{JT}
\ee

\subsubsection{Eye-patch graphs}
\label{SecEye}

It follows from the explicit construction of generalized eigenvectors below that 
the eigenvalues of $\Ec=\Ec_{\rho}$, $\rho>1$, are given by (\ref{laj}),
while the minimal and characteristic polynomials of $\Ec$ are given by
\bea
  &m(\Ec)\;=\;(\Ec-\la_0I)(\Ec-\la_\rho I)
       \displaystyle{\prod_{k=1}^{\rho-1}(\Ec-\la_k I)^3}
   \;=\;(\Ec^2-4I)U_{\rho-1}^3(\tfrac{\Ec}{2})\nn
 &\det(\la I-\Ec)\;=\;(\la-\la_0)(\la-\la_\rho)\displaystyle{\prod_{k=1}^{\rho-1}(\la-\la_k)^4}
   \;=\; (\la^2-4)U_{\rho-1}^4(\tfrac{\la}{2})
\label{charP}
\eea
This implies that the Jordan canonical form of $\Ec$ consists of $\rho-1$ rank-3 blocks
associated to the eigenvalues $\la_k$,  
and $\rho+1$ rank-1 blocks associated to the eigenvalues $\la_j$.
The number of linearly independent eigenvectors of $\Ec$ is thus $2\rho$. Since 
the null-space of $\Ec$ is empty for $\rho$ odd but two-dimensional for $\rho$ even, 
the rank of $\Ec$ is
\be
 \mathrm{rank}(\Ec)\;=\;4\rho-2-2\eps(\rho-1)\;=\;4(\rho-1)+2\eps(\rho)
\label{rankP}
\ee

To establish these results on $\Ec$, we now discuss the $\rho-1$ Jordan chains of length
3 associated to $\la_k$, and the additional $\rho+1$ eigenvectors 
corresponding to $\la_j$. Using the various properties of the functions $f_h(x)$ discussed
above, it is straightforward to verify that 
\be
 E_j\;=\;\left(\!\!\!\begin{array}{c}  
  f_1(\la_j)\\ \vdots\\ f_{\rho-1}(\la_j)
  \\[.15cm] \hline\\[-.3cm]  
  f_{\rho}(\la_j)\\ \vdots\\ f_{2\rho-1}(\la_j)
  \\[.15cm] \hline\\[-.3cm]  
  (-1)^j f_1(\la_j)\\ \vdots\\ (-1)^j f_{\rho-1}(\la_j)
  \\[.15cm] \hline\\[-.3cm]  
  (-1)^j f_{\rho}(\la_j)\\ \vdots\\ (-1)^j f_{2\rho-1}(\la_j)
  \end{array}\!\!\!\right),\quad j=0,\rho;\qquad
 E_k\;=\;\left(\!\!\!\begin{array}{c}  
  f_1(\la_k)\\ \vdots\\ f_{\rho-1}(\la_k)
  \\[.15cm] \hline\\[-.3cm]  
  f_{\rho}(\la_k)\\ \vdots\\ f_{2\rho-1}(\la_k)
  \\[.15cm] \hline\\[-.3cm]  
  (-1)^{k-1} f_{1}(\la_k)\\ \vdots\\ (-1)^{k-1} f_{\rho-1}(\la_k)
  \\[.15cm] \hline\\[-.3cm]  
  (-1)^{k-1} f_{\rho}(\la_k)\\ \vdots\\ (-1)^{k-1} f_{2\rho-1}(\la_k)
  \end{array}\!\!\!\right)
  \;=\;\left(\!\!\!\begin{array}{c}  
  f_1(\la_k)\\ \vdots\\ f_{\rho-1}(\la_k)
  \\[.15cm] \hline\\[-.3cm]  
  0\\ \vdots\\ 0
  \\[.15cm] \hline\\[-.3cm]  
  (-1)^{k-1}f_1(\la_k)\\ \vdots\\ (-1)^{k-1}f_{\rho-1}(\la_k)
  \\[.15cm] \hline\\[-.3cm]  
  0\\ \vdots\\ 0
  \end{array}\!\!\!\right)
\label{Ej}
\ee
and 
\be
 E_k^{(0)}\;=\;\left(\!\!\!\begin{array}{c}  
  f_1(\la_k)\\ \vdots\\ f_{\rho-1}(\la_k)
  \\[.15cm] \hline\\[-.3cm]  
  f_{\rho}(\la_k)\\ \vdots\\ f_{2\rho-1}(\la_k)
  \\[.15cm] \hline\\[-.3cm]  
  (-1)^{k} f_{1}(\la_k)\\ \vdots\\ (-1)^{k} f_{\rho-1}(\la_k)
  \\[.15cm] \hline\\[-.3cm]  
  (-1)^{k} f_{\rho}(\la_k)\\ \vdots\\ (-1)^{k} f_{2\rho-1}(\la_k)
  \end{array}\!\!\!\right)
  \;=\;\left(\!\!\!\begin{array}{c}  
  f_1(\la_k)\\ \vdots\\ f_{\rho-1}(\la_k)
  \\[.15cm] \hline\\[-.3cm]  
  0\\ \vdots\\ 0
  \\[.15cm] \hline\\[-.3cm]  
  (-1)^{k}f_1(\la_k)\\ \vdots\\ (-1)^{k}f_{\rho-1}(\la_k)
  \\[.15cm] \hline\\[-.3cm]  
  0\\ \vdots\\ 0
  \end{array}\!\!\!\right)
\label{Ek0}
\ee
are eigenvectors of $\Ec$
\be
 \Ec E_0\;=\;\la_0 E_0,\qquad 
 \Ec E_k\;=\;\la_k E_k,\qquad 
 \Ec E_k^{(0)}\;=\;\la_k E_k^{(0)},\qquad 
 \Ec E_\rho\;=\;\la_\rho E_\rho 
\ee
The vectors $E_k$ and $E_k^{(0)}$ are readily seen to be linearly independent.
For every $k\in\mathbb{Z}_{1,\rho-1}$, it is likewise verified that $E_k^{(0)}$ together with 
\be
 E_k^{(1)}\;=\;\left(\!\!\!\begin{array}{c}  
  f'_1(\la_k)\\ \vdots\\ f'_{\rho-1}(\la_k)
  \\[.15cm] \hline\\[-.3cm]  
  f'_{\rho}(\la_k)\\ \vdots\\ f'_{2\rho-1}(\la_k)
  \\[.15cm] \hline\\[-.3cm]  
  (-1)^k f'_1(\la_k)\\ \vdots\\ (-1)^k f'_{\rho-1}(\la_k)
  \\[.15cm] \hline\\[-.3cm]  
  (-1)^k f'_{\rho}(\la_k)\\ \vdots\\ (-1)^k f'_{2\rho-1}(\la_k)
  \end{array}\!\!\!\right),\qquad
 E_k^{(2)}\;=\;\left(\!\!\!\begin{array}{c}  
  \tfrac{1}{2}f''_1(\la_k)\\ \vdots\\ \tfrac{1}{2}f''_{\rho-1}(\la_k)
  \\[.15cm] \hline\\[-.3cm]  
  \tfrac{1}{2}f''_{\rho}(\la_k)\\ \vdots\\ \tfrac{1}{2}f''_{2\rho-1}(\la_k)
  \\[.15cm] \hline\\[-.3cm]  
  \tfrac{1}{2}(-1)^k f''_1(\la_k)\\ \vdots\\ \tfrac{1}{2}(-1)^k f''_{\rho-1}(\la_k)
  \\[.15cm] \hline\\[-.3cm]  
  \tfrac{1}{2}(-1)^k f''_{\rho}(\la_k)\\ \vdots\\ \tfrac{1}{2}(-1)^k f''_{2\rho-1}(\la_k)
  \end{array}\!\!\!\right)
\label{Ek12}
\ee
form the Jordan chain
\be
 \Ec E_k^{(0)}\;=\;\la_k E_k^{(0)},\qquad
 \Ec E_k^{(1)}\;=\;\la_k E_k^{(1)}+E_k^{(0)},\qquad
 \Ec E_k^{(2)}\;=\;\la_k E_k^{(2)}+E_k^{(1)}
\ee
 
Finally, the $(4\rho-2)$-dimensional matrix $Q_\Ec$ is constructed by concatenating 
the generalized eigenvectors (\ref{Ej}), (\ref{Ek0}) and (\ref{Ek12})
\be
 Q_\Ec\;=\;\left(\!\!\begin{array}{c|cccc|c|cccc|c} E_0&E_1&E_1^{(0)}&E_1^{(1)}&E_1^{(2)}&\ldots&
    E_{\rho-1}&E_{\rho-1}^{(0)}&E_{\rho-1}^{(1)}&E_{\rho-1}^{(2)}&E_\rho  \end{array}\!\!\right)
\label{QE}
\ee
By a similarity transformation, this matrix converts $\Ec$ into its Jordan canonical form
\be
 J_\Ec\;=\;Q_\Ec^{-1}\Ec Q_\Ec
   \;=\;\mathrm{diag}\big(\la_0;\la_1,\Jc_{\la_1,3};\ldots;\la_{\rho-1},\Jc_{\la_{\rho-1},3};
      \la_\rho\big)
\label{JE}
\ee

\section{Fundamental and auxiliary fusion graphs}
\label{SecFundAux}

\subsection{Fundamental fusion graphs}
\label{SecFundFusGra}

The graphs associated to the two fundamental modules are called 
{\em fundamental fusion graphs} and consist of certain connected components.
Every such subgraph is a tadpole graph or an eye-patch graph.

The fundamental fusion graph, whose adjacency matrix is 
given by $X$, consists of $p'-1$ tadpole graphs and $p'$ 
eye-patch graphs, all with order parameter $\rho=p$. For every $b\in\mathbb{Z}_{1,p'-1}$, 
the $3p-1$ vertices of the tadpole graphs are given by 
\be
 \Tc_p:\quad 
  \big(L_r,U_a,R,D_a\big)\;=\;\big(\ketw{r,b},\ketw{\R_{p,b}^{a,0}},\ketw{2p,b},
    \ketw{\R_{2p,b}^{a,0}}\big),
    \qquad r\in\mathbb{Z}_{1,p},\quad a\in\mathbb{Z}_{1,p-1}
\ee
while for every $\beta\in\mathbb{Z}_{0,p'-1}$, the $4p-2$ vertices of the eye-patch graphs
are given by
\be
 \Ec_p:\quad
  \big(L_r,U_a,R_r,D_a\big)\;=\;\big(\ketw{\R_{r,p'}^{0,\beta}},\ketw{\R_{p,p'}^{a,\beta}},
    \ketw{\R_{r,2p'}^{0,\beta}},\ketw{\R_{2p,p'}^{a,\beta}}\big),
    \qquad r\in\mathbb{Z}_{1,p},\quad a\in\mathbb{Z}_{1,p-1}
\ee
Likewise, the fundamental fusion graph, whose adjacency matrix is 
given by $Y$, consists of  $p-1$ tadpole graphs and $p$ 
eye-patch graphs, all with order parameter $\rho=p'$. For every $a\in\mathbb{Z}_{1,p-1}$, 
the $3p'-1$ vertices of the tadpole graphs are given by 
\be
 \Tc_{p'}:\quad 
  \big(L_s,U_b,R,D_b\big)\;=\;\big(\ketw{a,s},\ketw{\R_{a,p'}^{0,b}},\ketw{a,2p'},
    \ketw{\R_{a,2p'}^{0,b}}\big),
    \qquad s\in\mathbb{Z}_{1,p'},\quad b\in\mathbb{Z}_{1,p'-1}
\ee
while for every $\al\in\mathbb{Z}_{0,p-1}$, the $4p'-2$ vertices of the eye-patch graphs
are given by
\be
 \Ec_{p'}:\quad
  \big(L_s,U_b,R_s,D_b\big)\;=\;\big(\ketw{\R_{p,s}^{\al,0}},\ketw{\R_{p,p'}^{\al,b}},
    \ketw{\R_{2p,s}^{\al,0}},\ketw{\R_{p,2p'}^{\al,b}}\big),
    \qquad s\in\mathbb{Z}_{1,p'},\quad b\in\mathbb{Z}_{1,p'-1}
\ee

In accord with the results obtained in~\cite{Ras0908} on ${\cal WLM}(1,p')$, the graph
corresponding to $X$ consists of $2p'-1$ order-2 graphs (\ref{Gp1})
\be
 \big(L,R\big)\;\in\;\Big\{\big(\ketw{1,b},\ketw{2,b}\big),
 \big(\Wc(\D_{1,p'}),\Wc(\D_{2,p'})\big),
 \big(\ketw{\R_{1,p'}^{0,b}},\ketw{\R_{1,2p'}^{0,b}}\big)\Big\},\qquad
 b\in\mathbb{Z}_{1,p'-1}
\ee
where $\D_{2,p'}=\D_{1,2p'}$, while there is a single eye-patch graph associated to $Y$
\be
 \big(L_s,U_b,R_s,D_b\big)\;=\;\big(\Wc(\D_{1,s}),\ketw{\R_{1,p'}^{0,b}},
    \Wc(\D_{2,s}),\ketw{\R_{1,2p'}^{0,b}}\big),
    \qquad s\in\mathbb{Z}_{1,p'},\quad b\in\mathbb{Z}_{1,p'-1}
\ee

\subsubsection{${\cal W}$-extended critical percolation ${\cal WLM}(2,3)$}

${\cal W}$-extended critical percolation is described by ${\cal WLM}(2,3)$ where
$p=2$ and $p'=3$. In this case,
the fundamental fusion graph, whose adjacency matrix is given by $X$, consists 
of five connected components, all with order parameter 2. The order of the graph is 28.
The connected components are the two tadpole graphs 
\be
 \mbox{
 \begin{picture}(100,80)(100,-35)
    \unitlength=0.75cm
  \thinlines
\put(2.8,0){$\ketw{1,1}$}
\put(4.55,0.15){\vector(1,0){0.75}}
\put(5.45,0){$\ketw{2,1}$}
\put(7.45,1){\vector(-1,-1){0.5}}
\put(7.3,0.85){\vector(-1,-1){0.5}}
\put(7.05,0.6){\vector(1,1){0.5}}
\put(7.25,1.4){$\ketw{\R_{2,1}^{1,0}}$}
\put(8.4,1.1){\vector(1,-1){0.6}}
\put(8.55,0.95){\vector(1,-1){0.6}}
\put(7.55,-1){\vector(-1,1){0.6}}
\put(7.4,-0.85){\vector(-1,1){0.6}}
\put(7.25,-1.6){$\ketw{\R_{4,1}^{1,0}}$}
\put(8.5,-0.85){\vector(1,1){0.5}}
\put(8.65,-0.7){\vector(1,1){0.5}}
\put(8.9,-0.45){\vector(-1,-1){0.5}}
\put(9.25,-0.05){$\ketw{4,1}$}
 \end{picture}
}
\hspace{4cm}
 \mbox{
 \begin{picture}(100,80)(100,-35)
    \unitlength=0.75cm
  \thinlines
\put(2.8,0){$\ketw{1,2}$}
\put(4.55,0.15){\vector(1,0){0.75}}
\put(5.45,0){$\ketw{2,2}$}
\put(7.45,1){\vector(-1,-1){0.5}}
\put(7.3,0.85){\vector(-1,-1){0.5}}
\put(7.05,0.6){\vector(1,1){0.5}}
\put(7.25,1.4){$\ketw{\R_{2,2}^{1,0}}$}
\put(8.4,1.1){\vector(1,-1){0.6}}
\put(8.55,0.95){\vector(1,-1){0.6}}
\put(7.55,-1){\vector(-1,1){0.6}}
\put(7.4,-0.85){\vector(-1,1){0.6}}
\put(7.25,-1.6){$\ketw{\R_{4,2}^{1,0}}$}
\put(8.5,-0.85){\vector(1,1){0.5}}
\put(8.65,-0.7){\vector(1,1){0.5}}
\put(8.9,-0.45){\vector(-1,-1){0.5}}
\put(9.25,-0.05){$\ketw{4,2}$}
 \end{picture}
}
\ee
and the three eye-patch graphs
\be
 \mbox{
 \begin{picture}(100,80)(120,-35)
    \unitlength=0.75cm
  \thinlines
\put(2.8,0){$\ketw{1,3}$}
\put(4.55,0.15){\vector(1,0){0.75}}
\put(5.45,0){$\ketw{2,3}$}
\put(7.45,1){\vector(-1,-1){0.5}}
\put(7.3,0.85){\vector(-1,-1){0.5}}
\put(7.05,0.6){\vector(1,1){0.5}}
\put(7.25,1.4){$\ketw{\R_{2,3}^{1,0}}$}
\put(8.4,1.1){\vector(1,-1){0.6}}
\put(8.55,0.95){\vector(1,-1){0.6}}
\put(7.55,-1){\vector(-1,1){0.6}}
\put(7.4,-0.85){\vector(-1,1){0.6}}
\put(7.25,-1.6){$\ketw{\R_{4,3}^{1,0}}$}
\put(8.5,-0.85){\vector(1,1){0.5}}
\put(8.65,-0.7){\vector(1,1){0.5}}
\put(8.9,-0.45){\vector(-1,-1){0.5}}
\put(9.25,-0.05){$\ketw{2,6}$}
\put(11.7,0.1){\vector(-1,0){0.75}}
\put(11.85,0){$\ketw{1,6}$}
 \end{picture}
}
\label{X23E1}
\ee
and
\be
 \mbox{
 \begin{picture}(100,80)(120,-35)
    \unitlength=0.75cm
  \thinlines
\put(2.4,0){$\ketw{\R_{1,3}^{0,1}}$}
\put(4.35,0.15){\vector(1,0){0.75}}
\put(5.25,0){$\ketw{\R_{2,3}^{0,1}}$}
\put(7.45,1){\vector(-1,-1){0.5}}
\put(7.3,0.85){\vector(-1,-1){0.5}}
\put(7.05,0.6){\vector(1,1){0.5}}
\put(7.25,1.4){$\ketw{\R_{2,3}^{1,1}}$}
\put(8.4,1.1){\vector(1,-1){0.6}}
\put(8.55,0.95){\vector(1,-1){0.6}}
\put(7.55,-1){\vector(-1,1){0.6}}
\put(7.4,-0.85){\vector(-1,1){0.6}}
\put(7.25,-1.6){$\ketw{\R_{4,3}^{1,1}}$}
\put(8.5,-0.85){\vector(1,1){0.5}}
\put(8.65,-0.7){\vector(1,1){0.5}}
\put(8.9,-0.45){\vector(-1,-1){0.5}}
\put(9.25,-0.05){$\ketw{\R_{2,6}^{0,1}}$}
\put(11.9,0.1){\vector(-1,0){0.75}}
\put(12.05,0){$\ketw{\R_{1,6}^{0,1}}$}
 \end{picture}
}
\label{X23E2}
\ee
and
\be
 \mbox{
 \begin{picture}(100,80)(120,-35)
    \unitlength=0.75cm
  \thinlines
\put(2.4,0){$\ketw{\R_{1,3}^{0,2}}$}
\put(4.35,0.15){\vector(1,0){0.75}}
\put(5.25,0){$\ketw{\R_{2,3}^{0,2}}$}
\put(7.45,1){\vector(-1,-1){0.5}}
\put(7.3,0.85){\vector(-1,-1){0.5}}
\put(7.05,0.6){\vector(1,1){0.5}}
\put(7.25,1.4){$\ketw{\R_{2,3}^{1,2}}$}
\put(8.4,1.1){\vector(1,-1){0.6}}
\put(8.55,0.95){\vector(1,-1){0.6}}
\put(7.55,-1){\vector(-1,1){0.6}}
\put(7.4,-0.85){\vector(-1,1){0.6}}
\put(7.25,-1.6){$\ketw{\R_{4,3}^{1,2}}$}
\put(8.5,-0.85){\vector(1,1){0.5}}
\put(8.65,-0.7){\vector(1,1){0.5}}
\put(8.9,-0.45){\vector(-1,-1){0.5}}
\put(9.25,-0.05){$\ketw{\R_{2,6}^{0,2}}$}
\put(11.9,0.1){\vector(-1,0){0.75}}
\put(12.05,0){$\ketw{\R_{1,6}^{0,2}}$}
 \end{picture}
}
\label{X23E3}
\ee
Likewise, the fundamental fusion graph, whose adjacency matrix is given by $Y$, consists 
of three connected components, all with order parameter 3. The order of the graph is 28.
The connected components are the single tadpole graph
\be
 \mbox{
 \begin{picture}(100,80)(100,-35)
    \unitlength=0.75cm
  \thinlines
\put(0,0){$\ketw{1,1}$}
\put(2.1,0.15){\vector(-1,0){0.4}}
\put(2.1,0.15){\vector(1,0){0.45}}
\put(2.75,0){$\ketw{1,2}$}
\put(4.45,0.15){\vector(1,0){0.85}}
\put(5.4,0){$\ketw{1,3}$}
\put(7.45,1){\vector(-1,-1){0.5}}
\put(7.3,0.85){\vector(-1,-1){0.5}}
\put(7.05,0.6){\vector(1,1){0.5}}
\put(7.55,1.4){$\ketw{\R_{1,3}^{0,1}}$}
\put(9.9,1.55){\vector(-1,0){0.45}}
\put(9.9,1.55){\vector(1,0){0.45}}
\put(10.5,1.4){$\ketw{\R_{1,3}^{0,2}}$}
\put(12,1.1){\vector(1,-1){0.6}}
\put(12.15,0.95){\vector(1,-1){0.6}}
\put(7.55,-1){\vector(-1,1){0.6}}
\put(7.4,-0.85){\vector(-1,1){0.6}}
\put(7.55,-1.4){$\ketw{\R_{1,6}^{0,2}}$}
\put(9.9,-1.25){\vector(-1,0){0.45}}
\put(9.9,-1.25){\vector(1,0){0.45}}
\put(10.5,-1.4){$\ketw{\R_{1,6}^{0,1}}$}
\put(12.1,-0.85){\vector(1,1){0.5}}
\put(12.25,-0.7){\vector(1,1){0.5}}
\put(12.5,-0.45){\vector(-1,-1){0.5}}
\put(12.8,-0.05){$\ketw{1,6}$}
 \end{picture}
}
\ee
and the two eye-patch graphs
\be
 \mbox{
 \begin{picture}(100,80)(160,-35)
    \unitlength=0.75cm
  \thinlines
\put(0,0){$\ketw{2,1}$}
\put(2.1,0.15){\vector(-1,0){0.4}}
\put(2.1,0.15){\vector(1,0){0.45}}
\put(2.75,0){$\ketw{2,2}$}
\put(4.45,0.15){\vector(1,0){0.85}}
\put(5.4,0){$\ketw{2,3}$}
\put(7.45,1){\vector(-1,-1){0.5}}
\put(7.3,0.85){\vector(-1,-1){0.5}}
\put(7.05,0.6){\vector(1,1){0.5}}
\put(7.55,1.4){$\ketw{\R_{2,3}^{0,1}}$}
\put(9.9,1.55){\vector(-1,0){0.45}}
\put(9.9,1.55){\vector(1,0){0.45}}
\put(10.5,1.4){$\ketw{\R_{2,3}^{0,2}}$}
\put(12,1.1){\vector(1,-1){0.6}}
\put(12.15,0.95){\vector(1,-1){0.6}}
\put(7.55,-1){\vector(-1,1){0.6}}
\put(7.4,-0.85){\vector(-1,1){0.6}}
\put(7.55,-1.4){$\ketw{\R_{2,6}^{0,2}}$}
\put(9.9,-1.25){\vector(-1,0){0.45}}
\put(9.9,-1.25){\vector(1,0){0.45}}
\put(10.5,-1.4){$\ketw{\R_{2,6}^{0,1}}$}
\put(12.1,-0.85){\vector(1,1){0.5}}
\put(12.25,-0.7){\vector(1,1){0.5}}
\put(12.5,-0.45){\vector(-1,-1){0.5}}
\put(12.8,-0.05){$\ketw{4,3}$}
\put(15.3,0.1){\vector(-1,0){0.85}}
\put(15.45,-0.05){$\ketw{4,2}$}
\put(17.55,0.1){\vector(-1,0){0.4}}
\put(17.55,0.1){\vector(1,0){0.45}}
\put(18.15,-0.05){$\ketw{4,1}$}
 \end{picture}
}
\label{Y23E1}
\ee
and
\be
 \mbox{
 \begin{picture}(100,80)(160,-35)
    \unitlength=0.75cm
  \thinlines
\put(-0.4,0){$\ketw{\R_{2,1}^{1,0}}$}
\put(1.85,0.15){\vector(-1,0){0.4}}
\put(1.85,0.15){\vector(1,0){0.45}}
\put(2.4,0){$\ketw{\R_{2,2}^{1,0}}$}
\put(4.3,0.15){\vector(1,0){0.85}}
\put(5.25,0){$\ketw{\R_{2,3}^{1,0}}$}
\put(7.45,1){\vector(-1,-1){0.5}}
\put(7.3,0.85){\vector(-1,-1){0.5}}
\put(7.05,0.6){\vector(1,1){0.5}}
\put(7.55,1.4){$\ketw{\R_{2,3}^{1,1}}$}
\put(9.9,1.55){\vector(-1,0){0.45}}
\put(9.9,1.55){\vector(1,0){0.45}}
\put(10.5,1.4){$\ketw{\R_{2,3}^{1,2}}$}
\put(12,1.1){\vector(1,-1){0.6}}
\put(12.15,0.95){\vector(1,-1){0.6}}
\put(7.55,-1){\vector(-1,1){0.6}}
\put(7.4,-0.85){\vector(-1,1){0.6}}
\put(7.55,-1.4){$\ketw{\R_{2,6}^{1,2}}$}
\put(9.9,-1.25){\vector(-1,0){0.45}}
\put(9.9,-1.25){\vector(1,0){0.45}}
\put(10.5,-1.4){$\ketw{\R_{2,6}^{1,1}}$}
\put(12.1,-0.85){\vector(1,1){0.5}}
\put(12.25,-0.7){\vector(1,1){0.5}}
\put(12.5,-0.45){\vector(-1,-1){0.5}}
\put(12.8,-0.05){$\ketw{\R_{4,3}^{1,0}}$}
\put(15.5,0.1){\vector(-1,0){0.85}}
\put(15.65,-0.05){$\ketw{\R_{4,2}^{1,0}}$}
\put(17.9,0.1){\vector(-1,0){0.4}}
\put(17.9,0.1){\vector(1,0){0.45}}
\put(18.5,-0.05){$\ketw{\R_{4,1}^{1,0}}$}
 \end{picture}
}
\label{Y23E2}
\ee
We recall that 
\be
 \ketw{4,3}\;=\;\Wc(\D_{4,3})\;=\;\Wc(\D_{2,6})\;=\;\ketw{2,6},\quad
 \ketw{\R_{4,3}^{1,1}}\;=\;\ketw{\R_{2,6}^{1,1}},\quad
 \ketw{\R_{4,3}^{1,2}}\;=\;\ketw{\R_{2,6}^{1,2}}
\ee

\subsection{Fundamental fusion matrices}

We choose to work with the basis
\bea
 &&\Big\{\ketw{1,1},\ldots,\ketw{1,p'},\ketw{\R_{1,p'}^{0,1}},\ldots,\ketw{\R_{1,p'}^{0,p'-1}},\nn
    &&\hspace{4.5cm}\ketw{1,2p'},\ketw{\R_{1,2p'}^{0,1}},\ldots,\ketw{\R_{1,2p'}^{0,p'-1}},\nn
 &&\vdots\nn
 &&  \ketw{p-1,1},\ldots,\ketw{p-1,p'},\ketw{\R_{p-1,p'}^{0,1}},\ldots,\ketw{\R_{p-1,p'}^{0,p'-1}},\nn
   &&\hspace{4.5cm}\ketw{p-1,2p'},\ketw{\R_{p-1,2p'}^{0,1}},\ldots,\ketw{\R_{p-1,2p'}^{0,p'-1}},\nn
 &&\ketw{p,1},\ldots,\ketw{p,p'},\ketw{\R_{p,p'}^{0,1}},\ldots,\ketw{\R_{p,p'}^{0,p'-1}},\nn
   &&\hspace{4.5cm}\ketw{2p,1},\ldots,\ketw{2p,p'},
    \ketw{\R_{p,2p'}^{0,1}},\ldots,\ketw{\R_{p,2p'}^{0,p'-1}},\nn
 &&\ketw{\R_{p,1}^{1,0}},\ldots,\ketw{\R_{p,p'}^{1,0}},\ketw{\R_{p,p'}^{1,1}},\ldots,
    \ketw{\R_{p,p'}^{1,p'-1}},\nn
   &&\hspace{4.5cm}\ketw{\R_{2p,1}^{1,0}},\ldots,\ketw{\R_{2p,p'}^{1,0}},
    \ketw{\R_{p,2p'}^{1,1}},\ldots,\ketw{\R_{p,2p'}^{1,p'-1}},\nn
 &&\vdots\nn
 &&\ketw{\R_{p,1}^{p-1,0}},\ldots,\ketw{\R_{p,p'}^{p-1,0}},\ketw{\R_{p,p'}^{p-1,1}},\ldots,
    \ketw{\R_{p,p'}^{p-1,p'-1}},\nn
   &&\hspace{4.5cm}\ketw{\R_{2p,1}^{p-1,0}},\ldots,\ketw{\R_{2p,p'}^{p-1,0}},
    \ketw{\R_{p,2p'}^{p-1,1}},\ldots,\ketw{\R_{p,2p'}^{p-1,p'-1}}\Big\}
\label{Ybasis}
\eea
in which $Y$ has the simple form
\be
 Y\;=\;\mathrm{diag}\big(
   \underbrace{\Tc_{p'},\ldots,\Tc_{p'}}_{p-1},\,\underbrace{\Ec_{p'},\ldots,\Ec_{p'}}_{p}\big)
\label{Ydiag}
\ee
while $X$ is the matrix
\be
 X\;=\;\Pc^{-1}\mathrm{diag}\big(
   \underbrace{\Tc_{p},\ldots,\Tc_{p}}_{p'-1},\,\underbrace{\Ec_{p},\ldots,\Ec_{p}}_{p'}\big)\Pc\;=\;
\left(\!\!
\begin{array}{ccccc|c|cccccc}
  0&I&&&&&&&&& \\
  I&0&&&&&&&&& \\
  &&\ddots&&&&&&&& \\
  &&&0&I&&&&&& \\
  &&&I&0&\It&&&&& \\
\hline
  &&&&0&0&I&&&& \\
\hline
  &&&&&2I&0&I&&& \\
  &&&&&&I&0&&& \\
  &&&&&&&&\ddots& \\
  &&&&&&&&&0&I \\
  &&&&&2C&&&&I&0
\end{array}
\!\!\right)
\label{X}
\ee
Here $\Pc$ is a permutation matrix, while the $(4p'-2)\times(4p'-2)$-dimensional matrix $C$ 
and the $(3p'-1)\times(4p'-2)$-dimensional matrix $\It$ are given by
\be
 C\;=\;\left(\!\!
\begin{array}{cc}
  0&I \\
  I&0    \end{array}  \!\!\right),\qquad
 \It\;=\;\left(\!\!
\begin{array}{ccc}
  I_{(2p'-1)\times(2p'-1)}&0_{(2p'-1)\times(p'-1)}&0_{(2p'-1)\times(p')} \\
  0_{(p')\times(2p'-1)}&0_{(p')\times(p'-1)}&I_{(p')\times(p')}
\end{array}
\!\!\right)
\ee
In (\ref{X}), $X$ is written as a $(2p-1)\times(2p-1)$-dimensional matrix whose entries
are blocks. Every block (indicated by $0$ or $I$) to the left of the leftmost vertical delimiter 
has $3p'-1$ columns, while every block 
(indicated by $0$, $I$, $\It$, $2I$ or $2C$) to the right of this delimiter has $4p'-2$ columns.
Likewise, every block (indicated by $0$, $I$ or $\It$) above the upper vertical delimiter has 
$3p'-1$ rows, while every block (indicated by $0$, $I$, $2I$ or $2C$)
below this delimiter has $4p'-2$ rows.
For small values of $p$, the matrix $X$ in (\ref{X}) is meant to reduce to
\be
 X\big|_{p=2}\;=\;
\left(\!\!
\begin{array}{c|c|c}
  0&\It&0 \\
\hline
  0&0&I \\
\hline
  0&2I+2C&0
\end{array}
\!\!\right),\qquad
 X\big|_{p=1}\;=\;2C
\label{X2X1}
\ee
with $p'$-dependent dimensions of the blocks given as above.

For later convenience, it is noted that 
\be
 \It E_0\;=\;T_0,\qquad \It E_k\;=\;T_k^{(0)},
   \qquad \It E_k^{(\ell)}\;=\;T_k^{(\ell)},\qquad \It E_{p'}\;=\;T_{p'}
\label{ItE}
\ee
and 
\be
 C E_0\;=\;E_0,\qquad C E_k\;=\;(-1)^{k-1}C_k,
   \qquad C E_k^{(\ell)}\;=\;(-1)^k E_k^{(\ell)},\qquad C E_{p'}\;=\;(-1)^{p'}E_{p'}
\label{CE}
\ee
where $k\in\mathbb{Z}_{1,p'-1}$ and $\ell\in\mathbb{Z}_{0,2}$, while the order parameter 
appearing in the entries of the vectors is $\rho=p'$. We also note that the basis used 
in~\cite{Ras0908} on ${\cal WLM}(1,p')$ is different from (\ref{Ybasis}) for $p=1$
\bea
 \!\!\ketw{p,1},\ldots,\ketw{p,p'},\ketw{\R_{p,p'}^{0,1}},\ldots,\ketw{\R_{p,p'}^{0,p'-1}},
   \ketw{2p,1},\ldots,\ketw{2p,p'},
    \ketw{\R_{p,2p'}^{0,1}},\ldots,\ketw{\R_{p,2p'}^{0,p'-1}}
\eea
The two bases are related by a permutation.

\subsection{Auxiliary fusion graphs}

The two auxiliary fusion graphs consist of certain connected components.
Every such subgraph is a cycle graph or an eye-patch graph.

The auxiliary fusion graph, whose adjacency matrix is 
given by $\Xh$, consists of  $p'-1$ cycle graphs and $p'$ 
eye-patch graphs, all with order parameter $\rho=p$. For every $b\in\mathbb{Z}_{1,p'-1}$, 
the $2p$ vertices of the cycle graphs are given by 
\be
 \Cc_p:\quad 
  \big(L,U_a,R,D_a\big)\;=\;\big(\ketw{p,b},\ketw{\R_{p,b}^{a,0}},\ketw{2p,b},
    \ketw{\R_{2p,b}^{a,0}}\big),
    \qquad a\in\mathbb{Z}_{1,p-1}
\ee
while for every $\beta\in\mathbb{Z}_{0,p'-1}$, the $4p-2$ vertices of the eye-patch graphs
are given by
\be
 \Ec_p:\quad
  \big(L_r,U_a,R_r,D_a\big)\;=\;\big(\ketw{\R_{r,p'}^{0,\beta}},\ketw{\R_{p,p'}^{a,\beta}},
    \ketw{\R_{r,2p'}^{0,\beta}},\ketw{\R_{2p,p'}^{a,\beta}}\big),
    \qquad r\in\mathbb{Z}_{1,p},\quad a\in\mathbb{Z}_{1,p-1}
\ee
Likewise,the auxiliary fusion graph, whose adjacency matrix is 
given by $\Yh$, consists of $p-1$ cycle graphs and $p$ 
eye-patch graphs, all with order parameter $\rho=p'$. For every $a\in\mathbb{Z}_{1,p-1}$, 
the $2p'$ vertices of the cycle graphs are given by 
\be
 \Cc_{p'}:\quad 
  \big(L,U_b,R,D_b\big)\;=\;\big(\ketw{a,p'},\ketw{\R_{a,p'}^{0,b}},\ketw{a,2p'},
    \ketw{\R_{a,2p'}^{0,b}}\big),
    \qquad b\in\mathbb{Z}_{1,p'-1}
\ee
while for every $\al\in\mathbb{Z}_{0,p-1}$, the $4p'-2$ vertices of the eye-patch graphs
are given by
\be
 \Ec_{p'}:\quad
  \big(L_s,U_b,R_s,D_b\big)\;=\;\big(\ketw{\R_{p,s}^{\al,0}},\ketw{\R_{p,p'}^{\al,b}},
    \ketw{\R_{2p,s}^{\al,0}},\ketw{\R_{p,2p'}^{\al,b}}\big),
    \qquad s\in\mathbb{Z}_{1,p'},\quad b\in\mathbb{Z}_{1,p'-1}
\ee

\subsubsection{${\cal W}$-extended critical percolation ${\cal WLM}(2,3)$}

In the case of ${\cal W}$-extended critical percolation ${\cal WLM}(2,3)$, there are exactly 
two indecomposable modules (\ref{ab}) in the fundamental fusion algebra {\em not}
associated with boundary conditions, namely the identity $\ketw{1,1}$ and
the fundamental module $\ketw{1,2}$.
The auxiliary fusion graph, whose adjacency matrix is given by $\Xh$, consists 
of five connected components, all with order parameter 2. The order of the graph is 26.
The connected components are the two cycle graphs 
\be
 \mbox{
 \begin{picture}(100,80)(120,-35)
    \unitlength=0.75cm
  \thinlines
\put(5.45,0){$\ketw{2,1}$}
\put(7.45,1){\vector(-1,-1){0.5}}
\put(7.3,0.85){\vector(-1,-1){0.5}}
\put(7.05,0.6){\vector(1,1){0.5}}
\put(7.25,1.4){$\ketw{\R_{2,1}^{1,0}}$}
\put(8.4,1.1){\vector(1,-1){0.6}}
\put(8.55,0.95){\vector(1,-1){0.6}}
\put(7.55,-1){\vector(-1,1){0.6}}
\put(7.4,-0.85){\vector(-1,1){0.6}}
\put(7.25,-1.6){$\ketw{\R_{4,1}^{1,0}}$}
\put(8.5,-0.85){\vector(1,1){0.5}}
\put(8.65,-0.7){\vector(1,1){0.5}}
\put(8.9,-0.45){\vector(-1,-1){0.5}}
\put(9.25,-0.05){$\ketw{4,1}$}
 \end{picture}
}
\hspace{3cm}
 \mbox{
 \begin{picture}(100,80)(120,-35)
    \unitlength=0.75cm
  \thinlines
\put(5.45,0){$\ketw{2,2}$}
\put(7.45,1){\vector(-1,-1){0.5}}
\put(7.3,0.85){\vector(-1,-1){0.5}}
\put(7.05,0.6){\vector(1,1){0.5}}
\put(7.25,1.4){$\ketw{\R_{2,2}^{1,0}}$}
\put(8.4,1.1){\vector(1,-1){0.6}}
\put(8.55,0.95){\vector(1,-1){0.6}}
\put(7.55,-1){\vector(-1,1){0.6}}
\put(7.4,-0.85){\vector(-1,1){0.6}}
\put(7.25,-1.6){$\ketw{\R_{4,2}^{1,0}}$}
\put(8.5,-0.85){\vector(1,1){0.5}}
\put(8.65,-0.7){\vector(1,1){0.5}}
\put(8.9,-0.45){\vector(-1,-1){0.5}}
\put(9.25,-0.05){$\ketw{4,2}$}
 \end{picture}
}
\ee
and the three eye-patch graphs (\ref{X23E1}), (\ref{X23E2}) and (\ref{X23E3}).
Likewise, the auxiliary fusion graph, whose adjacency matrix is given by $\Yh$, consists 
of three connected components, all with order parameter 3. The order of the graph is 26.
The connected components are the single cycle graph
\be
 \mbox{
 \begin{picture}(100,80)(160,-35)
    \unitlength=0.75cm
  \thinlines
\put(5.4,0){$\ketw{1,3}$}
\put(7.45,1){\vector(-1,-1){0.5}}
\put(7.3,0.85){\vector(-1,-1){0.5}}
\put(7.05,0.6){\vector(1,1){0.5}}
\put(7.55,1.4){$\ketw{\R_{1,3}^{0,1}}$}
\put(9.9,1.55){\vector(-1,0){0.45}}
\put(9.9,1.55){\vector(1,0){0.45}}
\put(10.5,1.4){$\ketw{\R_{1,3}^{0,2}}$}
\put(12,1.1){\vector(1,-1){0.6}}
\put(12.15,0.95){\vector(1,-1){0.6}}
\put(7.55,-1){\vector(-1,1){0.6}}
\put(7.4,-0.85){\vector(-1,1){0.6}}
\put(7.55,-1.4){$\ketw{\R_{1,6}^{0,2}}$}
\put(9.9,-1.25){\vector(-1,0){0.45}}
\put(9.9,-1.25){\vector(1,0){0.45}}
\put(10.5,-1.4){$\ketw{\R_{1,6}^{0,1}}$}
\put(12.1,-0.85){\vector(1,1){0.5}}
\put(12.25,-0.7){\vector(1,1){0.5}}
\put(12.5,-0.45){\vector(-1,-1){0.5}}
\put(12.8,-0.05){$\ketw{1,6}$}
 \end{picture}
}
\ee
and the two eye-patch graphs (\ref{Y23E1}) and (\ref{Y23E2}).

\subsection{Auxiliary fusion matrices}

We choose to work with the basis
\bea
 &&\Big\{\ketw{1,p'},\ketw{\R_{1,p'}^{0,1}},\ldots,\ketw{\R_{1,p'}^{0,p'-1}},
      \ketw{1,2p'},\ketw{\R_{1,2p'}^{0,1}},\ldots,\ketw{\R_{1,2p'}^{0,p'-1}},\nn
 &&\vdots\nn
 &&  \ketw{p-1,p'},\ketw{\R_{p-1,p'}^{0,1}},\ldots,\ketw{\R_{p-1,p'}^{0,p'-1}},
      \ketw{p-1,2p'},\ketw{\R_{p-1,2p'}^{0,1}},\ldots,\ketw{\R_{p-1,2p'}^{0,p'-1}},\nn
 &&\ketw{p,1},\ldots,\ketw{p,p'},\ketw{\R_{p,p'}^{0,1}},\ldots,\ketw{\R_{p,p'}^{0,p'-1}},\nn
   &&\hspace{4.5cm}\ketw{2p,1},\ldots,\ketw{2p,p'},
    \ketw{\R_{p,2p'}^{0,1}},\ldots,\ketw{\R_{p,2p'}^{0,p'-1}},\nn
 &&\ketw{\R_{p,1}^{1,0}},\ldots,\ketw{\R_{p,p'}^{1,0}},\ketw{\R_{p,p'}^{1,1}},\ldots,
    \ketw{\R_{p,p'}^{1,p'-1}},\nn
   &&\hspace{4.5cm}\ketw{\R_{2p,1}^{1,0}},\ldots,\ketw{\R_{2p,p'}^{1,0}},
    \ketw{\R_{p,2p'}^{1,1}},\ldots,\ketw{\R_{p,2p'}^{1,p'-1}},\nn
 &&\vdots\nn
 &&\ketw{\R_{p,1}^{p-1,0}},\ldots,\ketw{\R_{p,p'}^{p-1,0}},\ketw{\R_{p,p'}^{p-1,1}},\ldots,
    \ketw{\R_{p,p'}^{p-1,p'-1}},\nn
   &&\hspace{4.5cm}\ketw{\R_{2p,1}^{p-1,0}},\ldots,\ketw{\R_{2p,p'}^{p-1,0}},
    \ketw{\R_{p,2p'}^{p-1,1}},\ldots,\ketw{\R_{p,2p'}^{p-1,p'-1}}\Big\}
\label{Yhbasis}
\eea
in which $\Yh$ has the simple form
\be
 \Yh\;=\;\mathrm{diag}\big(
   \underbrace{\Cc_{p'},\ldots,\Cc_{p'}}_{p-1},\,\underbrace{\Ec_{p'},\ldots,\Ec_{p'}}_{p}\big)
\label{Yhdiag}
\ee
while $\Xh$ is the matrix
\be
 \Xh\;=\;\Pch^{-1}\mathrm{diag}\big(
   \underbrace{\Cc_{p},\ldots,\Cc_{p}}_{p'-1},\,\underbrace{\Ec_{p},\ldots,\Ec_{p}}_{p'}\big)\Pch\;=\;
\left(\!\!
\begin{array}{ccccc|c|cccccc}
  0&I&&&&&&&&& \\
  I&0&&&&&&&&& \\
  &&\ddots&&&&&&&& \\
  &&&0&I&&&&&& \\
  &&&I&0&\Ih&&&&& \\
\hline
  &&&&0&0&I&&&& \\
\hline
  &&&&&2I&0&I&&& \\
  &&&&&&I&0&&& \\
  &&&&&&&&\ddots& \\
  &&&&&&&&&0&I \\
  &&&&&2C&&&&I&0
\end{array}
\!\!\right)
\label{Xh}
\ee
Here $\Pch$ is a permutation matrix, while the $(4p'-2)\times(4p'-2)$-dimensional matrix $C$ and
the $(2p')\times(4p'-2)$-dimensional matrix $\Ih$ are given by
\be
 C\;=\;\left(\!\!
\begin{array}{cc}
  0&I \\
  I&0    \end{array}  \!\!\right),\qquad
 \Ih\;=\;\left(\!\!
\begin{array}{cccc}
  0_{(p')\times(p'-1)}&I_{(p')\times(p')}&0_{(p')\times(p'-1)}&0_{(p')\times(p')} \\
  0_{(p')\times(p'-1)}&0_{(p')\times(p')}&0_{(p')\times(p'-1)}&I_{(p')\times(p')}
\end{array}
\!\!\right)
\ee
In (\ref{Xh}), $\Xh$ is written as a $(2p-1)\times(2p-1)$-dimensional matrix whose entries
are blocks. Every block (indicated by $0$ or $I$) to the left of the leftmost vertical delimiter 
has $2p'$ columns, while every block 
(indicated by $0$, $I$, $\Ih$, $2I$ or $2C$) to the right of this delimiter has $4p'-2$ columns.
Likewise, every block (indicated by $0$, $I$ or $\Ih$) above the upper vertical delimiter has 
$2p'$ rows, while every block (indicated by $0$, $I$, $2I$ or $2C$)
below this delimiter has $4p'-2$ rows.
For small values of $p$, the matrix $\Xh$ in (\ref{Xh}) is meant to reduce to
\be
 \Xh\big|_{p=2}\;=\;
\left(\!\!
\begin{array}{c|c|c}
  0&\Ih&0 \\
\hline
  0&0&I \\
\hline
  0&2I+2C&0
\end{array}
\!\!\right),\qquad
 \Xh\big|_{p=1}\;=\;2C
\label{Xh2Xh1}
\ee
with $p'$-dependent dimensions of the blocks given as above.
For later convenience, it is noted that 
\be
 \Ih E_0\;=\;C_0,\qquad \Ih E_k\;=\;\Ih E_k^{(0)}\;=\;0,
   \qquad \Ih E_k^{(1)}\;=\;C_k^{(0)},\qquad \Ih E_k^{(2)}\;=\;C_k^{(1)},\qquad \Ih E_{p'}\;=\;C_{p'}
\label{IhE}
\ee
where $k\in\mathbb{Z}_{1,p'-1}$, while the order parameter appearing in the entries of
the vectors is $\rho=p'$.

\section{Spectral decomposition of fusion matrices}
\label{SecSpectral}

The objective here is to examine to what extent the fusion matrices $N_\mu$ (or $\Nh_\mu$)
can be simultaneously brought to Jordan form. Our first goal is thus to
devise a similarity transformation in the form of a
matrix $Q$ ($\Qh$) which simultaneously brings the fundamental (auxiliary) fusion matrices 
$X$ and $Y$ ($\Xh$ and $\Yh$) to Jordan form. For $p>1$, this is only
possible modulo permutation similarity.
With the Jordan decompositions of $Y$ and $\Yh$ implemented, the best we can do is therefore 
\be
 Q^{-1}XQ\;=\;P^{-1}J_XP,\qquad\quad Q^{-1}YQ\;=\; J_Y
\label{QXQ}
\ee
and 
\be
 \Qh^{-1}\Xh\Qh\;=\;\Ph^{-1}J_{\Xh}\Ph,\qquad\quad \Qh^{-1}\Yh\Qh\;=\; J_{\Yh}
\label{QhXhQh}
\ee
where $J_X$ and $J_Y$ ($J_{\Xh}$ and $J_{\Yh}$) are {\em Jordan canonical forms} of 
$X$ and $Y$ ($\Xh$ and $\Yh$), while $P$ ($\Ph$) is a permutation matrix.
For every fusion matrix $N_\mu$ in (\ref{NNN}), it then follows that
\bea
 Q^{-1}N_\mu Q&=&Q^{-1}\mathrm{pol}_{\mu}(X,Y)Q
  \;=\;\mathrm{pol}_{\mu}(Q^{-1}XQ,Q^{-1}YQ)
  \;=\;\mathrm{pol}_{\mu}(P^{-1}J_XP,J_Y)\nn
  &=&P^{-1}\mathrm{pol}_{\mu}^{(x)}(J_X)P\,\mathrm{pol}_{\mu}^{(y)}(J_Y)
\label{QNQ}
\eea
where we have used that the polynomials $\mathrm{pol}_{\mu}(X,Y)$ (\ref{abR4}) 
factorize and thus can be written as
\be
 \mathrm{pol}_{\mu}(X,Y)\;=\;\mathrm{pol}_{\mu}^{(x)}(X)\,\mathrm{pol}_{\mu}^{(y)}(Y)
\label{polXY}
\ee
As we will demonstrate, $Q^{-1}N_\mu Q$ is a {\em block-diagonal matrix whose blocks
are upper-triangular matrices of dimension 1, 3, 5 or 9}, while $P$ is a {\em symmetric}
permutation matrix.
Likewise, for every $\Nh_\mu$ in (\ref{NNNb}), it follows that
\bea
 \Qh^{-1}\Nh_\mu\Qh&=&\Qh^{-1}\mathrm{pol}_{\mu}(\Xh,\Yh)\Qh
  \;=\;\mathrm{pol}_{\mu}(\Qh^{-1}\Xh\Qh,\Qh^{-1}\Yh\Qh)
  \;=\;\mathrm{pol}_{\mu}(\Ph^{-1}J_{\Xh}\Ph,J_{\Yh})\nn
  &=&\Ph^{-1}\mathrm{pol}_{\mu}^{(x)}(J_{\Xh})\Ph\,\mathrm{pol}_{\mu}^{(y)}(J_{\Yh})
\label{QhNhQh}
\eea
where $\Qh^{-1}\Nh_\mu \Qh$ turns out to be a {\em block-diagonal matrix whose blocks
are upper-triangular matrices of dimension 1, 2, 3 or 8}, while $\Ph$ is a {\em symmetric}
permutation matrix.
By reversing the conjugation in (\ref{QNQ}) or (\ref{QhNhQh}), one obtains an explicit
expression for the given fusion matrix.
We will describe the relations (\ref{QNQ}) and (\ref{QhNhQh}) in detail in
Section~\ref{SecGeneral} and Section~\ref{SecGeneralHat}.

\subsection{Jordan webs}

As discussed in Appendix~\ref{AppJordan}, two commuting matrices need not share
a complete set of common generalized eigenvectors. However, we 
will demonstrate that the two fundamental (or the two auxiliary) adjacency matrices $X$ and $Y$ 
($\Xh$ and $\Yh$) {\em do} have a common complete set of generalized eigenvectors. 
These generalized eigenvectors
are organized as a web constructed by interlacing the Jordan chains of the two matrices.
We refer to such a web as a {\em Jordan web}. It consists of a number of 
connected components or subwebs which we will characterize in the following.

\subsubsection{Fundamental fusion matrices}

With respect to $X$ or $Y$ separately, we only encounter Jordan chains of length 1 or 3.
As we will demonstrate in Section~\ref{SecFundWebs}, five different types of connected 
Jordan webs arise in the description of the common generalized eigenvectors 
$G_{\la,\la'}^{(\ell,\ell')}$ of $X$ and $Y$
\bea
 XG_{\la,\la'}^{(0,\ell')}\;=\;\la G_{\la,\la'}^{(0,\ell')},\qquad
 & XG_{\la,\la'}^{(1,\ell')}\;=\;\la G_{\la,\la'}^{(1,\ell')}+G_{\la,\la'}^{(0,\ell')},&\qquad
  XG_{\la,\la'}^{(2,\ell')}\;=\;\la G_{\la,\la'}^{(2,\ell')}+G_{\la,\la'}^{(1,\ell')}\nn
 YG_{\la,\la'}^{(\ell,0)}\;=\;\la' G_{\la,\la'}^{(\ell,0)},\qquad
 & YG_{\la,\la'}^{(\ell,1)}\;=\;\la' G_{\la,\la'}^{(\ell,1)}+G_{\la,\la'}^{(\ell,0)},&\qquad
  YG_{\la,\la'}^{(\ell,2)}\;=\;\la' G_{\la,\la'}^{(\ell,2)}+G_{\la,\la'}^{(\ell,1)}
\label{XYG}
\eea
These Jordan webs are
\be
\mbox{
 \begin{picture}(100,100)(40,0)
    \unitlength=0.75cm
  \thinlines
\put(-0.2,2){$W_{\la,\la'}^{(1,1)}:$}
\put(2,2){$G_{\la,\la'}^{(0,0)}$}
 \end{picture}
}
\hspace{.4cm}
\mbox{
 \begin{picture}(100,100)(40,0)
    \unitlength=0.75cm
  \thinlines
\put(-0.2,2){$W_{\la,\la'}^{(1,3)}:$}
\put(2,0){$G_{\la,\la'}^{(0,0)}$}
\put(2,2){$G_{\la,\la'}^{(0,1)}$}
\put(2,4){$G_{\la,\la'}^{(0,2)}$}
\put(2.4,1.6){\vector(0,-1){0.8}}
\put(2.4,3.6){\vector(0,-1){0.8}}
 \end{picture}
}
\hspace{.4cm}
 \mbox{
 \begin{picture}(100,100)(40,0)
    \unitlength=0.75cm
  \thinlines
\put(-0.2,2){$W_{\la,\la'}^{(3,1)}:$}
\put(2,2){$G_{\la,\la'}^{(0,0)}$}
\put(5,2){$G_{\la,\la'}^{(1,0)}$}
\put(8,2){$G_{\la,\la'}^{(2,0)}$}
\put(4.5,2.1){\vector(-1,0){1}}
\put(7.5,2.1){\vector(-1,0){1}}
 \end{picture}
}
\label{web13}
\ee
and
\be
\mbox{
 \begin{picture}(100,100)(60,0)
    \unitlength=0.75cm
  \thinlines
\put(-0.2,2){$W_{\la,\la'}^{(3,3)^\dagger}:$}
\put(2,0){$G_{\la,\la'}^{(0,0)}$}
\put(5,0){$G_{\la,\la'}^{(1,0)}$}
\put(8,0){$G_{\la,\la'}^{(2,0)}$}
\put(2,2){$G_{\la,\la'}^{(0,1)}$}
\put(2,4){$G_{\la,\la'}^{(0,2)}$}
\put(4.5,0.1){\vector(-1,0){1}}
\put(7.5,0.1){\vector(-1,0){1}}
\put(2.4,1.6){\vector(0,-1){0.8}}
\put(2.4,3.6){\vector(0,-1){0.8}}
 \end{picture}
}
\hspace{3.5cm}
 \mbox{
 \begin{picture}(100,100)(40,0)
    \unitlength=0.75cm
  \thinlines
\put(-0.2,2){$W_{\la,\la'}^{(3,3)}:$}
\put(2,0){$G_{\la,\la'}^{(0,0)}$}
\put(5,0){$G_{\la,\la'}^{(1,0)}$}
\put(8,0){$G_{\la,\la'}^{(2,0)}$}
\put(2,2){$G_{\la,\la'}^{(0,1)}$}
\put(5,2){$G_{\la,\la'}^{(1,1)}$}
\put(8,2){$G_{\la,\la'}^{(2,1)}$}
\put(2,4){$G_{\la,\la'}^{(0,2)}$}
\put(5,4){$G_{\la,\la'}^{(1,2)}$}
\put(8,4){$G_{\la,\la'}^{(2,2)}$}
\put(4.5,0.1){\vector(-1,0){1}}
\put(7.5,0.1){\vector(-1,0){1}}
\put(4.5,2.1){\vector(-1,0){1}}
\put(7.5,2.1){\vector(-1,0){1}}
\put(4.5,4.1){\vector(-1,0){1}}
\put(7.5,4.1){\vector(-1,0){1}}
\put(2.4,1.6){\vector(0,-1){0.8}}
\put(5.4,1.6){\vector(0,-1){0.8}}
\put(8.4,1.6){\vector(0,-1){0.8}}
\put(2.4,3.6){\vector(0,-1){0.8}}
\put(5.4,3.6){\vector(0,-1){0.8}}
\put(8.4,3.6){\vector(0,-1){0.8}}
 \end{picture}
}
\label{web59}
\ee
\\[-.35cm]
where a horizontal arrow from $G_{\la,\la'}$ to $G'_{\la,\la'}$ indicates that 
$XG_{\la,\la'}=\la G_{\la,\la'}+G'_{\la,\la'}$, while a vertical arrow from 
$G_{\la,\la'}$ to $G'_{\la,\la'}$ indicates that $YG_{\la,\la'}=\la' G_{\la,\la'}+G'_{\la,\la'}$.
We note that these five connected Jordan webs are all subwebs of $W_{\la,\la'}^{(3,3)}$. 
A web of the type $W_{\la,\la'}^{(1,1)}$ is merely a common eigenvector of 
$X$ and $Y$ not appearing in any non-trivial Jordan chain.

To describe the matrix realizations of the restrictions of $X$ and $Y$ to these connected
Jordan webs, we introduce an ordered basis $B_{\la,\la'}^{(\ell,\ell')}$ of generalized eigenvectors 
associated to $W_{\la,\la'}^{(\ell,\ell')}$. As always, we favour $Y$ and thus introduce
\bea
 &B_{\la,\la'}^{(1,1)}\;=\;\big\{G_{\la,\la'}^{(0,0)}\big\},\qquad
 B_{\la,\la'}^{(1,3)}\;=\;\big\{G_{\la,\la'}^{(0,0)},G_{\la,\la'}^{(0,1)},G_{\la,\la'}^{(0,2)}\big\},\qquad
 B_{\la,\la'}^{(3,1)}\;=\;\big\{G_{\la,\la'}^{(0,0)},G_{\la,\la'}^{(1,0)},G_{\la,\la'}^{(2,0)}\big\}\nn
 &B_{\la,\la'}^{(3,3)^\dagger}\;=\;\big\{G_{\la,\la'}^{(0,0)},G_{\la,\la'}^{(0,1)},G_{\la,\la'}^{(0,2)},
     G_{\la,\la'}^{(1,0)},G_{\la,\la'}^{(2,0)}\big\}\nn
 &B_{\la,\la'}^{(3,3)}\;=\;\big\{G_{\la,\la'}^{(0,0)},G_{\la,\la'}^{(0,1)},G_{\la,\la'}^{(0,2)},
     G_{\la,\la'}^{(1,0)},G_{\la,\la'}^{(1,1)},G_{\la,\la'}^{(1,2)},
     G_{\la,\la'}^{(2,0)},G_{\la,\la'}^{(2,1)},G_{\la,\la'}^{(2,2)}\big\}
\label{B}
\eea
The corresponding matrix realizations are denoted by $X_{\la,\la'}^{(\ell,\ell')}$
and $Y_{\la,\la'}^{(\ell,\ell')}$ and are given by
\bea
 &X_{\la,\la'}^{(1,1)}\;=\;\la I_{1\times1},\qquad
 X_{\la,\la'}^{(1,3)}\;=\;\la I_{3\times3},\qquad
 X_{\la,\la'}^{(3,1)}\;=\; \Jc_{\la,3}\nn
 &X_{\la,\la'}^{(3,3)^\dagger}\;=\;\begin{pmatrix} 
    \la&0&0&1&0\\ &\la&0&0&0\\ &&\la&0&0\\ &&&\la&1\\ &&&&\la\end{pmatrix},\qquad
 X_{\la,\la'}^{(3,3)}
  \;=\;\begin{pmatrix} \la I&I&0\\ 0&\la I&I\\   0&0&\la I \end{pmatrix}
\label{Xweb}
\eea
and
\bea
 &Y_{\la,\la'}^{(1,1)}\;=\;\la' I_{1\times1},\qquad
 Y_{\la,\la'}^{(1,3)}\;=\; \Jc_{\la',3},\qquad
 Y_{\la,\la'}^{(3,1)}\;=\;\la' I_{3\times3} \nn
 &Y_{\la,\la'}^{(3,3)^\dagger}\;=\;\mathrm{diag}\big(\Jc_{\la',3},\la',\la'\big),\qquad
 Y_{\la,\la'}^{(3,3)}  \;=\;\mathrm{diag}\big(\Jc_{\la',3},\Jc_{\la',3},\Jc_{\la',3}\big)
\label{Yweb}
\eea
In (\ref{Xweb}), the nine-dimensional matrix $X_{\la,\la'}^{(3,3)}$ is written as a three-dimensional
matrix whose entries are three-dimensional matrices.

\subsubsection{Auxiliary fusion matrices}
\label{SecAuxFusMat}

With respect to $\Xh$ or $\Yh$ separately, we only encounter Jordan chains of length 1, 2 or 3.
As we will demonstrate in Section~\ref{SecAuxWebs}, six different types of 
connected Jordan webs 
arise in the description of the common generalized eigenvectors of $\Xh$ and $\Yh$
\bea
 \Xh\Gh_{\la,\la'}^{(0,\ell')}\;=\;\la \Gh_{\la,\la'}^{(0,\ell')},\qquad
 & \Xh\Gh_{\la,\la'}^{(1,\ell')}\;=\;\la \Gh_{\la,\la'}^{(1,\ell')}+\Gh_{\la,\la'}^{(0,\ell')},&\qquad
  \Xh\Gh_{\la,\la'}^{(2,\ell')}\;=\;\la \Gh_{\la,\la'}^{(2,\ell')}+\Gh_{\la,\la'}^{(1,\ell')}\nn
 \Yh\Gh_{\la,\la'}^{(\ell,0)}\;=\;\la' \Gh_{\la,\la'}^{(\ell,0)},\qquad
 & \Yh\Gh_{\la,\la'}^{(\ell,1)}\;=\;\la' \Gh_{\la,\la'}^{(\ell,1)}+\Gh_{\la,\la'}^{(\ell,0)},&\qquad
  \Yh\Gh_{\la,\la'}^{(\ell,2)}\;=\;\la' \Gh_{\la,\la'}^{(\ell,2)}+\Gh_{\la,\la'}^{(\ell,1)}
\label{XYGh}
\eea
Three of these Jordan webs are inherited from $X$ and $Y$ as
$W_{\la,\la'}^{(1,1)}\to\Wh_{\la,\la'}^{(1,1)}$, 
$W_{\la,\la'}^{(3,1)}\to\Wh_{\la,\la'}^{(3,1)}$ and
$W_{\la,\la'}^{(1,3)}\to\Wh_{\la,\la'}^{(1,3)}$, that is,
\be
\mbox{
 \begin{picture}(100,100)(40,0)
    \unitlength=0.75cm
  \thinlines
\put(-0.2,2){$\Wh_{\la,\la'}^{(1,1)}:$}
\put(2,2){$\Gh_{\la,\la'}^{(0,0)}$}
 \end{picture}
}
\hspace{.4cm}
\mbox{
 \begin{picture}(100,100)(40,0)
    \unitlength=0.75cm
  \thinlines
\put(-0.2,2){$\Wh_{\la,\la'}^{(1,3)}:$}
\put(2,0){$\Gh_{\la,\la'}^{(0,0)}$}
\put(2,2){$\Gh_{\la,\la'}^{(0,1)}$}
\put(2,4){$\Gh_{\la,\la'}^{(0,2)}$}
\put(2.4,1.6){\vector(0,-1){0.8}}
\put(2.4,3.6){\vector(0,-1){0.8}}
 \end{picture}
}
\hspace{.4cm}
 \mbox{
 \begin{picture}(100,100)(40,0)
    \unitlength=0.75cm
  \thinlines
\put(-0.2,2){$\Wh_{\la,\la'}^{(3,1)}:$}
\put(2,2){$\Gh_{\la,\la'}^{(0,0)}$}
\put(5,2){$\Gh_{\la,\la'}^{(1,0)}$}
\put(8,2){$\Gh_{\la,\la'}^{(2,0)}$}
\put(4.5,2.1){\vector(-1,0){1}}
\put(7.5,2.1){\vector(-1,0){1}}
 \end{picture}
}
\label{webh13}
\ee
The Jordan web $W_{\la,\la'}^{(3,3)^\dagger}$\!, on the other hand, breaks down as only
the quotient $W_{\la,\la'}^{(3,3)^\dagger}\!/G_{\la,\la'}^{(0,0)}$ survives the reduction to 
$\Xh$ and $\Yh$ in the sense that
$W_{\la,\la'}^{(3,3)^\dagger}\to\Wh_{\la,\la'}^{(2,1)}\cup\Wh_{\la,\la'}^{(1,2)}$, where
\be
\mbox{
 \begin{picture}(100,60)(10,-5)
    \unitlength=0.75cm
  \thinlines
\put(-0.2,1.15){$\Wh_{\la,\la'}^{(1,2)}:$}
\put(2,0){$\Gh_{\la,\la'}^{(0,0)}$}
\put(2,2){$\Gh_{\la,\la'}^{(0,1)}$}
\put(2.4,1.6){\vector(0,-1){0.8}}
 \end{picture}
}
\hspace{.6cm}
 \mbox{
 \begin{picture}(100,60)(10,-5)
    \unitlength=0.75cm
  \thinlines
\put(-0.2,1.15){$\Wh_{\la,\la'}^{(2,1)}:$}
\put(2,1.15){$\Gh_{\la,\la'}^{(0,0)}$}
\put(5,1.15){$\Gh_{\la,\la'}^{(1,0)}$}
\put(4.5,1.25){\vector(-1,0){1}}
 \end{picture}
}
\label{webh1221}
\ee
Finally, only the quotient $W_{\la,\la'}^{(3,3)}/G_{\la,\la'}^{(0,0)}$ survives
the reduction of the Jordan web $W_{\la,\la'}^{(3,3)}$ to $\Xh$ and $\Yh$. 
This eight-dimensional connected Jordan web $\Wh_{\la,\la'}^{(3,3)}$ is given by
\be
 \mbox{
 \begin{picture}(100,100)(40,0)
    \unitlength=0.75cm
  \thinlines
\put(-0.2,2){$\Wh_{\la,\la'}^{(3,3)}:$}
\put(5,0){$\Gh_{\la,\la'}^{(1,0)}$}
\put(8,0){$\Gh_{\la,\la'}^{(2,0)}$}
\put(2,2){$\Gh_{\la,\la'}^{(0,1)}$}
\put(5,2){$\Gh_{\la,\la'}^{(1,1)}$}
\put(8,2){$\Gh_{\la,\la'}^{(2,1)}$}
\put(2,4){$\Gh_{\la,\la'}^{(0,2)}$}
\put(5,4){$\Gh_{\la,\la'}^{(1,2)}$}
\put(8,4){$\Gh_{\la,\la'}^{(2,2)}$}
\put(7.5,0.1){\vector(-1,0){1}}
\put(4.5,2.1){\vector(-1,0){1}}
\put(7.5,2.1){\vector(-1,0){1}}
\put(4.5,4.1){\vector(-1,0){1}}
\put(7.5,4.1){\vector(-1,0){1}}
\put(5.4,1.6){\vector(0,-1){0.8}}
\put(8.4,1.6){\vector(0,-1){0.8}}
\put(2.4,3.6){\vector(0,-1){0.8}}
\put(5.4,3.6){\vector(0,-1){0.8}}
\put(8.4,3.6){\vector(0,-1){0.8}}
 \end{picture}
}
\label{webh8}
\ee
where, with reference to (\ref{XYGh}), $\Gh_{\la,\la'}^{(0,0)}\equiv0$.

To describe the matrix realizations of the restrictions of $\Xh$ and $\Yh$ to these connected
Jordan webs, we introduce the $\Yh$-favouring ordered bases $\Bh_{\la,\la'}^{(\ell,\ell')}$ 
of generalized eigenvectors associated to $\Wh_{\la,\la'}^{(\ell,\ell')}$ by
\bea
 &\Bh_{\la,\la'}^{(1,1)}\;=\;\big\{\Gh_{\la,\la'}^{(0,0)}\big\},\qquad
 \Bh_{\la,\la'}^{(1,3)}\;=\;\big\{\Gh_{\la,\la'}^{(0,0)},\Gh_{\la,\la'}^{(0,1)},
    \Gh_{\la,\la'}^{(0,2)}\big\},\qquad
 \Bh_{\la,\la'}^{(3,1)}\;=\;\big\{\Gh_{\la,\la'}^{(0,0)},\Gh_{\la,\la'}^{(1,0)},\Gh_{\la,\la'}^{(2,0)}\big\}\nn
 &\Bh_{\la,\la'}^{(1,2)}\;=\;\big\{\Gh_{\la,\la'}^{(0,0)},\Gh_{\la,\la'}^{(0,1)}\big\},\qquad
 \Bh_{\la,\la'}^{(2,1)}\;=\;\big\{\Gh_{\la,\la'}^{(0,0)},\Gh_{\la,\la'}^{(1,0)}\big\}\nn
 &\Bh_{\la,\la'}^{(3,3)}\;=\;\big\{\Gh_{\la,\la'}^{(0,1)},\Gh_{\la,\la'}^{(0,2)},
     \Gh_{\la,\la'}^{(1,0)},\Gh_{\la,\la'}^{(1,1)},\Gh_{\la,\la'}^{(1,2)},
     \Gh_{\la,\la'}^{(2,0)},\Gh_{\la,\la'}^{(2,1)},\Gh_{\la,\la'}^{(2,2)}\big\}
\label{Bh}
\eea
The corresponding matrix realizations are denoted by $\Xh_{\la,\la'}^{(\ell,\ell')}$
and $\Yh_{\la,\la'}^{(\ell,\ell')}$ and are given by
\bea
 &\Xh_{\la,\la'}^{(1,1)}\;=\;\la I_{1\times1},\qquad
 \Xh_{\la,\la'}^{(1,3)}\;=\;\la I_{3\times3},\qquad
 \Xh_{\la,\la'}^{(3,1)}\;=\; \Jc_{\la,3}\nn
 &\Xh_{\la,\la'}^{(1,2)}\;=\;\la I_{2\times2},\qquad
 \Xh_{\la,\la'}^{(2,1)}\;=\; \Jc_{\la,2},\qquad
 \Xh_{\la,\la'}^{(3,3)}
  \;=\;\left(\!\!
\begin{array}{cc|ccc|ccc}
  \la&0&0&1&0&0&0&0 \\ 
  &\la&0&0&1&0&0&0 \\ 
\hline
  &&\la&0&0&1&0&0 \\ 
  &&&\la&0&0&1&0 \\ 
  &&&&\la&0&0&1 \\ 
\hline
  &&&&&\la&0&0 \\ 
  &&&&&&\la&0 \\ 
  &&&&&&&\la
\end{array}
\!\!\right)
\label{Xhweb}
\eea
and
\bea
 &\Yh_{\la,\la'}^{(1,1)}\;=\;\la' I_{1\times1},\qquad
 \Yh_{\la,\la'}^{(1,3)}\;=\; \Jc_{\la',3},\qquad
 \Yh_{\la,\la'}^{(3,1)}\;=\;\la' I_{3\times3}\nn
 &\Yh_{\la,\la'}^{(1,2)}\;=\; \Jc_{\la',2},\qquad
 \Yh_{\la,\la'}^{(2,1)}\;=\;\la' I_{2\times2},\qquad
 \Yh_{\la,\la'}^{(3,3)}  \;=\;\mathrm{diag}\big(\Jc_{\la',2},\Jc_{\la',3},\Jc_{\la',3}\big)
\label{Yhweb}
\eea
The eight-dimensional matrix $\Xh_{\la,\la'}^{(3,3)}$ in (\ref{Xhweb}) is obtained from the 
nine-dimensional matrix $X_{\la,\la'}^{(3,3)}$ in (\ref{Xweb}) by elimination of the first row 
and column. By similar eliminations, the eight-dimensional matrix $\Yh_{\la,\la'}^{(3,3)}$ 
in (\ref{Yhweb}) is obtained from the nine-dimensional matrix $Y_{\la,\la'}^{(3,3)}$ in (\ref{Yweb}),
while the four-dimensional matrices $\mathrm{diag}(\Xh_{\la,\la'}^{(1,2)},\Xh_{\la,\la'}^{(2,1)})$
and $\mathrm{diag}(\Yh_{\la,\la'}^{(1,2)},\Yh_{\la,\la'}^{(2,1)})$
follow from the five-dimensional matrices $X_{\la,\la'}^{(3,3)^\dagger}$
and $Y_{\la,\la'}^{(3,3)^\dagger}$\!, respectively.

\subsection{Fundamental fusion algebra}

In the following, we write
\be
 \beta_i\;=\;2\cos\frac{i\pi}{p},\quad i\in\mathbb{Z}_{0,p},\qquad\quad
 \beta_j'\;=\;2\cos\frac{j\pi}{p'},\quad j\in\mathbb{Z}_{0,p'}
\label{betabeta}
\ee
We also recall our label conventions $a\in\mathbb{Z}_{1,p-1}$ and 
$b\in\mathbb{Z}_{1,p'-1}$ introduced in (\ref{kkabrs}).

\subsubsection{Fundamental fusion matrices}
\label{SecFundWebs}

Due to the block-diagonal structure (\ref{Ydiag}) of the fundamental fusion matrix $Y$, 
its spectral decomposition
follows readily from the spectral decompositions of $\Tc_{p'}$ and $\Ec_{p'}$ discussed
in Section~\ref{SecTad} and Section~\ref{SecEye}, respectively. The Jordan canonical form of
$Y$ thus consists of $2p-1$ rank-1 blocks of eigenvalue $\beta'_j$ for every $j\in\{0,p'\}$, 
$p$ rank-1 blocks of eigenvalue $\beta'_b$ for every $b\in\mathbb{Z}_{1,p'-1}$, 
and $2p-1$ rank-3 blocks of eigenvalue $\beta'_b$ for every $b\in\mathbb{Z}_{1,p'-1}$.
Likewise, the Jordan canonical form of
$X$ consists of $2p'-1$ rank-1 blocks of eigenvalue $\beta_i$ for every $i\in\{0,p\}$, 
$p'$ rank-1 blocks of eigenvalue $\beta_a$ for every $a\in\mathbb{Z}_{1,p-1}$, 
and $2p'-1$ rank-3 blocks of eigenvalue $\beta_a$ for every $a\in\mathbb{Z}_{1,p-1}$.

To characterize the connected components of the Jordan web of the complete set of
common generalized eigenvectors of $X$ and $Y$, we choose to work in the 
$Y$-favouring basis (\ref{Ybasis}). A generalized vector $G_{\la,\la'}^{(\ell,\ell')}$ can thus
be written as a $(2p-1)$-dimensional vector whose $p-1$ upper entries are 
$(3p'-1)$-dimensional vectors of the type $T$ appearing in Section~\ref{SecTad},
while the $p$ lower entries are $(4p'-2)$-dimensional vectors of the type $E$
appearing in Section~\ref{SecEye}.

The connected subwebs of the type $W_{\la,\la'}^{(1,1)}$ are given by the following eigenvectors
\be
 W_{\beta_i,\beta'_j}^{(1,1)}:\quad G_{\beta_i,\beta'_j}^{(0,0)}\;=\;\left(\!\!\!\begin{array}{c}  
  f_{1}(\beta_i)T_j\\ \vdots\\ f_{p-1}(\beta_i)T_j
  \\[.15cm] \hline\\[-.3cm]  
  f_{p}(\beta_i)E_j\\ \vdots\\ f_{2p-1}(\beta_i)E_j
  \end{array}\!\!\!\right),\qquad 
 i+j\ \mathrm{even},\quad i\in\{0,p\},\ j\in\{0,p'\}
\label{Wij11}
\ee
or
\be
 W_{\beta_a,\beta'_j}^{(1,1)}:\quad G_{\beta_a,\beta'_j}^{(0,0)}\;=\;\left(\!\!\!\begin{array}{c}  
  f_{1}(\beta_a)T_j\\ \vdots\\ f_{p-1}(\beta_a)T_j
  \\[.15cm] \hline\\[-.3cm]  
  f_{p}(\beta_a)E_j\\ \vdots\\ f_{2p-1}(\beta_a)E_j
  \end{array}\!\!\!\right)\;=\;
 \left(\!\!\!\begin{array}{c}  
  f_{1}(\beta_a)T_j\\ \vdots\\ f_{p-1}(\beta_a)T_j
  \\[.15cm] \hline\\[-.3cm]  
  0\\ \vdots\\ 0
  \end{array}\!\!\!\right),\qquad 
 a+j\ \mathrm{odd},\quad j\in\{0,p'\}
\ee
or
\be
 W_{\beta_i,\beta'_b}^{(1,1)}:\quad G_{\beta_i,\beta'_b}^{(0,0)}\;=\;\left(\!\!\!\begin{array}{c}  
  f_{1}(\beta_i)T_b^{(0)}\\ \vdots\\ f_{p-1}(\beta_i)T_b^{(0)}
  \\[.15cm] \hline\\[-.3cm]  
  f_{p}(\beta_i)E_b\\ \vdots\\ f_{2p-1}(\beta_i)E_b
  \end{array}\!\!\!\right),\qquad 
 i+b\ \mathrm{odd},\quad i\in\{0,p\}
\ee
The connected subwebs of the type $W_{\la,\la'}^{(1,3)}$ consist of the following 
generalized eigenvectors
\be
 W_{\beta_i,\beta'_b}^{(1,3)}:\quad G_{\beta_i,\beta'_b}^{(0,\ell')}\;=\;\left(\!\!\!\begin{array}{c}  
  f_{1}(\beta_i)T_b^{(\ell')}\\ \vdots\\ f_{p-1}(\beta_i)T_b^{(\ell')}
  \\[.15cm] \hline\\[-.3cm]  
  f_{p}(\beta_i)E_b^{(\ell')}\\ \vdots\\ f_{2p-1}(\beta_i)E_b^{(\ell')}
  \end{array}\!\!\!\right),\qquad 
 i+b\ \mathrm{even},\quad i\in\{0,p\},\quad \ell'\in\mathbb{Z}_{0,2}
\ee
The connected subwebs of the type $W_{\la,\la'}^{(3,1)}$ consist of the following 
generalized eigenvectors
\be
 W_{\beta_a,\beta'_j}^{(3,1)}:\quad G_{\beta_a,\beta'_j}^{(\ell,0)}\;=\;\left(\!\!\!\begin{array}{c}  
  \tfrac{1}{\ell!}f_{1}^{(\ell)}(\beta_a)T_j\\ \vdots\\ \tfrac{1}{\ell!}f_{p-1}^{(\ell)}(\beta_a)T_j
  \\[.15cm] \hline\\[-.3cm]  
  \tfrac{1}{\ell!}f_{p}^{(\ell)}(\beta_a)E_j\\ \vdots\\ \tfrac{1}{\ell!}f_{2p-1}^{(\ell)}(\beta_a)E_j
  \end{array}\!\!\!\right),\qquad 
 a+j\ \mathrm{even},\quad j\in\{0,p'\},\quad \ell\in\mathbb{Z}_{0,2}
\ee
The connected subwebs of the type $W_{\la,\la'}^{(3,3)^\dagger}$ consist of the following 
generalized eigenvectors
\be
 W_{\beta_a,\beta'_b}^{(3,3)^\dagger}:\quad 
 G_{\beta_a,\beta'_b}^{(\ell,0)}\;=\;\left(\!\!\!\begin{array}{c}  
  \tfrac{1}{\ell!}f_{1}^{(\ell)}(\beta_a)T_b^{(0)}\\ \vdots\\ \tfrac{1}{\ell!}f_{p-1}^{(\ell)}(\beta_a)T_b^{(0)}
  \\[.15cm] \hline\\[-.3cm]  
  \tfrac{1}{\ell!}f_{p}^{(\ell)}(\beta_a)E_b\\ \vdots\\ \tfrac{1}{\ell!}f_{2p-1}^{(\ell)}(\beta_a)E_b
  \end{array}\!\!\!\right),\quad
 G_{\beta_a,\beta'_b}^{(0,\ell')}\;=\;
   \left(\!\!\!\begin{array}{c}  
  f_{1}(\beta_a)T_b^{(\ell')}\\ \vdots\\ f_{p-1}(\beta_a)T_b^{(\ell')}
  \\[.15cm] \hline\\[-.3cm]  
  f_{p}(\beta_a)E_b^{(\ell')}\\ \vdots\\ f_{2p-1}(\beta_a)E_b^{(\ell')}
  \end{array}\!\!\!\right),\qquad 
      a+b\ \mathrm{odd}, \quad \ell,\ell'\in\mathbb{Z}_{0,2}
\ee
where (\ref{fhlk}) ensures consistency of the two expressions for $G_{\beta_a,\beta'_b}^{(0,0)}$.
Finally, the connected subwebs of the type $W_{\la,\la'}^{(3,3)}$ consist of the following 
generalized eigenvectors
\be
 W_{\beta_a,\beta'_b}^{(3,3)}:\quad G_{\beta_a,\beta'_b}^{(\ell,\ell')}\;=\;\left(\!\!\!\begin{array}{c}  
  \tfrac{1}{\ell!}f_{1}^{(\ell)}(\beta_a)T_b^{(\ell')}\\ \vdots\\ 
      \tfrac{1}{\ell!}f_{p-1}^{(\ell)}(\beta_a)T_b^{(\ell')}
  \\[.15cm] \hline\\[-.3cm]  
  \tfrac{1}{\ell!}f_{p}^{(\ell)}(\beta_a)E_b^{(\ell')}\\ \vdots\\ 
      \tfrac{1}{\ell!}f_{2p-1}^{(\ell)}(\beta_a)E_b^{(\ell')}
  \end{array}\!\!\!\right),\qquad 
 a+b\ \mathrm{even},\quad \ell,\ell'\in\mathbb{Z}_{0,2}
\label{Wab33}
\ee
Using properties of the $T$ and $E$ vectors as generalized eigenvectors of 
$\Tc_{p'}$ and $\Ec_{p'}$, together with (\ref{ItE}) and (\ref{CE}), in particular,
it is straightforward to prove that the vectors given in (\ref{Wij11}) through (\ref{Wab33})
indeed correspond to the Jordan webs (\ref{web13}) and (\ref{web59}) 
consistent with (\ref{XYG}).
We also note that the number $\Nc^{(\ell,\ell')}$ of connected Jordan webs of the type 
$W^{(\ell,\ell')}$ is given by
\be
 \Nc^{(1,1)}=\;p+p',\quad
 \Nc^{(1,3)}=\;p'-1,\quad
 \Nc^{(3,1)}=\;p-1,\quad
 \Nc^{(3,3)^\dagger}=\;\Nc^{(3,3)}=\;\tfrac{1}{2}(p-1)(p'-1)
\ee
consistent with the total number (\ref{CardJFund}) of generalized eigenvectors.
In Appendix~\ref{AppJordanWeb}, 
we list the connected Jordan subwebs $W_{\beta_i,\beta'_j}^{(\ell,\ell')}$ with respect
to the labeling $i,j$ of the corresponding eigenvalues.

The similarity matrix $Q$ appearing in (\ref{QXQ}) is constructed by concatenating 
the common generalized eigenvectors of $X$ and $Y$ according to any ordering of the 
ordered bases (\ref{B}). The permutation matrix $P$
depends on this choice of ordering, and the degree of
convenience of such a choice depends on the intended application. Here we consider 
a general ordering reflecting the partitioning
\be
 \big\{B_{\la,\la'}^{(1,1)}\big\}\cup
 \big\{B_{\la,\la'}^{(1,3)}\big\}\cup
 \big\{B_{\la,\la'}^{(3,1)}\big\}\cup
 \big\{B_{\la,\la'}^{(3,3)^\dagger}\big\}\cup
 \big\{B_{\la,\la'}^{(3,3)}\big\}
\label{BBBBB}
\ee
such that every set (of generalized eigenvectors) of the type $B_{\la,\la'}^{(1,1)}$
comes before every set (of generalized eigenvectors) of the type $B_{\la,\la'}^{(1,3)}$,
and so on.
The Jordan canonical forms $J_X$ and  $J_Y$ in (\ref{QXQ}) are then of the form
\bea
 J_X\!\!&=&\!\!\mathrm{diag}\Big(
   \underbrace{\la,\ldots}_{p+p'},\,
   \underbrace{\la,\ldots}_{3p'-3},\,
   \underbrace{\Jc_{\la,3},\ldots}_{p-1},\,
   \underbrace{\mathrm{diag}\big(\Jc_{\la,3},\la,\la\big),\ldots}_{\tfrac{1}{2}(p-1)(p'-1)},\,
   \underbrace{\mathrm{diag}\big(\Jc_{\la,3},\Jc_{\la,3},\Jc_{\la,3}\big),
       \ldots}_{\tfrac{1}{2}(p-1)(p'-1)}\Big)
\label{JX}
\\
 J_Y\!\!&=&\!\!\mathrm{diag}\Big(
   \underbrace{\la',\ldots}_{p+p'},\,
   \underbrace{\Jc_{\la',3},\ldots}_{p'-1},\,
   \underbrace{\la',\ldots}_{3p-3},\,
   \underbrace{\mathrm{diag}\big(\Jc_{\la',3},\la',\la'\big),\ldots}_{\tfrac{1}{2}(p-1)(p'-1)},\,
   \underbrace{\mathrm{diag}\big(\Jc_{\la',3},\Jc_{\la',3},\Jc_{\la',3}\big),
       \ldots}_{\tfrac{1}{2}(p-1)(p'-1)}\Big)
\eea
We stress that the eigenvalues $\la$ and $\la'$ vary in these expressions but are always of the
form (\ref{betabeta}).
The corresponding permutation matrix $P$ is a block-diagonal matrix whose blocks are 
of dimension 1, 3, 3, 5 or 9, corresponding to the dimensions of the sets 
$B_{\la,\la'}^{(\ell,\ell')}$. By a similarity transformation (\ref{QXQ}), these $P$-blocks convert the 
blocks in $J_X$ into the corresponding upper-triangular matrices $X_{\la,\la'}^{(\ell,\ell')}$ 
in (\ref{Xweb}). The $P$-blocks of dimension 1 or 3 are identity matrices, 
while the $P$-blocks of dimension 5 or 9 are the symmetric permutation matrices
\be
 P_5\;=\;\begin{pmatrix} 1&0&0&0&0\\ 0&0&0&1&0\\ 0&0&0&0&1\\ 0&1&0&0&0\\ 0&0&1&0&0
   \end{pmatrix},\qquad
 P_9\;=\;\begin{pmatrix} 
    1&0&0&0&0&0&0&0&0\\
    0&0&0&1&0&0&0&0&0\\
    0&0&0&0&0&0&1&0&0\\
    0&1&0&0&0&0&0&0&0\\
    0&0&0&0&1&0&0&0&0\\
    0&0&0&0&0&0&0&1&0\\
    0&0&1&0&0&0&0&0&0\\
    0&0&0&0&0&1&0&0&0\\
    0&0&0&0&0&0&0&0&1  \end{pmatrix}
\label{P5P9}
\ee
where it is recalled that a symmetric permutation matrix equals its inverse.
The symmetric permutation matrix $P$ is thus given by
\be
 P\;=\;\mathrm{diag}\Big(\underbrace{1,\ldots,1}_{4p+4p'-6},\,
   \underbrace{P_5,\ldots,P_5}_{\tfrac{1}{2}(p-1)(p'-1)},\,
   \underbrace{P_9,\ldots,P_9}_{\tfrac{1}{2}(p-1)(p'-1)}\Big)
\ee
As actions on the connected Jordan webs $W_{\la,\la'}^{(3,3)^\dagger}$ and $W_{\la,\la'}^{(3,3)}$,
these permutations reflect the vertices (generalized eigenvectors) with respect to the line from
south-west to north-east through $G_{\la,\la'}^{(0,0)}$. 
$P_5$ and $P_9$ thus have one and three fix-points, respectively, in accord with the numbers of
units on their diagonals.

\subsubsection{General fusion matrices}
\label{SecGeneral}

Here we determine the upper-triangular block-diagonal
matrix $Q^{-1}N_\mu Q$ obtained from the general fusion matrix $N_\mu$ by
a similarity transformation with respect to $Q$ defined according to (\ref{BBBBB}).
{}From (\ref{QNQ}) and Section~\ref{SecFundWebs}, we have that 
\bea
 Q^{-1}N_\mu Q\!\!&=&\!\!\mathrm{diag}\Big(
   \underbrace{g(X_{\la,\la'}^{(1,1)})h(Y_{\la,\la'}^{(1,1)}),\ldots}_{p+p'},\,
   \underbrace{g(X_{\la,\la'}^{(1,3)})h(Y_{\la,\la'}^{(1,3)}),\ldots}_{p'-1},\,
   \underbrace{g(X_{\la,\la'}^{(3,1)})h(Y_{\la,\la'}^{(3,1)}),\ldots}_{p-1},\nn
 &&\qquad\qquad\qquad
   \underbrace{g(X_{\la,\la'}^{(3,3)^\dagger})h(Y_{\la,\la'}^{(3,3)^\dagger}),
          \ldots}_{\tfrac{1}{2}(p-1)(p'-1)},\,
   \underbrace{g(X_{\la,\la'}^{(3,3)})h(Y_{\la,\la'}^{(3,3)}),\ldots}_{\tfrac{1}{2}(p-1)(p'-1)}\Big)
\eea
where, for simplicity, $g(z)=\mathrm{pol}_{\mu}^{(x)}(z)$ and 
$h(z)=\mathrm{pol}_{\mu}^{(y)}(z)$. 
In a given block $g(X_{\la,\la'}^{(\ell,\ell')})h(Y_{\la,\la'}^{(\ell,\ell')})$, the pairs of labels $\la,\la'$ 
(eigenvalues (\ref{betabeta}) of $X$ and $Y$) of $X_{\la,\la'}^{(\ell,\ell')}$ and 
$Y_{\la,\la'}^{(\ell,\ell')}$ are the same, while they generally vary from block to block.
For the five types of blocks, we have
\bea
 g(X_{\la,\la'}^{(1,1)})h(Y_{\la,\la'}^{(1,1)})\!\!&=&\!\! g(\la)h(\la')\nn
 g(X_{\la,\la'}^{(1,3)})h(Y_{\la,\la'}^{(1,3)})\!\!&=&\!\! g(\la)h(\Jc_{\la',3}),\qquad
 g(X_{\la,\la'}^{(3,1)})h(Y_{\la,\la'}^{(3,1)})\;=\;g(\Jc_{\la,3})h(\la')\nn
  g(X_{\la,\la'}^{(3,3)^\dagger})h(Y_{\la,\la'}^{(3,3)^\dagger})\!\!&=&\!\!\begin{pmatrix} 
     g(\la)h(\la')&g(\la)h'(\la')&\tfrac{1}{2}g(\la)h''(\la')&g'(\la)h(\la')&\tfrac{1}{2}g''(\la)h(\la')  \\
     0&g(\la)h(\la')&g(\la)h'(\la')&0&0  \\
     0&0&g(\la)h(\la')&0&0  \\
     0&0&0&g(\la)h(\la')&g'(\la)h(\la')  \\
     0&0&0&0&g(\la)h(\la')
    \end{pmatrix}\nn
 g(X_{\la,\la'}^{(3,3)})h(Y_{\la,\la'}^{(3,3)})\!\!&=&\!\! g(\Jc_{\la,3})\times h(\Jc_{\la',3})
\label{gh}
\eea
where $g(\Jc_{\la,3})\times h(\Jc_{\la',3})$ denotes the nine-dimensional Kronecker product of 
the two three-dimensional matrices $g(\Jc_{\la,3})$ and $h(\Jc_{\la',3})$.
It is recalled that, for a function $f$ expandable as a power series in its argument, we have
\be
 f(\Jc_{\la,2})
    \;=\;\begin{pmatrix} f(\la)&f'(\la)
     \\ 0&f(\la)\end{pmatrix} ,\qquad\quad
 f(\Jc_{\la,3})
    \;=\;\begin{pmatrix} f(\la)&f'(\la)&\tfrac{1}{2}f''(\la) 
     \\ 0&f(\la)&f'(\la) \\ 0&0&f(\la) \end{pmatrix}
\ee
whose ranks depend on $f'(\la)$ and $f''(\la)$.
The first of these matrix expressions will be relevant in (\ref{ghhh}) below.
This completes the description of the upper-triangular block-diagonal matrix 
$Q^{-1}N_\mu Q$.

\subsubsection{${\cal W}$-extended critical percolation ${\cal WLM}(2,3)$}

In the case of ${\cal WLM}(2,3)$, the eigenvalues of $X$ and $Y$ are 
\be
 \beta_i\;=\;2\cos\frac{i\pi}{2},\quad i\in\{0,1,2\},\qquad\qquad
 \beta'_j\;=\;2\cos\frac{j\pi}{3},\quad j\in\{0,1,2,3\} 
\label{beta23}
\ee
respectively. As displayed in Figure~\ref{web23}, the connected components 
$W_{\beta_i,\beta'_j}^{(\ell,\ell')}$ of the Jordan web associated to the fundamental
fusion algebra are neatly organized with respect to the labels $i$ and $j$.
\psset{unit=1cm}
\begin{figure}
$$
\renewcommand{\arraystretch}{1.5}
\begin{array}{c|cccc}
 \mbox{}_i\diagdown\mbox{}^j&0&1&2&3\\[4pt]
\hline
\rule{0pt}{16pt}
 0&W_{2,2}^{(1,1)}&W_{2,1}^{(1,1)}&W_{2,-1}^{(1,3)}&$\O$
    \\[4pt]
 1&W_{0,2}^{(1,1)}&W_{0,1}^{(3,3)}&W_{0,-1}^{(3,3)^\dagger}&W_{0,-2}^{(3,1)}
    \\[4pt]
 2&W_{-2,2}^{(1,1)}&W_{-2,1}^{(1,1)}&W_{-2,-1}^{(1,3)}&$\O$
\end{array}
$$
\caption{The connected components $W_{\beta_i,\beta'_j}^{(\ell,\ell')}$ of the Jordan web
associated to the fundamental fusion algebra of ${\cal W}$-extended critical percolation 
${\cal WLM}(2,3)$. The two \O's
indicate that there are no common generalized eigenvectors corresponding to the pair 
$(\beta_0,\beta'_3)=(2,-2)$ or to the pair $(\beta_2,\beta'_3)=(-2,-2)$ of eigenvalues of the 
fundamental fusion matrices $X$ and $Y$.}
\label{web23}
\end{figure}
An example of an ordering of the common generalized eigenvectors of $X$ and $Y$
respecting (\ref{BBBBB}) is
\bea
 &\!\!G_{2,2}^{(0,0)}; G_{0,2}^{(0,0)}; G_{-2,2}^{(0,0)}; G_{2,1}^{(0,0)}; G_{-2,1}^{(0,0)};
  G_{2,-1}^{(0,0)}, G_{2,-1}^{(0,1)}, G_{2,-1}^{(0,2)};
  G_{-2,-1}^{(0,0)}, G_{-2,-1}^{(0,1)}, G_{-2,-1}^{(0,2)};
  G_{0,-2}^{(0,0)}, G_{0,-2}^{(1,0)}, G_{0,-2}^{(2,0)};\nn
 &\!\!G_{0,-1}^{(0,0)}, G_{0,-1}^{(0,1)}, G_{0,-1}^{(0,2)}, G_{0,-1}^{(1,0)}, G_{0,-1}^{(2,0)};
  G_{0,1}^{(0,0)}, G_{0,1}^{(0,1)}, G_{0,1}^{(0,2)}, G_{0,1}^{(1,0)}, G_{0,1}^{(1,1)}, G_{0,1}^{(1,2)},
  G_{0,1}^{(2,0)}, G_{0,1}^{(2,1)}, G_{0,1}^{(2,2)}
\eea
We define the similarity matrix $Q$ by concatenating these vectors in the order given.
Modulo a similarity transformation, $Q$ converts $X$ and $Y$ into the Jordan canonical forms 
\bea
 &J_X\;=\;P^{-1}Q^{-1}XQP\;=\;\mathrm{diag}\big(
   2,0,-2,2,-2,2,2,2,-2,-2,-2,\Jc_{0,3},\Jc_{0,3},0,0,\Jc_{0,3},\Jc_{0,3},\Jc_{0,3}\big) \nn
 &J_Y\;=\;Q^{-1}YQ\;=\;\mathrm{diag}\big(
   2,2,2,1,1,\Jc_{-1,3},\Jc_{-1,3},-2,-2,-2,\Jc_{-1,3},-1,-1,\Jc_{1,3},\Jc_{1,3},\Jc_{1,3}\big)
\eea
where $P$ is the symmetric permutation matrix
\be
 P\;=\;\mathrm{diag}\big(I_{14\times14},P_5,P_9\big)
\ee
The fusion matrix $N_\mu$ associated to the general module $\mu\in\Ic_f$ 
is polynomial in $X$ and $Y$
\bea 
 &N_{1,1}\;=\;I,\quad N_{1,2}\;=\;Y,\quad N_{1,3}\;=\;Y^2-I,\quad 
    N_{2,1}\;=\;X,\quad N_{2,2}\;=\;XY,\quad N_{2,3}\;=\;X(Y^2-I)\nn
 &N_{1,6}\;=\;\tfrac{1}{2}Y(Y^2-I)(Y^2-3I),\quad 
    N_{2,6}\;=\;\tfrac{1}{2}XY(Y^2-I)(Y^2-3I)\nn
 &N_{4,1}\;=\;\tfrac{1}{2}X(X^2-2I),\quad 
    N_{4,2}\;=\;\tfrac{1}{2}X(X^2-2I)Y  \nn
 &N_{1,3}^{0,1}\;=\;Y(Y^2-I),\quad 
    N_{1,3}^{0,2}\;=\;(Y^2-I)(Y^2-2I)\nn
 &N_{1,6}^{0,1}\;=\;\tfrac{1}{2}Y^2(Y^2-I)(Y^2-3I),\quad 
    N_{1,6}^{0,2}\;=\;\tfrac{1}{2}Y(Y^2-I)(Y^2-2I)(Y^2-3I)\nn
 &N_{2,3}^{0,1}\;=\;XY(Y^2-I),\quad 
    N_{2,3}^{0,2}\;=\;X(Y^2-I)(Y^2-2I)\nn
 &N_{2,6}^{0,1}\;=\;\tfrac{1}{2}XY^2(Y^2-I)(Y^2-3I),\quad 
    N_{2,6}^{0,2}\;=\;\tfrac{1}{2}XY(Y^2-I)(Y^2-2I)(Y^2-3I)\nn
 &N_{2,1}^{1,0}\;=\;X^2,\quad 
    N_{2,2}^{1,0}\;=\;X^2Y,\quad 
    N_{2,3}^{1,0}\;=\;X^2(Y^2-I)  \nn
 &N_{4,1}^{1,0}\;=\;\tfrac{1}{2}X^2(X^2-2I),\quad
    N_{4,2}^{1,0}\;=\;\tfrac{1}{2}X^2(X^2-2I)Y,\quad
    N_{4,3}^{1,0}\;=\;\tfrac{1}{2}X^2(X^2-2I)(Y^2-I)\nn
 &N_{2,3}^{1,1}\;=\;X^2Y(Y^2-I),\quad 
    N_{2,3}^{1,2}\;=\;X^2(Y^2-I)(Y^2-2I)\nn
 &N_{4,3}^{1,1}\;=\;\tfrac{1}{2}X^2(X^2-2I)Y(Y^2-I),\quad 
    N_{4,3}^{1,2}\;=\;\tfrac{1}{2}X^2(X^2-2I)(Y^2-I)(Y^2-2I)
\label{N23}
\eea
where we have introduced the abbreviations $N_{r,s}=N_{\ketw{r,s}}$,
$N_{r,s}=N_{\Wc(\D_{r,s})}$ and
$N_{r,s}^{\al,\beta}=N_{\ketw{\R_{r,s}^{\al,\beta}}}$\!.
The similarity transformation of $N_\mu$ is the block-diagonal matrix $Q^{-1}N_\mu Q$ whose
blocks are upper-triangular matrices. As illustrations of such block-diagonal matrices, 
we here consider
\bea
  Q^{-1}N_{4,2}Q\!\!&=&\!\!\mathrm{diag}\Big(
    4,0,-4,2,-2,\begin{pmatrix} -2&2&0\\ 0&-2&2\\ 0&0&-2\end{pmatrix},
  \begin{pmatrix} 2&-2&0\\ 0&2&-2\\ 0&0&2\end{pmatrix},
  \begin{pmatrix} 0&2&0\\ 0&0&2\\ 0&0&0\end{pmatrix},\nn
 &&\hspace{3cm} \begin{pmatrix} 0&0&0&1&0\\ 0&0&0&0&0\\ 0&0&0&0&0\\ 
                                                             0&0&0&0&1\\ 0&0&0&0&0\end{pmatrix},
   \begin{pmatrix} 0&-1&0\\ 0&0&-1\\ 0&0&0\end{pmatrix}\times 
   \begin{pmatrix} 1&1&0\\ 0&1&1\\ 0&0&1\end{pmatrix}\Big)\nn
  Q^{-1}N_{2,3}^{1,1}Q\!\!&=&\!\!\mathrm{diag}\Big(
   24,0,24,0,0,\begin{pmatrix} 0&8&-12\\ 0&0&8\\ 0&0&0\end{pmatrix},
  \begin{pmatrix} 0&8&-12\\ 0&0&8\\ 0&0&0\end{pmatrix},
  \begin{pmatrix} 0&0&-6\\ 0&0&0\\ 0&0&0\end{pmatrix},\nn
  &&\hspace{5cm}0,0,0,0,0,
  \begin{pmatrix} 0&0&1\\ 0&0&0\\ 0&0&0\end{pmatrix}\times 
  \begin{pmatrix} 0&2&3\\ 0&0&2\\ 0&0&0\end{pmatrix}\Big)
\eea

\subsection{Fusion algebra associated with boundary conditions}

\subsubsection{Auxiliary fusion matrices}
\label{SecAuxWebs}

Due to the the block-diagonal structure (\ref{Yhdiag}) of the auxiliary fusion matrix
$\Yh$, its spectral decomposition
follows readily from the spectral decompositions of $\Cc_{p'}$ and $\Ec_{p'}$ discussed
in Section~\ref{SecCycle} and \ref{SecEye}, respectively. The Jordan canonical form of
$\Yh$ thus consists of $2p-1$ rank-1 blocks of eigenvalue $\beta'_j$ for every $j\in\{0,p'\}$, 
$p$ rank-1 blocks of eigenvalue $\beta'_b$ for every $b\in\mathbb{Z}_{1,p'-1}$, 
$p-1$ rank-2 blocks of eigenvalue $\beta'_b$ for every $b\in\mathbb{Z}_{1,p'-1}$,
and $p$ rank-3 blocks of eigenvalue $\beta'_b$ for every $b\in\mathbb{Z}_{1,p'-1}$.
Likewise, the Jordan canonical form of
$\Xh$ consists of $2p'-1$ rank-1 blocks of eigenvalue $\beta_i$ for every $i\in\{0,p\}$, 
$p'$ rank-1 blocks of eigenvalue $\beta_a$ for every $a\in\mathbb{Z}_{1,p-1}$, 
$p'-1$ rank-2 blocks of eigenvalue $\beta_a$ for every $a\in\mathbb{Z}_{1,p-1}$,
and $p'$ rank-3 blocks of eigenvalue $\beta_a$ for every $a\in\mathbb{Z}_{1,p-1}$.

To characterize the connected components of the Jordan web of the complete set of
common generalized eigenvectors of $\Xh$ and $\Yh$, we choose to work in the 
$\Yh$-favouring basis (\ref{Yhbasis}). A generalized vector $\Gh_{\la,\la'}^{(\ell,\ell')}$ can thus
be written as a $(2p-1)$-dimensional vector whose $p-1$ upper entries are 
$2p'$-dimensional vectors of the type $C$ appearing in Section~\ref{SecCycle},
while the $p$ lower entries are $(4p'-2)$-dimensional vectors of the type $E$
appearing in Section~\ref{SecEye}.

The connected subwebs of the type $\Wh_{\la,\la'}^{(1,1)}$ are given by the following eigenvectors
\be
 \Wh_{\beta_i,\beta'_j}^{(1,1)}:\quad \Gh_{\beta_i,\beta'_j}^{(0,0)}\;=\;\left(\!\!\!\begin{array}{c}  
  f_{1}(\beta_i)C_j\\ \vdots\\ f_{p-1}(\beta_i)C_j
  \\[.15cm] \hline\\[-.3cm]  
  f_{p}(\beta_i)E_j\\ \vdots\\ f_{2p-1}(\beta_i)E_j
  \end{array}\!\!\!\right),\qquad 
 i+j\ \mathrm{even},\quad i\in\{0,p\},\ j\in\{0,p'\}
\label{Whij11}
\ee
or
\be
 \Wh_{\beta_a,\beta'_j}^{(1,1)}:\quad \Gh_{\beta_a,\beta'_j}^{(0,0)}\;=\;\left(\!\!\!\begin{array}{c}  
  f_{1}(\beta_a)C_j\\ \vdots\\ f_{p-1}(\beta_a)C_j
  \\[.15cm] \hline\\[-.3cm]  
  f_{p}(\beta_a)E_j\\ \vdots\\ f_{2p-1}(\beta_a)E_j
  \end{array}\!\!\!\right)\;=\;
 \left(\!\!\!\begin{array}{c}  
  f_{1}(\beta_a)C_j\\ \vdots\\ f_{p-1}(\beta_a)C_j
  \\[.15cm] \hline\\[-.3cm]  
  0\\ \vdots\\ 0
  \end{array}\!\!\!\right),\qquad 
 a+j\ \mathrm{odd},\quad j\in\{0,p'\}
\ee
or
\be
 \Wh_{\beta_i,\beta'_b}^{(1,1)}:\quad \Gh_{\beta_i,\beta'_b}^{(0,0)}\;=\;\left(\!\!\!\begin{array}{c}  
  0\\ \vdots\\ 0
  \\[.15cm] \hline\\[-.3cm]  
  f_{p}(\beta_i)E_b\\ \vdots\\ f_{2p-1}(\beta_i)E_b
  \end{array}\!\!\!\right),\qquad 
 i+b\ \mathrm{odd},\quad i\in\{0,p\}
\ee
The connected subwebs of the type $\Wh_{\la,\la'}^{(1,3)}$ consist of the following 
generalized eigenvectors
\be
 \Wh_{\beta_i,\beta'_b}^{(1,3)}:\quad \Gh_{\beta_i,\beta'_b}^{(0,\ell')}\;=\;\left(\!\!\!\begin{array}{c}  
  f_{1}(\beta_i)C_b^{(\ell'-1)}\\ \vdots\\ f_{p-1}(\beta_i)C_b^{(\ell'-1)}
  \\[.15cm] \hline\\[-.3cm]  
  f_{p}(\beta_i)E_b^{(\ell')}\\ \vdots\\ f_{2p-1}(\beta_i)E_b^{(\ell')}
  \end{array}\!\!\!\right),\qquad 
 i+b\ \mathrm{even},\quad i\in\{0,p\},\quad \ell'\in\mathbb{Z}_{0,2}
\ee
where $C_b^{(-1)}\equiv0$.
The connected subwebs of the type $\Wh_{\la,\la'}^{(3,1)}$ consist of the following 
generalized eigenvectors
\be
 \Wh_{\beta_a,\beta'_j}^{(3,1)}:\quad \Gh_{\beta_a,\beta'_j}^{(\ell,0)}\;=\;\left(\!\!\!\begin{array}{c}  
  \tfrac{1}{\ell!}f_{1}^{(\ell)}(\beta_a)C_j\\ \vdots\\ \tfrac{1}{\ell!}f_{p-1}^{(\ell)}(\beta_a)C_j
  \\[.15cm] \hline\\[-.3cm]  
  \tfrac{1}{\ell!}f_{p}^{(\ell)}(\beta_a)E_j\\ \vdots\\ \tfrac{1}{\ell!}f_{2p-1}^{(\ell)}(\beta_a)E_j
  \end{array}\!\!\!\right),\qquad 
 a+j\ \mathrm{even},\quad j\in\{0,p'\},\quad \ell\in\mathbb{Z}_{0,2}
\ee
The connected subwebs of the type $\Wh_{\la,\la'}^{(1,2)}$ consist of the following 
generalized eigenvectors
\be
 \Wh_{\beta_a,\beta'_b}^{(1,2)}:\quad \Gh_{\beta_a,\beta'_b}^{(0,\ell')}\;=\;
   \left(\!\!\!\begin{array}{c}  
  f_{1}(\beta_a)C_b^{(\ell')}\\ \vdots\\ f_{p-1}(\beta_a)C_b^{(\ell')}
  \\[.15cm] \hline\\[-.3cm]  
  0\\ \vdots\\ 0
  \end{array}\!\!\!\right),\qquad 
 a+b\ \mathrm{odd},\quad \ell'\in\mathbb{Z}_{0,1}
\ee
The connected subwebs of the type $\Wh_{\la,\la'}^{(2,1)}$ consist of the following 
generalized eigenvectors
\be
 \Wh_{\beta_a,\beta'_b}^{(2,1)}:\quad  \Gh_{\beta_a,\beta'_b}^{(\ell,0)}\;=\;
 \left(\!\!\!\begin{array}{c}  
  0\\ \vdots\\ 0
  \\[.15cm] \hline\\[-.3cm]  
  \tfrac{1}{(\ell+1)!}f_{p}^{(\ell+1)}(\beta_a)E_b\\ \vdots\\ 
      \tfrac{1}{(\ell+1)!}f_{2p-1}^{(\ell+1)}(\beta_a)E_b
  \end{array}\!\!\!\right),\qquad 
 a+b\ \mathrm{odd},\quad \ell\in\mathbb{Z}_{0,1}
\ee
Finally, the connected subwebs of the type $\Wh_{\la,\la'}^{(3,3)}$ consist of the following 
generalized eigenvectors 
\be
 \Wh_{\beta_a,\beta'_b}^{(3,3)}:\quad \Gh_{\beta_a,\beta'_b}^{(\ell,\ell')}\;=\;\left(\!\!\!\begin{array}{c}  
  \tfrac{1}{\ell!}f_{1}^{(\ell)}(\beta_a)C_b^{(\ell'-1)}\\ \vdots\\ 
      \tfrac{1}{\ell!}f_{p-1}^{(\ell)}(\beta_a)C_b^{(\ell'-1)}
  \\[.15cm] \hline\\[-.3cm]  
  \tfrac{1}{\ell!}f_{p}^{(\ell)}(\beta_a)E_b^{(\ell')}\\ \vdots\\ 
      \tfrac{1}{\ell!}f_{2p-1}^{(\ell)}(\beta_a)E_b^{(\ell')}
  \end{array}\!\!\!\right),\qquad 
 a+b\ \mathrm{even},\quad \ell,\ell'\in\mathbb{Z}_{0,2},\quad (\ell,\ell')\neq(0,0)
\label{Whab33}
\ee
where $C_b^{(-1)}\equiv0$ as above.
Using properties of the $C$ and $E$ vectors as generalized eigenvectors of 
$\Cc_{p'}$ and $\Ec_{p'}$, together with (\ref{IhE}) and (\ref{CE}), in particular,
it is straightforward to prove that the vectors given in (\ref{Whij11}) through (\ref{Whab33})
indeed correspond to the Jordan webs (\ref{webh13}), (\ref{webh1221}) and (\ref{webh8}) 
consistent with (\ref{XYGh}).
We also note that the number $\Nch^{(\ell,\ell')}$ of connected Jordan webs of the type 
$\Wh^{(\ell,\ell')}$ is given by
\be
 \Nch^{(1,1)}=\;p+p',\quad
 \Nch^{(3,1)}=\;p-1,\quad
 \Nch^{(1,3)}=\;p'-1,\quad
 \Nch^{(2,1)}=\;\Nch^{(1,2)}=\;\Nch^{(3,3)}=\;\tfrac{1}{2}(p-1)(p'-1)
\ee
consistent with the total number (\ref{6pp}) of generalized eigenvectors.

The similarity matrix $\Qh$ appearing in (\ref{QhXhQh}) is constructed by concatenating 
the common generalized eigenvectors of $\Xh$ and $\Yh$ according to any ordering of the 
ordered bases (\ref{Bh}). The permutation matrix $\Ph$
depends on this choice of ordering, and the degree of
convenience of such a choice depends on the intended application. Here we consider 
a general ordering reflecting the partitioning
\be
 \big\{\Bh_{\la,\la'}^{(1,1)}\big\}\cup
 \big\{\Bh_{\la,\la'}^{(1,3)}\big\}\cup
 \big\{\Bh_{\la,\la'}^{(3,1)}\big\}\cup
 \Big(\big\{\Bh_{\la,\la'}^{(1,2)}\big\}\cup\big\{\Bh_{\la,\la'}^{(2,1)}\big\}\Big)\cup
 \big\{\Bh_{\la,\la'}^{(3,3)}\big\}
\label{BBBBBh}
\ee
such that every set (of generalized eigenvectors) of the type $\Bh_{\la,\la'}^{(1,1)}$
comes before every set (of generalized eigenvectors) of the type $\Bh_{\la,\la'}^{(1,3)}$,
and so on. In addition, for every pair $\la,\la'$ in $\{\Bh_{\la,\la'}^{(1,2)}\}$ 
(or equivalently in $\{\Bh_{\la,\la'}^{(2,1)}\}$), the two vectors $\Gh_{\la,\la'}^{(0,0)}$ 
and $\Gh_{\la,\la'}^{(0,1)}$ in $\Bh_{\la,\la'}^{(1,2)}$ are followed immediately by the two 
vectors $\Gh_{\la,\la'}^{(0,0)}$ and $\Gh_{\la,\la'}^{(1,0)}$ in $\Bh_{\la,\la'}^{(2,1)}$.
The Jordan canonical forms $J_{\Xh}$ and  $J_{\Yh}$ in (\ref{QhXhQh}) are then of the form
\bea
 J_{\Xh}\!\!&=&\!\!\mathrm{diag}\Big(
   \underbrace{\la,\ldots}_{p+p'},\,
   \underbrace{\la,\ldots}_{3p'-3},\,
   \underbrace{\Jc_{\la,3},\ldots}_{p-1},\,
   \underbrace{\mathrm{diag}\big(\Jc_{\la,2},\la,\la\big),\ldots}_{\tfrac{1}{2}(p-1)(p'-1)},\,
   \underbrace{\mathrm{diag}\big(\Jc_{\la,2},\Jc_{\la,3},\Jc_{\la,3}\big)
      \ldots}_{\tfrac{1}{2}(p-1)(p'-1)}\Big)
\label{JXh}
\\
 J_{\Yh}\!\!&=&\!\!\mathrm{diag}\Big(
   \underbrace{\la',\ldots}_{p+p'},\,
   \underbrace{\Jc_{\la',3},\ldots}_{p'-1},\,
   \underbrace{\la',\ldots}_{3p-3},\,
   \underbrace{\mathrm{diag}\big(\Jc_{\la',2},\la',\la'\big),\ldots}_{\tfrac{1}{2}(p-1)(p'-1)},\,
   \underbrace{\mathrm{diag}\big(\Jc_{\la',2},\Jc_{\la',3},\Jc_{\la',3}\big)
      \ldots}_{\tfrac{1}{2}(p-1)(p'-1)}\Big)
\eea
We stress that the eigenvalues $\la$ and $\la'$ vary in these expressions but are always of the
form (\ref{betabeta}).
The corresponding permutation matrix $\Ph$ is a block-diagonal matrix whose blocks are 
of dimension 1, 3, 3, 4 or 8, corresponding to the dimensions of the sets 
$\Bh_{\la,\la'}^{(1,1)}$, $\Bh_{\la,\la'}^{(1,3)}$, $\Bh_{\la,\la'}^{(3,1)}$, 
$\Bh_{\la,\la'}^{(1,2)}\cup\Bh_{\la,\la'}^{(2,1)}$ and $\Bh_{\la,\la'}^{(3,3)}$, respectively. 
By a similarity transformation (\ref{QhXhQh}), these $\Ph$-blocks convert the 
blocks in $J_{\Xh}$ into the corresponding upper-triangular matrices $\Xh_{\la,\la'}^{(\ell,\ell')}$ 
in (\ref{Xhweb}) (where $\Xh_{\la,\la'}^{(1,2)}$ and $\Xh_{\la,\la'}^{(2,1)}$ are viewed as the 
single four-dimensional matrix $\mathrm{diag}(\la,\la,\Jc_{\la,2})$). 
The $\Ph$-blocks of dimension 1 or 3 are identity matrices, 
while the $\Ph$-blocks of dimension 4 or 8 are the symmetric permutation matrices
\be
 \Ph_4\;=\;\begin{pmatrix} 0&0&1&0\\ 0&0&0&1\\ 1&0&0&0\\ 0&1&0&0
   \end{pmatrix},\qquad
 \Ph_8\;=\;\begin{pmatrix} 
    0&0&1&0&0&0&0&0\\
    0&0&0&0&0&1&0&0\\
    1&0&0&0&0&0&0&0\\
    0&0&0&1&0&0&0&0\\
    0&0&0&0&0&0&1&0\\
    0&1&0&0&0&0&0&0\\
    0&0&0&0&1&0&0&0\\
    0&0&0&0&0&0&0&1  \end{pmatrix}
\label{Ph4Ph8}
\ee
The symmetric permutation matrix $\Ph$ is thus given by
\be
 \Ph\;=\;\mathrm{diag}\Big(\underbrace{1,\ldots,1}_{4p+4p'-6},\,
   \underbrace{\Ph_4,\ldots,\Ph_4}_{\tfrac{1}{2}(p-1)(p'-1)},\,
   \underbrace{\Ph_8,\ldots,\Ph_8}_{\tfrac{1}{2}(p-1)(p'-1)}\Big)
\ee
Acting on the non-connected Jordan web $\Wh_{\la,\la'}^{(1,2)}\cup\Wh_{\la,\la'}^{(2,1)}$,
the permutation matrix $\Ph_4$ interchanges the two connected components. 
As an action on the connected Jordan webs $\Wh_{\la,\la'}^{(3,3)}$, $\Ph_8$ 
reflects the vertices (generalized eigenvectors) with respect to the line from
south-west to north-east through $\Gh_{\la,\la'}^{(1,1)}$ and $\Gh_{\la,\la'}^{(2,2)}$. 
$\Ph_8$ thus has two fix-points in accord with the two units on the diagonal.

By eliminating the first row and column of the permutation matrices $P_5$ and $P_9$ 
in (\ref{P5P9}), one obtains the permutation matrices $\Ph_4$ and $\Ph_8$, respectively. 
Likewise, the Jordan canonical forms $J_{\Xh}$ and $J_{\Yh}$ follow from the Jordan
canonical forms $J_X$ and $J_Y$ by elimination of the corresponding rows and columns.
Instead of preserving this elimination property, $\Ph_4$ could have been chosen as the 
four-dimensional identity matrix in which case the blocks $\mathrm{diag}(\Jc_{\la,2},\la,\la)$
in (\ref{JXh}) are replaced by $\mathrm{diag}(\la,\la,\Jc_{\la,2})$.

\subsubsection{General fusion matrices}
\label{SecGeneralHat}

Here we determine the upper-triangular block-diagonal
matrix $\Qh^{-1}\Nh_\mu \Qh$ obtained from the general fusion matrix $\Nh_\mu$ by
a similarity transformation with respect to $\Qh$ defined according to (\ref{BBBBBh}).
{}From (\ref{QhNhQh}) and Section~\ref{SecAuxWebs}, we have that 
\bea
 \Qh^{-1}\Nh_\mu \Qh\!\!&=&\!\!\mathrm{diag}\Big(
   \underbrace{g(\Xh_{\la,\la'}^{(1,1)})h(\Yh_{\la,\la'}^{(1,1)}),\ldots}_{p+p'},\,
   \underbrace{g(\Xh_{\la,\la'}^{(1,3)})h(\Yh_{\la,\la'}^{(1,3)}),\ldots}_{p'-1},\,
   \underbrace{g(\Xh_{\la,\la'}^{(3,1)})h(\Yh_{\la,\la'}^{(3,1)}),\ldots}_{p-1},\nn
 &&\qquad
   \underbrace{\mathrm{diag}\big(g(\Xh_{\la,\la'}^{(1,2)})h(\Yh_{\la,\la'}^{(1,2)}),
     g(\Xh_{\la,\la'}^{(2,1)})h(\Yh_{\la,\la'}^{(2,1)})\big),
          \ldots}_{\tfrac{1}{2}(p-1)(p'-1)},\,
   \underbrace{g(\Xh_{\la,\la'}^{(3,3)})h(\Yh_{\la,\la'}^{(3,3)}),\ldots}_{\tfrac{1}{2}(p-1)(p'-1)}\Big)
\eea
where, as before, $g(z)=\mathrm{pol}_{\mu}^{(x)}(z)$ and 
$h(z)=\mathrm{pol}_{\mu}^{(y)}(z)$, and where 
\bea
 g(\Xh_{\la,\la'}^{(1,1)})h(\Yh_{\la,\la'}^{(1,1)})\!\!&=&\!\! g(\la)h(\la')\nn
 g(\Xh_{\la,\la'}^{(1,3)})h(\Yh_{\la,\la'}^{(1,3)})\!\!&=&\!\! g(\la)h(\Jc_{\la',3}),\qquad
 g(\Xh_{\la,\la'}^{(3,1)})h(\Yh_{\la,\la'}^{(3,1)})\;=\;g(\Jc_{\la,3})h(\la')\nn
 g(\Xh_{\la,\la'}^{(1,2)})h(\Yh_{\la,\la'}^{(1,2)})\!\!&=&\!\! g(\la)h(\Jc_{\la',2}),\qquad
 g(\Xh_{\la,\la'}^{(2,1)})h(\Yh_{\la,\la'}^{(2,1)})\;=\;g(\Jc_{\la,2})h(\la')\nn
 g(\Xh_{\la,\la'}^{(3,3)})h(\Yh_{\la,\la'}^{(3,3)})\!\!&=&\!\! \left(\!\!\begin{array}{cccccccc}
  gh&gh'&0&g'h&g'h'&0&\tfrac{1}{2}g''h&\tfrac{1}{2}g''h'\\[3pt]
  0&gh&0&0&g'h&0&0&\tfrac{1}{2}g''h\\[3pt]
  0&0&gh&gh'&\tfrac{1}{2}gh''&g'h&g'h'&\tfrac{1}{2}g'h''\\[3pt]
  0&0&0&gh&gh'&0&g'h&g'h'\\[3pt]
  0&0&0&0&gh&0&0&g'h\\[3pt]
  0&0&0&0&0&gh&gh'&\tfrac{1}{2}gh''\\[3pt]
  0&0&0&0&0&0&gh&gh'\\[3pt]
  0&0&0&0&0&0&0&gh
   \end{array}\!\!\right)
\label{ghhh}
\eea
To simplify the notation, we have used the abbreviations $g=g(\la)$ and $h=h(\la')$.
The eight-dimensional matrix $g(\Xh_{\la,\la'}^{(3,3)})h(\Yh_{\la,\la'}^{(3,3)})$ in (\ref{ghhh})
is obtained from the nine-dimensional matrix $g(X_{\la,\la'}^{(3,3)})h(Y_{\la,\la'}^{(3,3)})$ 
given in (\ref{gh}) by elimination of the first row and column.
This completes the description of the upper-triangular block-diagonal matrix 
$\Qh^{-1}\Nh_\mu \Qh$.

\subsubsection{${\cal W}$-extended critical percolation ${\cal WLM}(2,3)$}

As for $X$ and $Y$, the eigenvalues of $\Xh$ and $\Yh$ are given in (\ref{beta23})
in the case of ${\cal WLM}(2,3)$. 
As displayed in Figure~\ref{web23h}, the connected components 
$\Wh_{\beta_i,\beta'_j}^{(\ell,\ell')}$ of the Jordan web 
associated to the fusion algebra of modules associated with boundary conditions 
are neatly organized with respect to the labels $i$ and $j$.
\psset{unit=1cm}
\begin{figure}
$$
\renewcommand{\arraystretch}{1.5}
\begin{array}{c|cccc}
 \mbox{}_i\diagdown\mbox{}^j&0&1&2&3\\[4pt]
\hline
\rule{0pt}{16pt}
 0&\Wh_{2,2}^{(1,1)}&\Wh_{2,1}^{(1,1)}&\Wh_{2,-1}^{(1,3)}&$\O$
    \\[4pt]
 1&\Wh_{0,2}^{(1,1)}&\Wh_{0,1}^{(3,3)}&\Wh_{0,-1}^{(1,2)}\cup\Wh_{0,-1}^{(2,1)}
    &\Wh_{0,-2}^{(3,1)}
    \\[4pt]
 2&\Wh_{-2,2}^{(1,1)}&\Wh_{-2,1}^{(1,1)}&\Wh_{-2,-1}^{(1,3)}&$\O$
\end{array}
$$
\caption{The connected components $\Wh_{\beta_i,\beta'_j}^{(\ell,\ell')}$ of the Jordan web
associated to the fusion algebra of modules associated with boundary conditions in
${\cal W}$-extended critical percolation ${\cal WLM}(2,3)$. The two \O's
indicate that there are no common generalized eigenvectors corresponding to the pair 
$(\beta_0,\beta'_3)=(2,-2)$ or to the pair $(\beta_2,\beta'_3)=(-2,-2)$ of eigenvalues of the 
auxiliary fusion matrices $\Xh$ and $\Yh$.}
\label{web23h}
\end{figure}
An example of an ordering of the common generalized eigenvectors of $\Xh$ and $\Yh$
respecting (\ref{BBBBBh}) is
\bea
 &\Gh_{2,2}^{(0,0)}; \Gh_{0,2}^{(0,0)}; \Gh_{-2,2}^{(0,0)}; \Gh_{2,1}^{(0,0)}; \Gh_{-2,1}^{(0,0)};
  \Gh_{2,-1}^{(0,0)}, \Gh_{2,-1}^{(0,1)}, \Gh_{2,-1}^{(0,2)};
  \Gh_{-2,-1}^{(0,0)}, \Gh_{-2,-1}^{(0,1)}, \Gh_{-2,-1}^{(0,2)};
  \Gh_{0,-2}^{(0,0)}, \Gh_{0,-2}^{(1,0)}, \Gh_{0,-2}^{(2,0)};\nn
 &\Gh_{0,-1}^{(0,1)}, \Gh_{0,-1}^{(0,2)}, \Gh_{0,-1}^{(1,0)}, \Gh_{0,-1}^{(2,0)};
  \Gh_{0,1}^{(0,1)}, \Gh_{0,1}^{(0,2)}, \Gh_{0,1}^{(1,0)}, \Gh_{0,1}^{(1,1)}, \Gh_{0,1}^{(1,2)},
  \Gh_{0,1}^{(2,0)}, \Gh_{0,1}^{(2,1)}, \Gh_{0,1}^{(2,2)}
\eea
We define the similarity matrix $\Qh$ by concatenating these vectors in the order given. Modulo 
a similarity transformation, $\Qh$ converts $\Xh$ and $\Yh$ into the Jordan canonical forms 
\bea
 &J_{\Xh}\;=\;\Ph^{-1}\Qh^{-1}\Xh\Qh\Ph\;=\;\mathrm{diag}\big(
   2,0,-2,2,-2,2,2,2,-2,-2,-2,\Jc_{0,3},\Jc_{0,2},0,0,\Jc_{0,2},\Jc_{0,3},\Jc_{0,3}\big) \nn
 &J_{\Yh}\;=\;\Qh^{-1}\Yh\Qh\;=\;\mathrm{diag}\big(
   2,2,2,1,1,\Jc_{-1,3},\Jc_{-1,3},-2,-2,-2,\Jc_{-1,2},-1,-1,\Jc_{1,2},\Jc_{1,3},\Jc_{1,3}\big)
\eea
where $\Ph$ is the symmetric permutation matrix
\be
 \Ph\;=\;\mathrm{diag}\big(I_{14\times14},\Ph_4,\Ph_8\big)
\ee
The fusion matrix $\Nh_\mu$ associated to the general module $\mu\in\Ic_b$ 
is polynomial in $\Xh$ and $\Yh$. It is given by the same polynomial as in (\ref{N23})
but as a function of $\Xh,\Yh$ instead of $X,Y$. We recall that the only two modules
in the fundamental fusion algebra
{\em not} associated with boundary conditions are $\ketw{1,1}$ and $\ketw{1,2}$, that is,
\be
 \Ic_f\setminus\Ic_b\;=\;\{\ketw{1,1},\ketw{1,2}\}
\ee
The similarity transformation of $\Nh_\mu$ is the block-diagonal matrix $\Qh^{-1}\Nh_\mu \Qh$
whose blocks are upper-triangular matrices. As illustrations of such block-diagonal matrices, 
we here consider
\bea
  \Qh^{-1}\Nh_{4,2}\Qh\!\!&=&\!\!\mathrm{diag}\Big(
    4,0,-4,2,-2,\begin{pmatrix} -2&2&0\\ 0&-2&2\\ 0&0&-2\end{pmatrix},
  \begin{pmatrix} 2&-2&0\\ 0&2&-2\\ 0&0&2\end{pmatrix},
  \begin{pmatrix} 0&2&0\\ 0&0&2\\ 0&0&0\end{pmatrix},\nn
 &&\hspace{4.5cm} 0,0,\Jc_{0,2},
   \begin{pmatrix} 
    0&0&0&-1&-1&0&0&0\\ 
    0&0&0&0&-1&0&0&0\\
    0&0&0&0&0&-1&-1&0\\
    0&0&0&0&0&0&-1&-1\\
    0&0&0&0&0&0&0&-1\\
    0&0&0&0&0&0&0&0\\
    0&0&0&0&0&0&0&0
    \end{pmatrix}\Big)\nn
  \Qh^{-1}\Nh_{2,3}^{1,1}\Qh\!\!&=&\!\!\mathrm{diag}\Big(
   24,0,24,0,0,\begin{pmatrix} 0&8&-12\\ 0&0&8\\ 0&0&0\end{pmatrix},
  \begin{pmatrix} 0&8&-12\\ 0&0&8\\ 0&0&0\end{pmatrix},
  \begin{pmatrix} 0&0&-6\\ 0&0&0\\ 0&0&0\end{pmatrix},\nn
  &&\hspace{4.5cm}0,0,0,0,
  \begin{pmatrix} 
    0&0&0&-1&-1&0&0&0\\ 
    0&0&0&0&-1&0&0&0\\
    0&0&0&0&0&-1&-1&0\\
    0&0&0&0&0&0&-1&-1\\
    0&0&0&0&0&0&0&-1\\
    0&0&0&0&0&0&0&0\\
    0&0&0&0&0&0&0&0
    \end{pmatrix}\Big)
\eea

\section{Conclusion}
\label{SecConclusion}

We have extended the work~\cite{Ras0908} on ${\cal WLM}(1,p')$ by considering the spectral
decompositions of the regular representations of the graph fusion algebras of the general
${\cal W}$-extended logarithmic minimal model ${\cal WLM}(p,p')$. 
In preparation therefore, we first defined and examined three types of directed and
connected graphs, here called cycle, tadpole and eye-patch graphs.
As in the rational minimal models, the fundamental fusion algebra of ${\cal WLM}(p,p')$
is described by a simple graph fusion algebra. 
The graphs associated with the two fundamental modules consist of a number of
tadpole and eye-patch graphs.
The corresponding adjacency matrices share a complete set of common generalized 
eigenvectors organized as a web. This Jordan web is constructed by interlacing the 
Jordan chains of the two matrices and consists of connected 
subwebs with 1, 3, 5 or 9 generalized eigenvectors.
The similarity matrix, formed by concatenating these vectors, simultaneously brings
the two fundamental adjacency matrices to Jordan canonical form modulo permutation similarity. 
By the same similarity transformation, the general fusion matrices are brought simultaneously to
block-diagonal forms whose blocks are upper-triangular matrices of dimension 1, 3, 5 or 9.
For $p>1$, only some of the modules in the fundamental fusion algebra of ${\cal WLM}(p,p')$ 
are associated with boundary conditions within our lattice approach. 
The regular representation of the corresponding fusion subalgebra has features similar to the
ones in the regular representation of the fundamental fusion algebra, but with dimensions 
of the connected Jordan-web components and upper-triangular blocks given by 1, 2, 3 or 8.
In addition to eye-patch graphs, cycle graphs appear as connected components of the two 
auxiliary fusion matrices obtained from the fundamental fusion matrices by elimination of certain
rows and columns. The general fusion matrices associated with boundary conditions 
are conveniently described in terms of the two auxiliary fusion matrices.
Some of the key results have been illustrated for ${\cal W}$-extended critical percolation 
${\cal WLM}(2,3)$.

There are several natural continuations of this work, all of which we hope to discuss elsewhere.
The first one concerns an algebraic extension of the fundamental fusion algebra of
${\cal WLM}(p,p')$ for $p>1$. It amounts to including all modules arising from fusions
of the complete set of irreducible modules in the model as discussed 
in~\cite{GRW0905,Ras0906,Wood0907}. 

The second continuation concerns the derivation of a generalized Verlinde formula from the
spectral decomposition of the various fusion matrices of ${\cal WLM}(p,p')$. 
This problem was solved in~\cite{Ras0908} for $p=1$.
Other approaches to a Verlinde-like formula for ${\cal WLM}(1,p')$ have been proposed
in~\cite{FHST0306,FK0705,GR0707,GT0711,PRR0907}.
In the case of the so-called {\em projective modules} in ${\cal WLM}(p,p')$
\be
 \big\{\ketw{\kappa p,p'},\ketw{\R_{\kappa p,p'}^{a,0}},\ketw{\R_{p,\kappa p'}^{0,b}},
    \ketw{\R_{\kappa p,p'}^{a,b}};\ \kappa\in\mathbb{Z}_{1,2},\ a\in\mathbb{Z}_{1,p-1},\ 
    b\in\mathbb{Z}_{1,p'-1}
 \big\}
\ee
of which there are $2pp'$~\cite{Ras0805},
the structure of the corresponding Verlinde-like formula~\cite{PR0912} resembles the ordinary
Verlinde formulas. This is intimately related to the observation that the
auxiliary fusion graphs underlying the restrictions of the fundamental matrices $X$ and $Y$
to the projective modules are simply given by $p'$ cycle graphs $\Cc_p$ and $p$ cycle
graphs $\Cc_{p'}$, respcetively. Their spectral decompositions are much simpler than the ones 
considered here as they only involve rank-1 and rank-2 blocks. The two matrices share a 
complete set of ($2pp'$) common generalized eigenvectors with the numbers of connected
Jordan webs given by
\be
 \Nc^{(1,1)}_{\mathrm{proj}}\,=\;2,\qquad
 \Nc^{(1,2)}_{\mathrm{proj}}\,=\;p'-1,\qquad
 \Nc^{(2,1)}_{\mathrm{proj}}\,=\;p-1,\qquad
 \Nc^{(2,2)}_{\mathrm{proj}}\,=\;\tfrac{1}{2}(p-1)(p'-1)
\label{Nproj}
\ee
As in the case of the connected Jordan webs associated with the auxiliary (boundary) fusion
matrices $\Xh$ and $\Yh$, cf. Section~\ref{SecAuxFusMat}, 
the connected Jordan webs (\ref{Nproj}) associated with the 
auxiliary (projective) fusion matrices can be viewed as quotients of the connected Jordan
webs associated with the fundamental fusion matrices $X$ and $Y$.

The third continuation concerns the spectral decomposition of the (matrix) generators
of the Grothendieck ring associated to ${\cal WLM}(p,p')$. For $p=1$, this ring is obtained
by elevating the various character identities to equivalence relations between the corresponding 
generators (modules) of the fusion algebra. For $p>1$, on the other hand, the 
situation is more complicated as also pointed out in~\cite{GRW0905,Wood0907}.
Partition functions only concern characters, not the full-fledged fusion algebra. It thus suffices 
to consider the Grothendieck ring instead of the fusion algebra when discussing partition 
functions. In such circumstances, one is simply not concerned with the reducible yet 
indecomposable module structures, only in their characters. 
Based on spectral decompositions of the regular representation of the Grothendieck ring of 
${\cal WLM}(1,p')$, a Verlinde-like formula was derived in~\cite{PRR0907}.
In~\cite{Ras0908}, a general framework is outlined within which it makes sense to discuss rings 
of equivalence classes of fusion-algebra generators. Together
with the insight we have just gained by studying the graph fusion
algebras and fusion graphs, this may provide the means to classify
Grothendieck-like rings associated to ${\cal WLM}(p,p')$.

%
%
%
\subsection*{Acknowledgments}
\vskip.1cm
\noindent
This work is supported by the Australian Research Council. 
The author thanks Paul A. Pearce for helpful discussions.

\appendix

\section{Commuting matrices and Jordan forms}
\label{AppJordan}

Given two commuting $n$-dimensional matrices, there exists a complete chain of subspaces
$0=M_0\subset M_1\subset\ldots\subset M_n=\mathbb{C}^n$, $\mathrm{dim}(M_j)=j$, such
that $M_j$ is invariant with respect to both matrices for all $j\in\mathbb{Z}_{0,n}$.
This fundamental result on commuting matrices readily extends to all finite
sets of commuting $n$-dimensional matrices, see \cite{GLR06}, for example.
It does not, however, imply that the two matrices share
a complete set of common generalized eigenvectors. Nor does it imply that the two
matrices can be simultaneously brought to Jordan form.
It does, on the other hand, imply that the two matrices can be simultaneously brought
to upper-block-triangular form.

To illustrate that two commuting matrices $A$ and $B$ do not necessarily share a complete
set of generalized eigenvectors, even if they can be simultaneously brought to Jordan form,
we consider
\be
 A\;=\;\begin{pmatrix} \la&1\\ 0&\la\end{pmatrix},\qquad\quad 
 B\;=\;\begin{pmatrix} \la&2\\ 0&\la\end{pmatrix}
\ee
These matrices are already in Jordan form, albeit $B$ not in Jordan canonical form.
The most general Jordan chain associated to $A$ is
\be
 A\begin{pmatrix}a\\ 0\end{pmatrix}\;=\;\la\begin{pmatrix}a\\ 0\end{pmatrix},\qquad
 A\begin{pmatrix}b\\ a\end{pmatrix}\;=\;\la\begin{pmatrix}b\\ a\end{pmatrix}
   +\begin{pmatrix}a\\ 0\end{pmatrix},\qquad a\neq0
\ee
However, since $a\neq0$, the evaluation
\be
 B\begin{pmatrix}b\\ a\end{pmatrix}\;=\;\la\begin{pmatrix}b\\ a\end{pmatrix}
   +2\begin{pmatrix}a\\ 0\end{pmatrix}
\ee
demonstrates that $A$ and $B$ do not share a complete set of generalized eigenvectors.

To illustrate that it is not always possible to bring a pair of commuting matrices $A$ and $B$
simultaneously to Jordan form, even if they share a complete set of generalized eigenvectors,
we consider
\be
 A\;=\;\begin{pmatrix} \la&1&0\\ 0&\la&0\\ 0&0&\la \end{pmatrix},\qquad
 B\;=\;\begin{pmatrix} \la&0&1\\ 0&\la&0\\ 0&0&\la \end{pmatrix},\qquad
 C\;=\;\begin{pmatrix} \la&c&0\\ 0&\la&0\\ 0&0&\la \end{pmatrix},\qquad
 D\;=\;\begin{pmatrix} \la&0&0\\ 0&\la&d\\ 0&0&\la \end{pmatrix}
\ee
Here we have also defined matrices $C$ and $D$, which, for $c,d\neq0$, 
represent general three-dimensional (upper) Jordan forms consisting of a rank-1 and a rank-2 
Jordan block with respect to the single eigenvalue $\la$. The three vectors 
\be
 \begin{pmatrix}1\\ 0\\ 0\end{pmatrix},\qquad
 \begin{pmatrix}0\\ 1\\ 0\end{pmatrix},\qquad
 \begin{pmatrix}0\\ 0\\ 1\end{pmatrix}
\ee
form a complete set of generalized eigenvectors of both $A$ and $B$. The most general 
similarity matrices $S$ and $\Sc$ bringing $A$ to the Jordan form $C$ or $D$
\be
 S^{-1}AS\;=\;C,\qquad \Sc^{-1}A\Sc\;=\;D
\ee
are given by
\be
 S\;=\;\begin{pmatrix} S_{1,1}&S_{1,2}&S_{1,3}\\ 0&cS_{1,1}&0\\ 0&S_{3,2}&S_{3,3}
   \end{pmatrix},\quad S_{1,1},S_{3,3}\neq0;\qquad
 \Sc\;=\;\begin{pmatrix} \Sc_{1,1}&\Sc_{1,2}&\Sc_{1,3}\\ 0&0&d\Sc_{1,2}\\ \Sc_{3,1}&0&\Sc_{3,3}
   \end{pmatrix},\quad \Sc_{1,2},\Sc_{3,1}\neq0
\ee
However, it is readily verified that neither $S^{-1}BS$ nor $\Sc^{-1}B\Sc$ is in 
(upper) Jordan form.

\section{Jordan webs}
\label{AppJordanWeb}

In the tables in Figure~\ref{webOddOdd}, \ref{webOddEven} and \ref{webEvenOdd}, 
we collect the connected Jordan subwebs $W_{\beta_i,\beta'_j}^{(\ell,\ell')}$
formed by the common generalized eigenvectors of the two fundamental fusion matrices 
$X$ and $Y$. There is a table for each of the three possible parity combinations of
$p$ and $p'$. In Figure~\ref{webOddOdd}, $p$ and $p'$ are both odd;
in Figure~\ref{webOddEven}, $p$ is odd and $p'$ is even; while
in Figure~\ref{webEvenOdd}, $p$ is even and $p'$ is odd. 
An {\O} in position $i,j$ indicates that there are no common generalized 
eigenvectors corresponding to the pair $\beta_i,\beta'_{j}$ of eigenvalues of $X$ and $Y$.
For every parity combination, there are exactly two \O's in the corresponding table. 
This reflects the rather trivial observation
\be
 \Nc^{(1,1)}+\Nc^{(1,3)}+\Nc^{(3,1)}+\Nc^{(3,3)^\dagger}+\Nc^{(3,3)}
 \;=\;(p+1)(p'+1)-2
\ee
Similar tables for the connected Jordan subwebs $\Wh_{\beta_i,\beta'_j}^{(\ell,\ell')}$ associated 
to the auxiliary fusion matrices $\Xh$ and $\Yh$ are obtained from the tables in 
Figure~\ref{webOddOdd}, \ref{webOddEven} and \ref{webEvenOdd} by the replacements
\bea
 &W_{\beta_i,\beta'_j}^{(1,1)}\;\to\;\Wh_{\beta_i,\beta'_j}^{(1,1)},\qquad
 W_{\beta_i,\beta'_j}^{(1,3)}\;\to\;\Wh_{\beta_i,\beta'_j}^{(1,3)},\qquad
 W_{\beta_i,\beta'_j}^{(3,1)}\;\to\;\Wh_{\beta_i,\beta'_j}^{(3,1)}\nn
 &W_{\beta_i,\beta'_j}^{(3,3)^\dagger}\;\to\;
    \Wh_{\beta_i,\beta'_j}^{(1,2)}\cup\Wh_{\beta_i,\beta'_j}^{(2,1)},\qquad
 W_{\beta_i,\beta'_j}^{(3,3)}\;\to\;\Wh_{\beta_i,\beta'_j}^{(3,3)}
\eea
\psset{unit=1cm}
\begin{figure}
$$
\renewcommand{\arraystretch}{1.5}
\begin{array}{c||c|ccccccc|c}
 \mbox{}_i\diagdown\mbox{}^j&0&1&2&3&\ldots&p'-3&p'-2&p'-1&p'\\[4pt]
\hline\hline
\rule{0pt}{16pt}
  0&(1,1)&(1,1)&(1,3)&(1,1)&\ldots&(1,3)&(1,1)&(1,3)&$\O$
\\[4pt]
\hline
\rule{0pt}{16pt}
  1&(1,1)&(3,3)&(3,3)^\dagger&(3,3)&\ldots&(3,3)^\dagger&(3,3)&(3,3)^\dagger&(3,1)
\\[4pt]
  2&(3,1)&(3,3)^\dagger&(3,3)&(3,3)^\dagger&\ldots&(3,3)&(3,3)^\dagger&(3,3)&(1,1)
\\[4pt]
  3&(1,1)&(3,3)&(3,3)^\dagger&(3,3)&\ldots&(3,3)^\dagger&(3,3)&(3,3)^\dagger&(3,1)
\\[4pt]
  \vdots&\vdots&\vdots&\vdots&\vdots&&\vdots&\vdots&\vdots&\vdots
\\[4pt]
  p-3&(3,1)&(3,3)^\dagger&(3,3)&(3,3)^\dagger&\ldots&(3,3)&(3,3)^\dagger&(3,3)&(1,1)
\\[4pt]
  p-2&(1,1)&(3,3)&(3,3)^\dagger&(3,3)&\ldots&(3,3)^\dagger&(3,3)&(3,3)^\dagger&(3,1)
\\[4pt]
  p-1&(3,1)&(3,3)^\dagger&(3,3)&(3,3)^\dagger&\ldots&(3,3)&(3,3)^\dagger&(3,3)&(1,1)
\\[4pt]
\hline
\rule{0pt}{16pt}
  p&$\O$&(1,3)&(1,1)&(1,3)&\ldots&(1,1)&(1,3)&(1,1)&(1,1)
\end{array}
$$
\caption{The connected components $W_{\beta_i,\beta'_j}^{(\ell,\ell')}$ of the Jordan web
associated to the fundamental fusion algebra of  ${\cal WLM}(p,p')$ for $p$ and $p'$ 
both odd. Since the eigenvalues $\beta_i,\beta'_j$ of $W_{\beta_i,\beta'_j}^{(\ell,\ell')}$
are given by the location in the table, it suffices to indicate 
the component by the ranks $(\ell,\ell')$.
The two \O's reflect that there are no common generalized eigenvectors corresponding to 
the pairs $\beta_p,\beta'_0$ and $\beta_0,\beta'_{p'}$.}
\label{webOddOdd}
\end{figure}
\psset{unit=1cm}
\begin{figure}
$$
\renewcommand{\arraystretch}{1.5}
\begin{array}{c||c|ccccccc|c}
 \mbox{}_i\diagdown\mbox{}^j&0&1&2&3&\ldots&p'-3&p'-2&p'-1&p'\\[4pt]
\hline\hline
\rule{0pt}{16pt}
  0&(1,1)&(1,1)&(1,3)&(1,1)&\ldots&(1,1)&(1,3)&(1,1)&(1,1)
\\[4pt]
\hline
\rule{0pt}{16pt}
  1&(1,1)&(3,3)&(3,3)^\dagger&(3,3)&\ldots&(3,3)&(3,3)^\dagger&(3,3)&(1,1)
\\[4pt]
  2&(3,1)&(3,3)^\dagger&(3,3)&(3,3)^\dagger&\ldots&(3,3)^\dagger&(3,3)&(3,3)^\dagger&(3,1)
\\[4pt]
  3&(1,1)&(3,3)&(3,3)^\dagger&(3,3)&\ldots&(3,3)&(3,3)^\dagger&(3,3)&(1,1)
\\[4pt]
  \vdots&\vdots&\vdots&\vdots&\vdots&&\vdots&\vdots&\vdots&\vdots
\\[4pt]
  p-3&(3,1)&(3,3)^\dagger&(3,3)&(3,3)^\dagger&\ldots&(3,3)^\dagger&(3,3)&(3,3)^\dagger&(3,1)
\\[4pt]
  p-2&(1,1)&(3,3)&(3,3)^\dagger&(3,3)&\ldots&(3,3)&(3,3)^\dagger&(3,3)&(1,1)
\\[4pt]
  p-1&(3,1)&(3,3)^\dagger&(3,3)&(3,3)^\dagger&\ldots&(3,3)^\dagger&(3,3)&(3,3)^\dagger&(3,1)
\\[4pt]
\hline
\rule{0pt}{16pt}
  p&$\O$&(1,3)&(1,1)&(1,3)&\ldots&(1,3)&(1,1)&(1,3)&$\O$
\end{array}
$$
\caption{The connected components $W_{\beta_i,\beta'_j}^{(\ell,\ell')}$ of the Jordan web
associated to the fundamental fusion algebra of  ${\cal WLM}(p,p')$ for $p$ odd and $p'$ 
even. Since the eigenvalues $\beta_i,\beta'_j$ of $W_{\beta_i,\beta'_j}^{(\ell,\ell')}$
are given by the location in the table, it suffices to indicate 
the component by the ranks $(\ell,\ell')$.
The two \O's reflect that there are no common generalized eigenvectors corresponding to 
the pairs $\beta_p,\beta'_0$ and $\beta_p,\beta'_{p'}$.}
\label{webOddEven}
\end{figure}
\psset{unit=1cm}
\begin{figure}
$$
\renewcommand{\arraystretch}{1.5}
\begin{array}{c||c|ccccccc|c}
 \mbox{}_i\diagdown\mbox{}^j&0&1&2&3&\ldots&p'-3&p'-2&p'-1&p'\\[4pt]
\hline\hline
\rule{0pt}{16pt}
  0&(1,1)&(1,1)&(1,3)&(1,1)&\ldots&(1,3)&(1,1)&(1,3)&$\O$
\\[4pt]
\hline
\rule{0pt}{16pt}
  1&(1,1)&(3,3)&(3,3)^\dagger&(3,3)&\ldots&(3,3)^\dagger&(3,3)&(3,3)^\dagger&(3,1)
\\[4pt]
  2&(3,1)&(3,3)^\dagger&(3,3)&(3,3)^\dagger&\ldots&(3,3)&(3,3)^\dagger&(3,3)&(1,1)
\\[4pt]
  3&(1,1)&(3,3)&(3,3)^\dagger&(3,3)&\ldots&(3,3)^\dagger&(3,3)&(3,3)^\dagger&(3,1)
\\[4pt]
  \vdots&\vdots&\vdots&\vdots&\vdots&&\vdots&\vdots&\vdots&\vdots
\\[4pt]
  p-3&(1,1)&(3,3)&(3,3)^\dagger&(3,3)&\ldots&(3,3)^\dagger&(3,3)&(3,3)^\dagger&(3,1)
\\[4pt]
  p-2&(3,1)&(3,3)^\dagger&(3,3)&(3,3)^\dagger&\ldots&(3,3)&(3,3)^\dagger&(3,3)&(1,1)
\\[4pt]
  p-1&(1,1)&(3,3)&(3,3)^\dagger&(3,3)&\ldots&(3,3)^\dagger&(3,3)&(3,3)^\dagger&(3,1)
\\[4pt]
\hline
\rule{0pt}{16pt}
  p&(1,1)&(1,1)&(1,3)&(1,1)&\ldots&(1,3)&(1,1)&(1,3)&$\O$
\end{array}
$$
\caption{The connected components $W_{\beta_i,\beta'_j}^{(\ell,\ell')}$ of the Jordan web
associated to the fundamental fusion algebra of  ${\cal WLM}(p,p')$ for $p$ even and $p'$ 
odd. Since the eigenvalues $\beta_i,\beta'_j$ of $W_{\beta_i,\beta'_j}^{(\ell,\ell')}$
are given by the location in the table, it suffices to indicate 
the component by the ranks $(\ell,\ell')$.
The two \O's reflect that there are no common generalized eigenvectors corresponding to 
the pairs $\beta_0,\beta'_{p'}$ and $\beta_p,\beta'_{p'}$.}
\label{webEvenOdd}
\end{figure}
%
%


\end{document}